\documentclass[10pt]{article}

\pdfoutput=1

\usepackage{amsmath,amssymb,color,graphicx,longtable,mathrsfs,times,url}
\usepackage[margin=0.7in]{geometry}
\usepackage[round]{natbib}
\newcommand{\cartwidth}{0.5\textwidth}
\newcommand{\eigwidthone}{0.55\textwidth}
\newcommand{\eigwidthtwo}{0.55\textwidth}
\newcommand{\globewidth}{0.7\textwidth}
\newcommand{\globeswidth}{0.7\textwidth}
\newcommand{\recswidth}{0.85\textwidth}
\newcommand{\fglobeswidth}{0.8\textwidth}
\newcommand{\specwidth}{0.4\textwidth}
\newcommand{\strutt}{\rule[0em]{0em}{1.25em}}
\newcommand{\Bfun}{\boldsymbol{B}}

\newcommand{\Cfun}{\boldsymbol{C}}

\newcommand{\Earthradinv}{{r_p^{-1}}}
\newcommand{\Earthrad}{{r_p}}
\newcommand{\Efunvec}{\boldsymbol{\mathcal{E}}}
\newcommand{\Efun}{\boldsymbol{E}}
\newcommand{\EpointsUP}{{\boldsymbol{\mathsf{E}}}_\Uparrow}
\newcommand{\Epoints}{\boldsymbol{\mathsf{E}}}
\newcommand{\Extsignal}{W}
\newcommand{\Ffunvec}{\boldsymbol{\mathcal{F}}}
\newcommand{\Ffun}{\boldsymbol{F}}
\newcommand{\FpointsUP}{{\boldsymbol{\mathsf{F}}}_\Uparrow}
\newcommand{\Fpoints}{\boldsymbol{\mathsf{F}}}
\newcommand{\GVSFs}{\GVSF{}s}
\newcommand{\GVSF}{gradient vector Slepian function}
\newcommand{\Gfunup}{\boldsymbol{\Gfun}}
\newcommand{\Gfunvecup}{\boldsymbol{\mathcal{G}}}
\newcommand{\Gfunvec}{\mathcal{G}}
\newcommand{\Gfun}{G}
\newcommand{\Gmat}{\mathbf{G}}
\newcommand{\GpointsUPJ}{{\boldsymbol{\mathsf{G}}}_{\Uparrow J}}
\newcommand{\Gpoints}{\boldsymbol{\mathsf{G}}}
\newcommand{\Hfunor}{\mathring{\boldsymbol{H}}}
\newcommand{\Hfuno}{\boldsymbol{H}}

\newcommand{\Hfunvec}{\boldsymbol{\mathcal{H}}}

\newcommand{\Imat}{\mathbf{I}}
\newcommand{\InUpelm}{A}
\newcommand{\InUpmat}{\mathbf{\InUpelm}}
\newcommand{\Intsignal}{V}
\newcommand{\Kmat}{\mathbf{K}}
\newcommand{\Lamat}{\mathbf{\Lambda}}
\newcommand{\Lin}{L}
\newcommand{\Lout}{{L_o}}
\newcommand{\Omat}{\mathbf{0}}
\newcommand{\OutGfunup}{\mathring{\Gfunup}}
\newcommand{\OutGfunvecup}{\mathring{\Gfunvecup}}
\newcommand{\OutGfunvec}{\mathring{\Gfunvec}}
\newcommand{\OutGfun}{\mathring{\Gfun}}
\newcommand{\OutGmat}{\mathring{\Gmat}}
\newcommand{\OutGpointsUPJ}{\mathring{\Gpoints}_{\Uparrow \nSlepfun}}

\newcommand{\OutHfunvec}{{\mathring{\Hfunvec}}}

\newcommand{\OutKmat}{\mathring{\Kmat}}
\newcommand{\OutLamat}{\mathring{\Lamat}}
\newcommand{\OutSlepcoefEarth}{\OutSlepcoef}
\newcommand{\OutSlepcoef}{\mathrm{t}}
\newcommand{\OutUpelm}{\breve{A}}
\newcommand{\OutUpmat}{\mathbf{\OutUpelm}}
\newcommand{\Outlambda}{\mathring{\lambda}}
\newcommand{\Outradinv}{{r_q^{-1}}}
\newcommand{\Outrad}{{r_q}}
\newcommand{\OutsKm}{\mathring{\sK}_m}
\newcommand{\Outslepfuncoef}{\mathring{\slepfuncoef}}
\newcommand{\Outvslepfuncoef}{{\mathring{\vslepfuncoef}}}
\newcommand{\Pfun}{\boldsymbol{P}}

\newcommand{\Qmat}{\mathbf{Q}}

\newcommand{\Slepcoef}{\mathrm{s}}
\newcommand{\Tit}{^T}
\newcommand{\Xlm}{X_{lm}}
\newcommand{\Xlpm}{X_{l'm}}
\newcommand{\Yfunvec}{\mathcal{Y}}
\newcommand{\Yfun}{Y}
\newcommand{\aGVSFs}{GVSF{}}
\newcommand{\aGVSF}{GVSF}
\newcommand{\also}{\quad\mbox{and}\quad}
\newcommand{\bch}{\mathbf{\hat{c}}}

\newcommand{\bnabla}{\mbox{\boldmath$\nabla$}}
\newcommand{\bphih}{\mbox{\boldmath$\hat{\phi}$}} 

\newcommand{\brh}{\mathbf{\hat{r}}}

\newcommand{\bthetah}{\mbox{\boldmath$\hat{\theta}$}}

\newcommand{\but}{\quad\mbox{but}\quad}
\newcommand{\dOmega}{\,d\Omega}
\newcommand{\dXlm}{X'_{lm}}
\newcommand{\dXlpm}{X'_{l'm}}
\newcommand{\datavec}{\boldsymbol{d}}
\newcommand{\deltamm}{\delta_{mm'}}
\newcommand{\dimin}{(\Lin+1)^2}
\newcommand{\dimout}{(\Lout+1)^2-1}
\newcommand{\divsinsq}{(\sin \theta)^{-2}}
\newcommand{\dpoints}{\mathsf{d}}
\newcommand{\efnormfactprod}{\sqrt{l(l'+1)(2l+1)(2l'+1)}}
\newcommand{\elements}{\quad\mbox{with elements}\quad}
\newcommand{\enormfactinv}{\frac{1}{\enormfact}}
\newcommand{\enormfactprod}{\sqrt{(l+1)(l'+1)(2l+1)(2l'+1)}}
\newcommand{\enormfact}{\sqrt{(l+1)(2l+1)}}
\newcommand{\estExtsignali}{\hat{\Extsignal}}
\newcommand{\estExtsignal}{\tilde{\Extsignal}}
\newcommand{\estIntsignali}{\hat{\Intsignal}}
\newcommand{\estIntsignal}{\tilde{\Intsignal}}
\newcommand{\estOutSlepcoefi}{\mathrm{\hat{t}}}
\newcommand{\estOutSlepcoef}{\mathrm{\tilde{t}}}
\newcommand{\estSlepcoefi}{\hat{\mathrm{s}}}
\newcommand{\estSlepcoef}{\tilde{\mathrm{s}}}
\newcommand{\estoutsphcoefEarth}{\estoutsphcoef_\Lout^\Outrad}
\newcommand{\estoutsphcoef}{\mathrm{\tilde{w}}}

\newcommand{\estsignal}{\tilde\signal}
\newcommand{\estsphcoefEarth}{\estsphcoef_\Lin^\Earthrad}
\newcommand{\estsphcoef}{\mathrm{\tilde v}}
\newcommand{\fnormfactinv}{\frac{1}{\fnormfact}}
\newcommand{\fnormfactprod}{\sqrt{l\hspace{0.1em}l'(2l+1)(2l'+1)}}
\newcommand{\fnormfact}{\sqrt{l(2l+1)}}
\newcommand{\funT}{\mathcal{T}}
\newcommand{\gpower}{\mathscr{G}}

\newcommand{\hsom}{\hspace{0.1em}}

\newcommand{\intO}{\int_\Omega}
\newcommand{\intR}{\int_\region}
\newcommand{\intoth}{\int_0^\Theta}
\newcommand{\into}{\quad\mbox{into}\quad}
\newcommand{\matT}{\mathrm{T}}

\newcommand{\nSlepfun}{{J}}
\newcommand{\noisepower}{\mathbf{\mathcal{N}}}
\newcommand{\npoints}{k}
\newcommand{\nummods}{k_m}
\newcommand{\orelse}{\quad\mbox{or}\quad}
\newcommand{\orq}{\quad\mbox{or, equivalently,}\quad}
\newcommand{\outSlepcoef}{\mathring{\mathrm{s}}}
\newcommand{\outsphcoefEarthall}{\wsphcoef^\Outrad_{>\Lout}}
\newcommand{\outsphcoefEarth}{\outsphcoef^\Outrad}
\newcommand{\outsphcoef}{\mathrm{w}_\Lout}
\newcommand{\outs}{\mathring{s}}
\newcommand{\phvec}{\boldsymbol{\hat\phi}}
\newcommand{\pl}{\partial}
\newcommand{\pointT}{{\sf T}}
\newcommand{\region}{R}
\newcommand{\renormprod}{\sqrt{l(l+1)l'(l'+1)}}
\newcommand{\resmat}{\boldsymbol{\mathsf{R}}}
\newcommand{\respoints}{\mathsf{r}}
\newcommand{\rit}{\boldsymbol{r}}
\newcommand{\rvec}{\boldsymbol{\hat{r}}}

\newcommand{\sKiom}{\sK^{io}_m}
\newcommand{\sKm}{\sK_m}
\newcommand{\sKom}{\sK^o_m}
\newcommand{\sK}{\mathbf{K}}
\newcommand{\sOutslepfuncoef}{\mathring{\sslepfuncoef}}

\newcommand{\satalt}{{r_s}}
\newcommand{\sestOutSlepcoefi}{\hat t}
\newcommand{\sestOutSlepcoef}{\tilde t}
\newcommand{\sestSlepcoefi}{\hat s}
\newcommand{\sestSlepcoef}{\tilde s}
\newcommand{\signal}{V}
\newcommand{\sigpowerw}{\mathscr{W}}
\newcommand{\sigpower}{\mathscr{V}}
\newcommand{\sindth}{\sin \theta \, d\theta}
\newcommand{\slepfuncoef}{\mathrm{g}}
\newcommand{\soutsphcoefEarth}{\soutsphcoef^\Outrad}
\newcommand{\soutsphcoef}{w}
\newcommand{\sphcoefEarthall}{\vsphcoef^\Earthrad_{>\Lin}}
\newcommand{\sphcoefEarth}{\sphcoef^\Earthrad}
\newcommand{\sphcoef}{\mathrm{v}_\Lin}
\newcommand{\sphshell}{\circledcirc}
\newcommand{\sslepfuncoef}{g}
\newcommand{\ssphcoefEarth}{\ssphcoef^\Earthrad}
\newcommand{\ssphcoef}{v}
\newcommand{\sumlmLin}{\sum_{l=0}^{\Lin}\sum_{m=-l}^l}
\newcommand{\sumlmLoutone}{\sum_{l=1}^{\Lout}\sum_{m=-l}^l}
\newcommand{\sumlminfone}{\sum_{l=1}^{\infty}\sum_{m=-l}^l}
\newcommand{\sumlminf}{\sum_{l=0}^{\infty}\sum_{m=-l}^l}
\newcommand{\thvec}{\boldsymbol{\hat\theta}}
\newcommand{\vecExtsignal}{\bnabla \Extsignal}
\newcommand{\vecIntsignal}{\bnabla \Intsignal}
\newcommand{\vecestExtsignali}{\bnabla \estExtsignali}

\newcommand{\vecestIntsignali}{\bnabla \estIntsignali}

\newcommand{\vecnoise}{\boldsymbol{n}}
\newcommand{\vslepfuncoef}{\mathrm{h}}
\newcommand{\vsphcoef}{\mathrm{v}}
\newcommand{\where}{\quad\mbox{where}\quad}
\newcommand{\with}{\quad\mbox{with}\quad}
\newcommand{\wsphcoef}{\mathrm{w}}
\newcommand{\zerovec}{\boldsymbol{0}}

\onecolumn

\title{Internal and
  external potential-field estimation from regional vector data at
  varying satellite altitude}
\author{Alain Plattner$^1$ and Frederik~J.~Simons$^{2,3}$\\
$^1$ Department of Earth and Environmental Sciences, California State
 University, Fresno,\\ Fresno, CA 93740, USA\\
$^2$ Department of Geosciences, Princeton University, Princeton,
 NJ 08544, USA\\
$^3$ Program in Applied \& Computational
Mathematics, Princeton University,\\ Princeton, NJ 08544, USA}

\begin{document}
\maketitle
\begin{abstract}
When modeling global satellite data to recover a planetary magnetic or
gravitational potential field and evaluate it elsewhere, the method of
choice remains their analysis in terms of spherical harmonics. When
only regional data are available, or when data quality varies strongly
with geographic location, the inversion problem becomes severely
ill-posed. In those cases, adopting explicitly local methods is to be
preferred over adapting global ones (e.g., by regularization). Here,
we develop the theory behind a procedure to invert for planetary
potential fields from vector observations collected within a spatially
bounded region at varying satellite altitude. Our method relies on the
construction of spatiospectrally localized bases of functions that
mitigate the noise amplification caused by downward continuation (from
the satellite altitude to the planetary surface) while balancing the
conflicting demands for spatial concentration and spectral
limitation. The `altitude-cognizant' \GVSFs{} (AC-\aGVSF) were first
employed in a preceding paper.  They enjoy a noise tolerance under
downward continuation that is much improved relative to the
`classical' \GVSFs{} (CL-\aGVSF), which do not factor satellite
altitude into their construction. Furthermore, venturing beyond the
realm of their first application, in the present article we extend the
theory to being able to handle both internal and external
potential-field estimation. Solving simultaneously for internal and
external fields in the same setting of regional data availability
reduces internal-field artifacts introduced by downward-continuing
unmodeled external fields, as we show with numerical examples. We
explain our solution strategies on the basis of analytic expressions
for the behavior of the estimation bias and variance of models for
which signal and noise are uncorrelated, (essentially) space- and
bandlimited, and spectrally (almost) white. The AC-\aGVSF{} are
optimal linear combinations of vector spherical harmonics. Their
construction is not altogether very computationally demanding when the
concentration domains (the regions of spatial concentration) have
circular symmetry, e.g., on spherical caps or rings --- even when the
spherical-harmonic bandwidth is large. Data inversion proceeds by
solving for the expansion coefficients of truncated function
sequences, by least-squares analysis in a reduced-dimensional
space. Hence, our method brings high-resolution regional
potential-field modeling from incomplete and noisy vector-valued
satellite data within reach of contemporary desktop machines.
\end{abstract}


\section{Introduction}

Potential fields such as gravity and magnetic fields provide
indispensable information about planetary or lunar structure and
evolution \cite[]{Kaula68,Lambeck88,Langel+98,Merrill+98}. At the
scale of the globe for Earth and Moon, and more generally for other
planets and their moons, the vast majority of the data is derived from
satellite missions \cite[]{Connerney2015,Wieczorek2015}. Recording
gravity and magnetic fields \textit{in} space is an engineering
problem of instrumentation. Mapping such fields \textit{from} space,
back onto the planetary surface where they originate, and separately
from any fields generated externally, down to the body of interest, is
a problem of inversion \cite[]{Plattner+2015c,Sabaka+2015}. Regional
modeling is predicated on the ability to include data collected at
varying satellite altitude, alleviating noise amplification under
`downward continuation', and, in particular in the case of magnetic
field modeling, taking external fields into account. The full
estimation problem as we consider it here consists in determining
`best' models --- suitable for evaluation at the surface of the
planetary body, and geological interpretation as far as accuracy and
resolution permit --- of an internally generated field noisily
observed at a scattered, areally-limited set of locations taken at
varying satellite altitude, in the presence of an external field.

Beginning with \cite{Gauss1839}, the parameterization of the solution
in terms of global basis functions, spherical \cite[]{Backus+96} or
ellipsoidal \cite[][]{Boelling+2005} harmonics, remains today a
popular practical approach \cite[]{Sneeuw94}. At the other end of the
modeling spectrum are local methods, specifically, those based on
gridded sets of monopoles \cite[e.g.][]{OBrien+94}, equivalent-source
dipoles \cite[e.g.][]{Langel+98}, or point masses
\cite[e.g.][]{Baur+2011}. In-between those extremes of spectral and
spatial selectivity \cite[for a classification,
  see][]{Freeden+99,Freeden+2016} lies a variety of methods that uses
functions such as radial basis functions \cite[e.g.][]{Schmidt+2007},
mascons \cite[e.g.][]{Watkins+2015} and spherical cap harmonics
\cite[e.g.][]{Thebault+2006,Langlais+2010}, spherical-harmonic splines
\cite[e.g.][]{Shure+82,Amirbekyan+2008b} and wavelets
\cite[][]{Holschneider+2003,Mayer+2006,Gerhards2012}. Among the
constructively `spatio-spectrally localized' spherical functions
\cite[e.g.][]{Lesur2006} features the general class of `Slepian
functions' \cite[]{Simons+2006a,Plattner+2014a,Simons+2015} upon which
we build our present work.

Building new bases (or `frames', in a wider sense) by the judiciously
weighted linear combination of spherical harmonics, which most of the
above localization methods have in common, provides a natural way to
respect the harmonicity of the potential fields under study. When the
spherical-harmonic expansion coefficients of a potential field at a
certain altitude are `known', downward continuation to the zero height
of the planetary surface, usually approximated by a sphere, amounts to
a simple reevaluation via multiplication of the coefficients with
factors that depend on the radii of the measurement sphere and the
planet \cite[e.g.][]{Blakely95,Backus+96,Dahlen+98}. In the case of
imperfect knowledge, however, numerical and statistical stability
limit the spatial resolution of the reevaluated fields that can be
obtained in this way, depending on the relative altitude and the
signal-to-noise ratios of the coefficients. Such difficulties are
exacerbated if the source of the uncertainty, fundamentally, lies in
the original data being available over an incomplete portion of the
measurement sphere
\cite[]{Kaula67a,Xu92b,Trampert+96b,Simons+2006b,Schachtschneider+2012}. For
such problems, inversion methods that rely on spherical-harmonics
based localized basis functions confer efficiency and stability,
dimensional reduction, and the overall ease and ability to produce and
downward-continue regional potential-field models with less
statistical \textit{a priori} information or numerical regularization.

Satellite data coverage is far from being always `global'. Coverage
may be only regional, as is the case over Mercury
\cite[][]{Solomon+2001,Solomon+2007}, or data quality may vary due to
spatial variations of signal-to-noise levels or satellite altitude,
rendering a geographical restriction of the area of interest
desirable. Such was the situation for Mars \cite[][]{Albee+2001},
where \cite{Plattner+2015a} selected low-altitude nighttime
magnetic-field data for inversion using the `altitude-cognizant
\GVSFs{}' that are the subject of this paper, resulting in a new
lithospheric magnetic-field model of the Martian South Pole. They
subtracted an external-field model made independently by
\cite{Olsen+2010a} from the data prior to inversion. In the present
paper, we treat the estimation of internally and externally generated
fields as an inverse problem that considers both jointly.

Our method traces its history to the one-dimensional theory of
`prolate spheroidal wave functions' by \cite{Slepian+61}, its
applications in signal processing \cite[]{Slepian83}, and especially
its extensions to scalar spherical fields by \cite{Simons+2006a} and
\cite{Simons+2006b}, to spherical vector fields by
\cite{Plattner+2014a}, and to gradient vector spherical functions
(curl-free potential fields) by \cite{Plattner+2015c}. In the above
cited works, satellite altitude, though explicitly considered within
the context of the inverse problem, was never a factor in the
optimization construction of the Slepian functions, and so we will
term them `canonical' or `classical'. In particular, the functions of
\cite{Plattner+2015c} will hereafter be known as `classical \GVSFs{}'
(CL-\aGVSF). In contrast, the construction by \cite{Plattner+2015a},
reformulated in the present paper, does incorporate satellite altitude
directly, hence their designation `altitude-cognizant \GVSFs{}'
(AC-\aGVSF). The CL-\aGVSF{} solve a spatial (surface) optimization
problem for bandlimited functions, while the AC-\aGVSF{} incorporate
optimization under downward-continuation from satellite altitude.
Using the AC-\aGVSF{} for satellite-data inversion is different than
using the CL-\aGVSF{} basis. In the latter case, vector-field
measurements are first inverted for a best-fitting model at altitude,
and the results are downward-continued afterwards. As
\cite{Plattner+2015c} already noted in their Sections~7.1--7.2, in
that case, the model at the planetary surface is potentially biased by
power in the high spherical-harmonic degrees leaking in through the
downward continuation. This bias is a consequence of using functions
that solely optimize spatial concentration within a given region. The
general method presented here aims at overcoming these issues.

We construct a basis of functions from linear combinations of gradient
vector spherical harmonics by solving an optimization problem that
incorporates the satellite altitude at which the data are acquired. We
present two versions of an inversion method that use different forms
of the altitude-cognizant \GVSFs{} (AC-\aGVSF). In the first method we
assume that external fields are not present, or have been removed from
the data by prior analysis. In our second method, we model external
fields simultaneously while solving for the internal field. Only the
first approach was used by \cite{Plattner+2015a}, and they did not
present a complete mathematical analysis, as we do here. Notation and
preliminary considerations can be found Section~\ref{sec:prelim}. A
statement of the problem that we solve is found in
Section~\ref{sec:problem}. The body of the paper is arranged around
the three questions~`what?', `how?',
and~`why?'. Sections~\ref{sec:innersource}--\ref{numexsect} cover the
question `what' and touch on the question
`how?'. Sections~\ref{sec:statsint} and~\ref{sec:statsintext} answer
the question~`why?'. Finally, the Appendix focuses again on the
question~`how?', in more detail. More precisely, in
Section~\ref{sec:innersource} we present the purely internal-field
AC-\aGVSFs{} that we will use in Section~\ref{sec:innersolution} to
solve for a potential-field model from purely internal-field regional
vector data. Section~\ref{sec:inoutSlep} describes the construction of
internal and external field AC-\aGVSFs{} that we use, in
Section~\ref{sec:intextest}, to solve simultaneously for the internal
and external potential field from regional satellite data with varying
altitude. We test both methods on a simulated data set in
Section~\ref{numexsect} and investigate the effect of neglecting to
account for an external field. In Sections~\ref{sec:statsint}
and~\ref{sec:statsintext} we provide a more in-depth analysis of the
relationship of our new Slepian functions to the generic vector
spherical Slepian functions presented by \cite{Plattner+2014a} and
showcase their mathematical and statistical properties. We summarize
our findings in Section~\ref{sec:conclusions} and explain methods to
significantly decrease the computational costs of high
spherical-harmonic degree calculations in the Appendix.

Compared to other regional methods, the AC-\aGVSF{} approach has the
overall advantage that all calculations happen within a space spanned
by bandlimited spherical-harmonics, the natural basis for source-free
potential fields outside a sphere (Section~\ref{subsscalsph}). Our
method can be interpreted as a computationally tractable approximation
to the truncated singular-value decomposition of the full
spherical-harmonic global problem focused on a chosen region. The
AC-\aGVSF{} are easy to use, computationally efficient, and work with
discrete data collected at varying satellite altitude. A benchmark
comparison test with other methods would be beyond the scope of this
article. Instead we summarize where we discern the main differences
with other regional methods. The popular Revised Spherical Cap
Harmonic Analysis (R-SCHA) method by \cite{Thebault+2006} fits data
using basis functions that solve Laplace's equation inside a cone
covering the chosen region with appropriate boundary conditions. Our
method allows for the separation of internal- and external fields
(with bias, as we show in eqs~\ref{bibiint1}--\ref{bibiext2}), which
can not be readily achieved using R-SCHA
\cite[]{Thebault+2006}. Potential fields obtained using AC-\aGVSF{}
can be expressed in a wavelength-dependent power spectrum, which
appears to not be possible for R-SCHA, at least for non-trivial
boundary conditions \cite[][]{Thebault+2006}. The spherical wavelet
methods by \cite{Mayer+2006} and \cite{Gerhards2011,
  Gerhards2012,Gerhards2014a} provide another powerful regional
approach. To the best of our knowledge these methods assume that the
satellite orbit is of constant altitude, which is not required in our
method. Discrete-source methods such as monopoles
\cite[][]{OBrien+94}, equivalent dipoles \cite[][]{Langel+98}, or
point-mass modeling \cite[][]{Baur+2011} require the assumption of a
known source depth. In special cases this may be justified but in
general, the choice of the source depth requires independent
consideration. By solving for the uniquely constrained potential field
on the planetary surface, instead of the non-uniquely constrained
physical sources themselves, our method avoids this problem.

We assume that the magnetic field that we solve for is static, in that
we do not incorporate a direct time dependence. To avoid temporal
aliasing, the data should be binned into episodic clusters before
inversion and then inverted individually using the same set of
AC-\aGVSFs{}.

\section{Preliminary considerations}
\label{sec:prelim}

In this section we establish the mathematical building blocks and
develop a consistent notation for the development in the rest of the
paper. 

\subsection{Scalar spherical harmonics for potential fields}
\label{subsscalsph}

\begin{figure}\centering
  \includegraphics[width=\cartwidth,angle=0,trim=2.5cm 1.6cm 2.5cm 2.7cm,clip]
                  {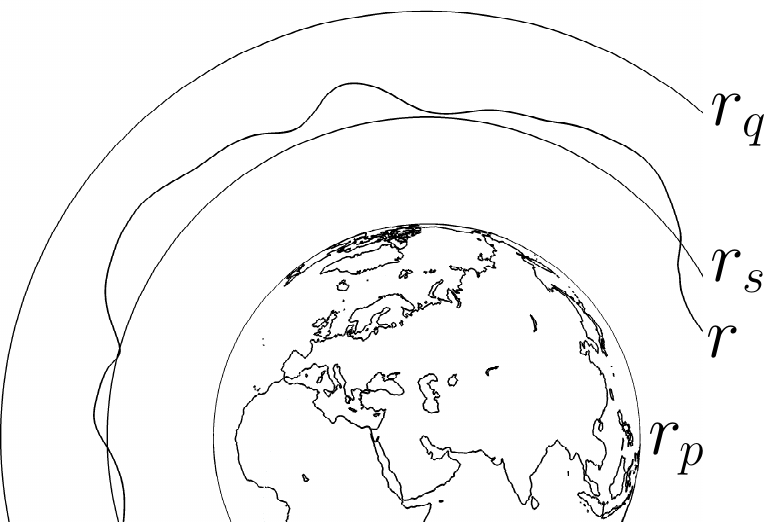}
  \caption{\label{daSituation}Schematic of the geometry considered in
    this paper (not to scale). The variable satellite locations~$r$,
    with a representative value~$\satalt$, lie between the planetary
    surface approximated by the sphere of radius~$\Earthrad$ and an
    outer sphere of radius~$\Outrad$. The spherical shell~$\sphshell$
    between~$\Earthrad$ and~$\Outrad$ is assumed source-free.}
\end{figure}

In a coordinate system originating at the planetary or lunar center we
define a spherical shell $\sphshell = \{\rit \mid \Earthrad \leq
\lVert \rit \rVert=r \leq \Outrad \}$ between an inner sphere with
radius~$\Earthrad$, an approximation of the planetary surface, and an
outer sphere defined by the radius~$\Outrad$, outside of the satellite
orbit~$\satalt$. Fig.~\ref{daSituation} shows the relative location of the
different radial positions considered as discussed in this section and
beyond. Our goal is to `map' a magnetic or gravity measurement made at
an altitude~$r-\Earthrad$ above the planet, $\Earthrad<r<\Outrad$,
onto a potential evaluated on the sphere of radius~$\Earthrad$. In the
absence of field sources within the spherical shell~$\sphshell$, the
only sources lying either within the ball of radius~$\Earthrad$ or
outside of the sphere of radius~$\Outrad$, the true ``full'' field
inside~$\sphshell$ is the superposition of two scalar potentials: an
`internal' field~$\Intsignal$, and an `external'
field~$\Extsignal$. Both fields are `harmonic': they solve the
spherical Laplace equation~\cite[]{Blakely95,Snieder2004,Newman2016}
\begin{equation}\label{scalarLaplaceequation}
\nabla^2 \left[\Intsignal(r\rvec) + \Extsignal(r\rvec)\right]=0,\qquad
\Earthrad<r,\satalt<\Outrad,
\end{equation}
where the usual Laplacian~$\nabla^2 = \pl^2_r + 2r^{-1}\pl_r + r^{-2}
\nabla_1^2$, and surface Laplacian~$\nabla_1^2 = \pl_\theta^2 +
\cot{\theta}\,\pl_\theta + (\sin{\theta})^{-2}\pl_\phi^2$ for
colatitude $0\leq \theta\leq \pi$ and longitude $0\leq \phi < 2\pi$.
For a point $\rvec$ on the surface of the unit sphere~$\Omega=\{\rvec
\mid \lVert \rvec \rVert = 1\}$, \linebreak we define the orthonormal set of
scalar spherical-harmonics as do \cite{Dahlen+98},
\cite{Simons+2006a}, and \linebreak
\cite{Plattner+2014a,Plattner+2015a,Plattner+2015c}. These are given
by
\begin{align}
\label{Y_definition}
\Yfun_{lm}(\rvec)=\Yfun_{lm}(\theta,\phi)&=
\begin{cases}  
\sqrt{2}X_{l \lvert m \rvert}(\theta) \cos m\phi &\text{if } -l\leq m<0,\\
X_{l 0}(\theta)&\text{if } m=0,\\
\sqrt{2}X_{lm}(\theta)\sin m\phi &\text{if } 0<m\leq l,
\end{cases}\\
\label{X_definition}
X_{lm}(\theta)&=(-1)^m\left(\frac{2l+1}{4\pi}\right)^{1/2}
\left[\frac{(l-m)!}{(l+m)!}\right]^{1/2}
\frac{1}{2^ll!}(1-\mu^2)^{m/2}\left(\frac{d}{d\mu}\right)^{l+m}(\mu^2-1)^l
,
\end{align}
where $0\le l\le\infty$ is the angular degree of the spherical
harmonic, and $-l\le m\le l$ its angular order. As shown by
\cite{Backus86} and \cite{Freeden+2009}, among others, the general
solution to eq.~(\ref{scalarLaplaceequation}) involves linear
combinations of the
functions~(\ref{Y_definition})--(\ref{X_definition}), the `inner'
solid harmonics $r^l\Yfun_{lm}$, and the `outer' solid harmonics,
$r^{-l-1}\Yfun_{lm}$. The outer harmonics extinguish at infinity and,
therefore, are a suitable basis for the functions~$\Intsignal$,
generated by internal sources. The inner harmonics vanish at the
center of the coordinate system, and, therefore, form the basis for
the functions~$\Extsignal$, generated by external sources. Inside the
shell~$\sphshell$, the individual fields are represented as
\begin{align}\label{Vrr}
\Intsignal(r\rvec)&=\sumlminf
\left( \frac{r}{\Earthrad}\right)
^{-l-1}\ssphcoefEarth_{lm} \Yfun_{lm}(\rvec),
\with
 \ssphcoefEarth_{lm}=\intO 
 \Intsignal(\Earthrad\rvec) \Yfun_{lm}
(\rvec) \dOmega,
\\
\label{Wrr}
\Extsignal(r\rvec)&=\sumlminfone
\left(\frac{r}{\Outrad}\right)
^{l}\soutsphcoefEarth_{lm} \Yfun_{lm}(\rvec),
\with
 \soutsphcoefEarth_{lm}=\intO 
 \Extsignal(\Outrad\rvec) \Yfun_{lm}
(\rvec) \dOmega.
\end{align}
In eqs~(\ref{Vrr})--(\ref{Wrr}) we selected $\Earthrad$ and $\Outrad$
as reference radii for the coefficients $\ssphcoefEarth_{lm}$ and
$\soutsphcoefEarth_{lm}$, respectively. We collect bandlimited subsets
of the spherical-harmonic coefficients~$\ssphcoefEarth_{lm}$, for the
internal field in a vector~$\sphcoefEarth$, and a set
of~$\soutsphcoefEarth_{lm}$ for the external field in a
vector~$\outsphcoefEarth$,
\begin{equation}\label{crud}
\sphcoefEarth=\begin{pmatrix}
\ssphcoefEarth_{00}&\cdots&\ssphcoefEarth_{lm}&\cdots&\ssphcoefEarth_{\Lin\Lin}\end{pmatrix}\Tit
\also\outsphcoefEarth=\begin{pmatrix}
\soutsphcoefEarth_{1\,-1}&\cdots&\soutsphcoefEarth_{lm}
&\cdots&\soutsphcoefEarth_{\Lout\Lout}\end{pmatrix}\Tit
.
\end{equation} 
We assemble the scalar spherical harmonics~$\Yfun_{lm}$ into an
infinite-dimensional column vector~$\Yfunvec$ that we think of as
consisting of two parts,
\begin{equation}\label{ass}
\Yfunvec=
\begin{pmatrix}
\Yfunvec_\Lin^\funT \quad  \Yfunvec_{>\Lin}^\funT
\end{pmatrix}^\funT
=\begin{pmatrix}
\Yfun_{00} \quad \Yfunvec_\Lout^\funT \quad \Yfunvec_{>\Lout}^\funT
\end{pmatrix}^\funT
.
\end{equation}
As to the bandlimited, $\dimin$ and $[\dimout]$-dimensional portions,
we have
\begin{equation}\label{yfunvec}
\Yfunvec_\Lin=
\begin{pmatrix}\Yfun_{00}&\cdots&\Yfun_{lm}&\cdots&\Yfun_{\Lin\Lin}\end{pmatrix}\Tit
\also
\Yfunvec_\Lout=
\begin{pmatrix}\Yfun_{1\,-1}&\cdots&\Yfun_{lm}&\cdots&\Yfun_{\Lout\Lout}\end{pmatrix}\Tit
.
\end{equation}
Similarly extending eq.~(\ref{crud}) leads to the attractive shorthand
for eqs~(\ref{Vrr})--(\ref{Wrr}) in the form
\begin{align}
\Intsignal(\Earthrad\rvec)&=
\Yfunvec_\Lin^\funT\sphcoefEarth+\Yfunvec_{>\Lin}^\funT\sphcoefEarthall
,\with  
\sphcoefEarth=\intO\Intsignal(\Earthrad\rvec)\,\Yfunvec_\Lin\dOmega,
\\
\Extsignal(\Outrad\rvec)&=\Yfunvec_\Lout^\funT\outsphcoefEarth+\Yfunvec_{>\Lout}^\funT\outsphcoefEarthall
,\with  
\outsphcoefEarth=\intO\Extsignal(\Outrad\rvec)\,\Yfunvec_\Lout\dOmega
.
\end{align}

Finite identity matrices will be subscripted by their dimension,
whereas the unadorned zero matrix will simply be as large as
required. For example, the orthonormality of the spherical harmonics
will be expressed as
\begin{equation}\label{ortho}
\intO\Yfunvec\,\Yfunvec^\funT d\Omega=\Imat,
\qquad
\intO\Yfunvec_\Lin\Yfunvec_\Lin^\funT d\Omega=\Imat_{\dimin},
\also
\intO\Yfunvec_\Lin\Yfunvec_{>\Lin}^\funT d\Omega=\Omat.
\end{equation}

\subsection{Vector spherical harmonics for vector fields}

Many satellite instruments do not measure the scalar potential fields
$\Intsignal$ and $\Extsignal$ directly, but rather the vector field
$\Bfun$ (e.g., the magnetic field or gravitational force) of their
superposition, namely
\begin{equation}\label{superposition}
\Bfun(r\rvec) = \vecIntsignal(r\rvec) + \vecExtsignal(r\rvec).
\end{equation}
Using $\bnabla = \rvec\hsom \pl_r + r^{-1}\bnabla_1$, and
$\bnabla_1= \thvec\hsom\pl_\theta + \phvec \hsom(\sin \theta)^{-1}
\pl_\phi$, we express the gradients of the potentials in
eqs~(\ref{Vrr})--(\ref{Wrr}) in eq.~(\ref{superposition}) as
\begin{align}\label{innersourcegradient}
  \vecIntsignal(r\rvec)&=\sumlminf-\Earthradinv
  \left(\frac{r}{\Earthrad}\right)^{-l-2}
\ssphcoefEarth_{lm}
\Big[\rvec\hsom(l+1)\Yfun_{lm}(\rvec) -\bnabla_1\Yfun_{lm}(\rvec) \Big]
,
\\\label{outersourcegradient}
\vecExtsignal(r\rvec)&=\sumlminfone \Outradinv
\left(\frac{r}{\Outrad}\right)^{l-1}
\soutsphcoefEarth_{lm}
\Big[ \rvec\hsom l\Yfun_{lm}(\rvec) +\bnabla_1\Yfun_{lm}(\rvec)\Big].
\end{align}
In eqs~(\ref{innersourcegradient})--(\ref{outersourcegradient}), the
expressions in the square brackets are vector-valued spherical
harmonic functions. We jointly orthonormalize them over the unit
sphere and name them
\begin{align}\label{gradientvectorsphericalharmonics}
\Efun_{lm}(\rvec)&=\enormfactinv\Big[
  \rvec\hsom(l+1)\Yfun_{lm}(\rvec)-\bnabla_1\Yfun_{lm}(\rvec)
  \Big],\\ \label{bobbly}
\Ffun_{lm}(\rvec)&=\fnormfactinv\Big[ 
  \rvec\hsom l\Yfun_{lm}(\rvec)+\bnabla_1\Yfun_{lm}(\rvec) 
\Big]. 
\end{align}
With the help of
eqs~(\ref{gradientvectorsphericalharmonics})--(\ref{bobbly}) we
rewrite eqs~(\ref{innersourcegradient})--(\ref{outersourcegradient})
succinctly in a form amenable to up- and downward continuation within
the shell~$\sphshell$,
\begin{align}\label{gradv}
 \vecIntsignal(r\rvec) &= 
\sumlminf \InUpelm_l(r)\,\ssphcoefEarth_{lm}\, \Efun_{lm}(\rvec)
\where
\InUpelm_l(r)= -\Earthradinv\enormfact  
\left(\frac{r}{\Earthrad}\right)^{-l-2}
,\\
  \label{gradw}
   \vecExtsignal(r\rvec) &= 
\sumlminfone \OutUpelm_l(r)\,\soutsphcoefEarth_{lm}\,
   \Ffun_{lm}(\rvec)
\where
\OutUpelm_l(r)=\Outradinv\fnormfact
\left(\frac{r}{\Outrad}\right)^{l-1}
.
\end{align}
None of our formulations used `toroidal' vector harmonics, the
$\Cfun_{lm}$ of \cite{Dahlen+98} or \linebreak 
\cite{Plattner+2014a,Plattner+2015c}. These do not result from
potential-field gradients and cannot be analytically continued as the
expressions containing $\Efun_{lm}$ and $\Ffun_{lm}$. Any of their
contributions to our data will be disregarded as unmodelable
components. Where necessary we expand the degree-dependent harmonic
continuation operators $\InUpelm_l(r)$ and $\OutUpelm_l(r)$ to a
`full' form, defining `stretchable' diagonal matrices, whose
dimensions depend on the context,
\begin{align}\label{InUpelements}
\InUpmat(r)&\elements\InUpelm_{lm,l'm'}(r)=
\InUpelm_l(r)\,\delta_{ll'}\delta_{mm'},\\
\label{OutUpelements}
\OutUpmat(r)&\elements\OutUpelm_{lm,l'm'}(r)=\OutUpelm_l(r)\,\delta_{ll'}\delta_{mm'},
\end{align}
and when we suppress their argument, we shall mean $r=\satalt$, using
the `silent' notation
\begin{align}\label{InUpels}
\InUpmat&\elements\InUpelm_{lm,l'm'}=
\InUpelm_l(\satalt)\,\delta_{ll'}\delta_{mm'},\\
\label{OutUpels}
\OutUpmat&\elements\OutUpelm_{lm,l'm'}=\OutUpelm_l(\satalt)\,\delta_{ll'}\delta_{mm'}.
\end{align}

As with eq.~(\ref{ass}), we define the infinite-dimensional
vectors~$\Efunvec$ and~$\Ffunvec$ to comprise all
functions~$\Efun_{lm}$ and~$\Ffun_{lm}$, but partition them into
bandlimited subsets, $\Efunvec_\Lin$ and $\Ffunvec_\Lout$, and their
infinite-dimensional complements, $\Efunvec_{>\Lin}$ and
$\Ffunvec_{>\Lout}$,
\begin{equation}\label{joob}
  \Efunvec=\begin{pmatrix}
    \Efunvec^\funT_\Lin\quad\Efunvec^\funT_{>\Lin}
  \end{pmatrix}^\funT
  \also\Ffunvec=\begin{pmatrix}
  \Ffunvec^\funT_\Lout\quad\Ffunvec^\funT_{>\Lout}
  \end{pmatrix}^\funT
,
\end{equation}
where, as in eq.~(\ref{yfunvec}), we write out the pieces containing
the $\Efun_{lm}$ for $0\leq l \leq \Lin$, and the $\Ffun_{lm}$ for
$1\leq l \leq \Lout$, as follows,
\begin{equation}\label{dotdefine}
  \Efunvec_\Lin=\begin{pmatrix}
  \Efun_{00}&\cdots&\Efun_{lm}&\cdots&\Efun_{\Lin\Lin} \end{pmatrix}\Tit
  \also
  \Ffunvec_\Lout=\begin{pmatrix}
  \Ffun_{1\,-1}&\cdots& 
  \Ffun_{lm}&\cdots&\Ffun_{\Lout\Lout} \end{pmatrix}\Tit.
\end{equation}
We arrive at the useful shorthand for
eqs~(\ref{gradv})--(\ref{gradw}), evaluated at a common radius~$\satalt$, in the form
\begin{align}
\vecIntsignal(\satalt\rvec)&=\Efunvec_\Lin^\funT\InUpmat^\matT\sphcoefEarth+\Efunvec_{>\Lin}^\funT\InUpmat^\matT\sphcoefEarthall,\\
\vecExtsignal(\satalt\rvec)&=\Efunvec_\Lout^\funT\OutUpmat^\matT\outsphcoefEarth+\Ffunvec_{>\Lout}^\funT\OutUpmat^\matT\outsphcoefEarthall
.\end{align}

We use the symbol $\cdot$ to denote the inner products applied to each
element pair of the vector or matrix of vector-valued functions, e.g.,
\begin{equation}\label{exto} 
\Efunvec_\Lin\cdot\Efunvec_\Lin^\funT = \begin{pmatrix} 
\Efun_{00}\cdot\Efun_{00}&\cdots&\Efun_{00}\cdot\Efun_{\Lin\Lin}\\ 
\vdots&&\vdots\\
\Efun_{\Lin\Lin}\cdot\Efun_{00}&\cdots&\Efun_{\Lin\Lin}\cdot\Efun_{\Lin\Lin}
\end{pmatrix}
\also
\Efunvec_\Lin\cdot\Ffunvec_\Lout^\funT = \begin{pmatrix} 
\Efun_{00}\cdot\Ffun_{1\,-1}&\cdots&\Efun_{00}\cdot\Ffun_{\Lout\Lout}\\ 
\vdots&&\vdots\\
\Efun_{\Lin\Lin}\cdot\Ffun_{1\,-1}&\cdots&\Efun_{\Lin\Lin}\cdot\Ffun_{\Lout\Lout}
\end{pmatrix}
.
\end{equation}
and likewise for future combinations of $\Efunvec$ and $\Ffunvec$,
subscripted or not. Hence, the joint orthonormality leads to
expressions of the form
\begin{equation}\label{ortho2}
\intO\Efunvec\cdot\Efunvec^\funT d\Omega=\Imat,
\qquad
\intO\Ffunvec\cdot\Ffunvec^\funT d\Omega=\Imat,
\also
\intO\Efunvec\cdot\Ffunvec^\funT d\Omega=\Omat.
\end{equation}

\section{Statement of the problem}
\label{sec:problem}

When estimating planetary gravity or magnetic fields from sources
within a planet we assume that the spherical shell~$\sphshell$ between
the planetary surface and the upper limit of our satellite data
altitude is free of any field sources. This allows us to
downward-continue the coefficients for the potential fields at~$r$
onto the planetary surface~$\Earthrad$, or upward-continue them to the
outer range of the spherical shell~$\Outrad$, per
eqs~(\ref{gradv})--(\ref{gradw}).

The measured data are a superposition of the fields from internal and
external sources, and contributions collected in a noise term,
\begin{equation}\label{dataonlyinregionout}
\datavec(r\rvec)=\begin{cases}
\vecIntsignal(r\rvec) + \vecExtsignal(r\rvec)
+ \vecnoise(r\rvec) &\text{if } \rvec \in \region\subset\Omega
,\\
\text{unknown} &\text{if } \rvec \in \Omega \setminus \region,
\end{cases}
\end{equation}
Our goal is to obtain, from discrete satellite data collected within a
confined region~$\region$ at varying altitude above the planetary
surface approximated by a sphere of radius~$\Earthrad$, the
bandlimited set of coefficients~$\ssphcoefEarth_{lm}\in\sphcoefEarth$,
$0\le l\le\Lin$, that describe~$\Intsignal$ on the planetary
surface. We will discuss two versions of this problem. In the first,
treated in Section~\ref{sec:innersource}, we remove the external field
from consideration. In the second, discussed in
Section~\ref{sec:inoutSlep}, we solve simultaneously also for the
external-field coefficients that describe the potential~$\Extsignal$,
recovering as best we can the
$\soutsphcoefEarth_{lm}\in\outsphcoefEarth$, $1\le l\le\Lout$. As made
explicit via eq.~(\ref{crud}), we will be forming strictly bandlimited
estimates of the broadband fields $\bnabla V$ and $\bnabla W$ in
eqs~(\ref{gradv})--(\ref{gradw}), up to a certain $\Lin$ and $\Lout$,
not necessarily identical, for the internal and external fields,
respectively. In Sections~\ref{sec:statsint} and \ref{sec:statsintext}
we will be quantifying the broadband bias that is the inevitable
result of such choices.

In our formulation of the problem in eq.~(\ref{dataonlyinregionout}),
we assume static data with any time variation collected in the noise
term. To avoid temporal aliasing the data might be binned by
time stamp, and each bin inverted individually.

\section{Solution by the internal-field altitude-cognizant GVSF method}
\label{sec:innersource}

When we ignore the external field, either because it is deemed
insignificant, or because we modeled and subtracted it from the data,
the term containing~$\Extsignal$ in eq.~(\ref{superposition})
vanishes. For example, \cite{Plattner+2015a} subtracted a model of the
external nighttime field for Mars obtained by \cite{Olsen+2010b} and
assumed no further external-field contribution. In this section we
detail the \cite{Plattner+2015a} construction of a Slepian basis of
functions to serve as an effective lesser-dimensional alternative to
the vector spherical harmonics for inverse modeling of the internal
potential field~$\Intsignal$ from vector data at satellite
altitude. These internal-field altitude-cognizant \GVSFs{}
(AC-\aGVSF{}) take into account the region of data
availability~$\region$, the radial coordinate of data acquisition~$r$,
and the resolvable spherical-harmonic bandwidth~$L$. The present
section is only concerned with the construction of the basis
functions; the next section deals with their usage in realistic data
analysis settings.

\subsection{Restatement of the inverse problem}
\label{sec:restatementInt}

For the construction of the localized function basis we assume that
the data exist as a continuous function on the sphere of
radius~$\satalt$. Avoiding for now the specification of discrete data
locations, the only manner in which the new basis construction depends
on the data is through the boundary of their geographic region of
availability, and the satellite altitude that is representative for
them. For regions with rotational symmetry the construction will then
be computationally very efficient, as we discuss in
Appendix~\ref{app:compint}. Within the region $\region$ the
vector-valued satellite data $\datavec(\satalt\rvec)$ are the sum of
the vector-valued gradient of the unknown scalar-potential internal
field and a noise term, 
\begin{equation}\label{dataonlyinregion}
\datavec(\satalt\rvec)=\begin{cases}
\vecIntsignal(\satalt\rvec) + \vecnoise(\satalt\rvec) &\text{if~}\rvec \in \region,\\
\text{unknown} &\text{if~}\rvec \in \Omega \setminus \region
.
\end{cases}
\end{equation} 
We will be using the `silent' notation whereby
$\datavec=\datavec(\satalt\rvec)$. As is clear from eq.~(\ref{gradv}),
in terms of the unit-sphere harmonics $\Efun_{lm}(\rvec)$, the
coefficients $\ssphcoefEarth_{lm}$ that describe the internal
potential field on the planetary surface~$\Earthrad$ relate to the
ones describing that same field at some average satellite
radius~$\satalt$ via a upward-continuation transformation that we
write in terms of the diagonal matrix~$\InUpmat$ defined in
eq.~(\ref{InUpels}). In the notation developed in
Section~\ref{sec:prelim}, this allows us to rewrite the bandlimited
portion of the vector field that we attempt to recover, for a fixed
average satellite orbital radius~$\satalt$, in terms of the vector
spherical harmonic basis functions $\InUpmat\Efunvec_\Lin$, and with
that, we define a solution as the least-squares minimizer
\begin{equation}\label{inner-sourceoptimizationproblem}
\estsphcoef^\Earthrad_\Lin=\arg\min_{\sphcoefEarth}\left\{ \intR
\left(\Efunvec_\Lin^\funT\InUpmat^\matT\sphcoefEarth-\datavec\right)^2 
d\Omega \right\}.
\end{equation}
We solve eq.~(\ref{inner-sourceoptimizationproblem}) by calculating
the derivative with respect to the coefficient vector~$\sphcoefEarth$
and setting the result equal to zero to obtain
\begin{equation}\label{innerlinearproblem}
\InUpmat
\left(\intR \Efunvec_\Lin \cdot \Efunvec_\Lin^\funT d\Omega\right)
\InUpmat^\matT\estsphcoefEarth
= \InUpmat \intR \Efunvec_\Lin\cdot \datavec \dOmega.
\end{equation} 
In the dot product notation of eq.~(\ref{exto}), 
\begin{equation}\label{pol}
\Efunvec_\Lin\cdot \datavec =\begin{pmatrix}
\Efun_{00}\cdot\datavec&\cdots&
\Efun_{lm}\cdot\datavec&\cdots&
\Efun_{\Lin\Lin}\cdot\datavec \end{pmatrix}\Tit
.
\end{equation}

\subsection{\label{sec:inneraltslepconstruction}A Slepian approach to
  the internal-field problem} 

Eq.~(\ref{innerlinearproblem}) is linear in the data. As did
\cite{Plattner+2015a} we define the symmetric positive-definite matrix
\begin{equation}\label{defKin}
\Kmat=\InUpmat
\left(\intR \Efunvec_\Lin \cdot \Efunvec_\Lin^\funT d\Omega\right)\,
\InUpmat^\matT.
\end{equation}
At large bandwidths~$\Lin$ and for high satellite altitudes~$\satalt$,
the $\dimin\times\dimin$-dimensional matrix $\Kmat$ is poorly
conditioned, and solving the system~(\ref{innerlinearproblem}) via
matrix inversion to obtain~$\estsphcoefEarth$ requires
regularization. Under a Tikhonov numerical scheme we would add a
suitably regular matrix to~$\Kmat$ prior to inversion, which, in a
Bayesian interpretation is akin to supplying a priori information
\cite[]{Aster+2013}. Approaching the problem via singular value
decomposition (SVD), we focus on solving for the well-conditioned
components of problem~(\ref{innerlinearproblem}). Working from the
eigenvector decomposition of~$\Kmat$ we will rewrite the solution in
terms of those eigenvectors that have relatively large eigenvalues,
and then we will solve for their expansion coefficients. When the
region~$\region$ is an arbitrary geographical domain, and when $\Lin$
is large, the eigendecomposition will be costly, but in
Appendix~\ref{app:compint} we show that for regions with special
symmetry, $\Kmat$ can be transformed into a block-diagonal matrix with
a maximal block size of $(\Lin+1)\times(\Lin+1)$, which significantly
reduces its diagonalization cost.

The eigenvectors of the real-valued Hermitian matrix~$\Kmat$ are
orthogonal. We orthonormalize the eigenvectors, hence we write
\begin{equation}\label{eigenprobleminner}
\Kmat\Gmat = \Gmat\Lamat
,\where
\Gmat^\matT\Gmat=\Imat_{\dimin}=
\Gmat\Gmat^\matT
,
\end{equation}
where $\Lamat$ is the diagonal matrix of sorted real-valued
eigenvalues $1>\lambda_1\geq \lambda_2\geq\cdots\geq
\lambda_{\dimin} >0$, and $\Imat$ the identity. The columns of the
$\dimin\times \dimin$-dimensional matrix $\Gmat$ are the
eigenvectors of~$\Kmat$. The matrix $\Gmat_\nSlepfun$ is the
$\dimin\times \nSlepfun$-dimensional restriction to its first
$\nSlepfun$ columns, thus the untruncated $\Gmat=\Gmat_{\dimin}$,
and $\Gmat_{>\nSlepfun}$ will be the complement to
$\Gmat_\nSlepfun$. We note right away that
\begin{equation}\label{ugh}
\Kmat\Gmat_\nSlepfun = \Gmat_\nSlepfun\Lamat_\nSlepfun
,\where
\Gmat_\nSlepfun^\matT\Gmat_\nSlepfun=\Imat_\nSlepfun
,\but
\left(\Gmat_\nSlepfun\Gmat_\nSlepfun^\matT\right)^{n}
=\Gmat_\nSlepfun\Gmat_\nSlepfun^\matT\ne\Imat_{\dimin}
,
\end{equation}
for any positive integer~$n$, is a non-invertible projection except for
$\region=\Omega$. We do note that
$\Gmat_\nSlepfun\Gmat_\nSlepfun^\matT
+\Gmat_{>\nSlepfun}\Gmat_{>\nSlepfun}^\matT =\Imat_{\dimin}$. Each of
the columns of $\Gmat$, which we denote $\slepfuncoef_\alpha$ and
arrange as
\begin{equation}\label{intslepcoef}
\Gmat=\begin{pmatrix} \slepfuncoef_1&\cdots&
\slepfuncoef_\alpha&\cdots &\slepfuncoef_{\dimin} \end{pmatrix}
=
\begin{pmatrix}\Gmat_\nSlepfun\quad \Gmat_{>\nSlepfun}\end{pmatrix}
,\where \Gmat_\nSlepfun=\begin{pmatrix} \slepfuncoef_1&\cdots&
  \slepfuncoef_\alpha&\cdots &\slepfuncoef_\nSlepfun \end{pmatrix} ,
\end{equation}
contains a spherical-harmonic coefficient set
$\slepfuncoef_\alpha =
\begin{pmatrix}
\sslepfuncoef_{00,\alpha} & \cdots &  \sslepfuncoef_{lm,\alpha} & \cdots &
  \sslepfuncoef_{\Lin\Lin,\alpha} 
\end{pmatrix}\Tit$, with a power spectrum or mean-squared value
\begin{equation}\label{intonlypowerspec}
  \gpower_\alpha(l)=\frac{1}{2l+1}\sum_{m=-l}^l \sslepfuncoef_{lm,\alpha}^2.
\end{equation}
In what follows we take pains to distinguish `light' and `bold',
uncapitalized and capitalized, roman ($\slepfuncoef$, $\Gmat$),
italicized ($\sslepfuncoef$, $\Gfun$, $\Gfunup$), calligraphic
($\Gfunvec$, $\Gfunvecup$) or script ($\gpower$) fonts, depending on
whether the quantity of interest is a column vector or a matrix, a
scalar function or a vector function, a column vector of scalar
functions or of vector functions, or a power-spectrum, respectively
--- exactly as in Section~\ref{sec:prelim}, where we had already
encountered $\vsphcoef$, $\InUpmat$, $\ssphcoef$, $\Intsignal$,
$\Efun$, $\Yfunvec$, and $\Efunvec$, for example. We will furthermore
see that~$\sslepfuncoef_{lm,\alpha}\in\slepfuncoef_\alpha\subset\Gmat$,
$\Gfun_\alpha\in\Gfunvec$, $\Gfunup_\alpha\in\Gfunvecup$, and so on.

We use the internal-field altitude-cognizant Slepian transform in three
distinct ways.

[1]~Expanded in scalar spherical harmonics in the coordinates of the
unit sphere, we define an altitude-cognizant scalar Slepian function,
\begin{equation}\label{intonlypotslepfun}
  \Gfun_\alpha(\rvec)=\sumlmLin\sslepfuncoef_{lm,\alpha}
  \Yfun_{lm}(\rvec)
.
\end{equation}
With the help of eq.~(\ref{yfunvec}), we use eqs~(\ref{intslepcoef})
and~(\ref{intonlypotslepfun}) to define the column vector of such
functions~$\Gfun_\alpha(\rvec)=\Yfunvec_\Lin^\funT\slepfuncoef_\alpha$,
\begin{equation}\label{cox}
\Gfunvec_\nSlepfun=\begin{pmatrix}
\Gfun_1&\cdots&\Gfun_\alpha&\cdots&\Gfun_\nSlepfun \end{pmatrix}\Tit
 =
\Gmat^\matT_\nSlepfun\Yfunvec_\Lin
,\with
1\le\nSlepfun\le\dimin
.
\end{equation} 
We partition the full set as
$\Gfunvec=\Gfunvec_{\dimin}=(\Gfunvec_\nSlepfun^\funT\quad
\Gfunvec_{>\nSlepfun}^\funT)^\funT$, whereby
$\Gfunvec^\funT\Gfunvec
= \Gfunvec_{\nSlepfun}^\funT\Gfunvec_{\nSlepfun} +
\Gfunvec_{>\nSlepfun}^\funT\Gfunvec_{>\nSlepfun}=
\Yfunvec^\funT_\Lin\Yfunvec_\Lin$.

[2]~When expanded into the vector spherical-harmonic basis on the
planetary sphere of radius~$\Earthrad$, every eigenvector
$\slepfuncoef_\alpha$ yields an altitude-cognizant \GVSF{}
(AC-\aGVSF{}),
\begin{equation}\label{intonlyvecslepfun}
  \Gfunup_\alpha(\Earthrad\rvec)=\sumlmLin
\InUpelm_l(\Earthrad)\,\sslepfuncoef_{lm,\alpha}\,
  \Efun_{lm}(\rvec)
.
\end{equation}
In the same vein as eq.~(\ref{cox}), using eqs~(\ref{dotdefine}),
(\ref{intslepcoef}) and~(\ref{intonlyvecslepfun}), we write the vector
containing those
functions~$\Gfunup_\alpha(\Earthrad\rvec)=\Efunvec_\Lin^\funT\InUpmat(\Earthrad)\,\slepfuncoef_\alpha$,
\begin{equation}\label{cix}
\Gfunvecup_\nSlepfun=\begin{pmatrix}
\Gfunup_1&\cdots&\Gfunup_\alpha&\cdots&\Gfunup_\nSlepfun \end{pmatrix}\Tit
=
\Gmat^\matT_\nSlepfun\InUpmat(\Earthrad)\,\Efunvec_\Lin
,\with
1\le\nSlepfun\le\dimin
,
\end{equation}
again partitioned as
$\Gfunvecup=\Gfunvecup_{\dimin}=(\Gfunvecup^\funT_\nSlepfun\quad
\Gfunvecup^\funT_{>\nSlepfun})^\funT$.

[3]~At satellite altitude~$\satalt$, we define a vector of
upward-continued AC-\aGVSF{} in the `silent' notation
$\Gfunup_{\uparrow\alpha}(\satalt\rvec)=\Efunvec_\Lin^\funT
\InUpmat\,\slepfuncoef_\alpha$,
\begin{equation}\label{maria}
\Gfunvecup_{\uparrow \nSlepfun} = \begin{pmatrix} 
\Gfunup_{\uparrow 1}&\cdots& 
\Gfunup_{\uparrow \alpha}
& \cdots &\Gfunup_{\uparrow\nSlepfun} 
\end{pmatrix}\Tit
= \Gmat_\nSlepfun^\matT \InUpmat\,\Efunvec_\Lin
,\with
1\le\nSlepfun\le\dimin
.
\end{equation}
We maintain the usual partition
$\Gfunvecup_\uparrow=\Gfunvecup_{\uparrow\dimin}=(\Gfunvecup^\funT_{\uparrow
  \nSlepfun}\quad \Gfunvecup^\funT_{\uparrow >\nSlepfun})^\funT$. 

Whatever the region of concentration~$R$, owing to the orthogonality
of the transformation~(\ref{eigenprobleminner})--(\ref{cox}), the
entire $\nSlepfun=\dimin$-dimensional untruncated set of scalar
Slepian functions remains a complete basis for bandlimited
internal-potential functions. On the planetary surface~$\Earthrad$,
\begin{equation}\label{transfo}
\Yfunvec_\Lin^\funT\sphcoefEarth=
\Yfunvec_\Lin^\funT\left(\Gmat
\Gmat^\matT\right)
\sphcoefEarth
=
\left(\Gmat^\matT\Yfunvec_\Lin\right)^\matT
\left(\Gmat^\matT\sphcoefEarth\right)
=
\Gfunvec^\funT\Slepcoef
.
\end{equation}
The expansion coefficients of a bandlimited potential field at
$\Earthrad$, e.g. $\ssphcoefEarth_{lm}$ in the scalar
spherical-harmonic basis $\Yfun_{lm}(\rvec)$, transform to the
coefficients of a scalar Slepian basis $\Gfun_\alpha(\rvec)$, designed
with the same bandwidth~$\Lin$ but for whichever region~$\region$, as
\begin{align}\label{otherneedit}
s_\alpha&=
\intO 
\Intsignal(\Earthrad\rvec)\,\Gfun_{\alpha}(\rvec)\dOmega
=\sumlmLin\sslepfuncoef_{lm,\alpha}\,\ssphcoefEarth_{lm}
\quad\text{or, in vector form,}\\\label{needit}
\Slepcoef_\nSlepfun&=
\intO 
\Intsignal(\Earthrad\rvec)\,\Gfunvec_{\nSlepfun}(\rvec)\dOmega
=\Gmat_\nSlepfun^\matT\sphcoefEarth
,\with
1\le\nSlepfun\le\dimin
.
\end{align}
The crux of our inversion method will rely on the property that a
$J<\dimin$-dimensional truncated subset of high-eigenvalue Slepian
functions will remain an approximate basis for functions considered
over a confined geographical domain~$R$, if it matches the region for
which the Slepian functions were designed, and for the same
bandlimit~$\Lin$, in the sense that
$\intR(\Gfunvec_\nSlepfun^\funT\Slepcoef_\nSlepfun-\Gfunvec^\funT\Slepcoef)^2\dOmega$
will be small for suitable~$\nSlepfun$.

\begin{figure}\centering
    \includegraphics[width=\globewidth,angle=0,trim=1.0cm 3.6cm 0.2cm 2.5cm,clip]
                    {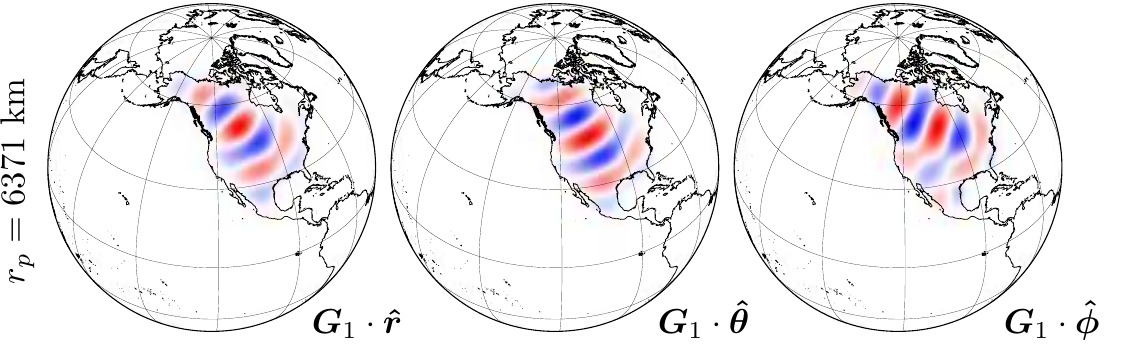}\\[3.5em]
    \includegraphics[width=0.45\textwidth,angle=0,trim= 0.8cm 9.2cm 2.2cm 10.8cm,clip]
                     {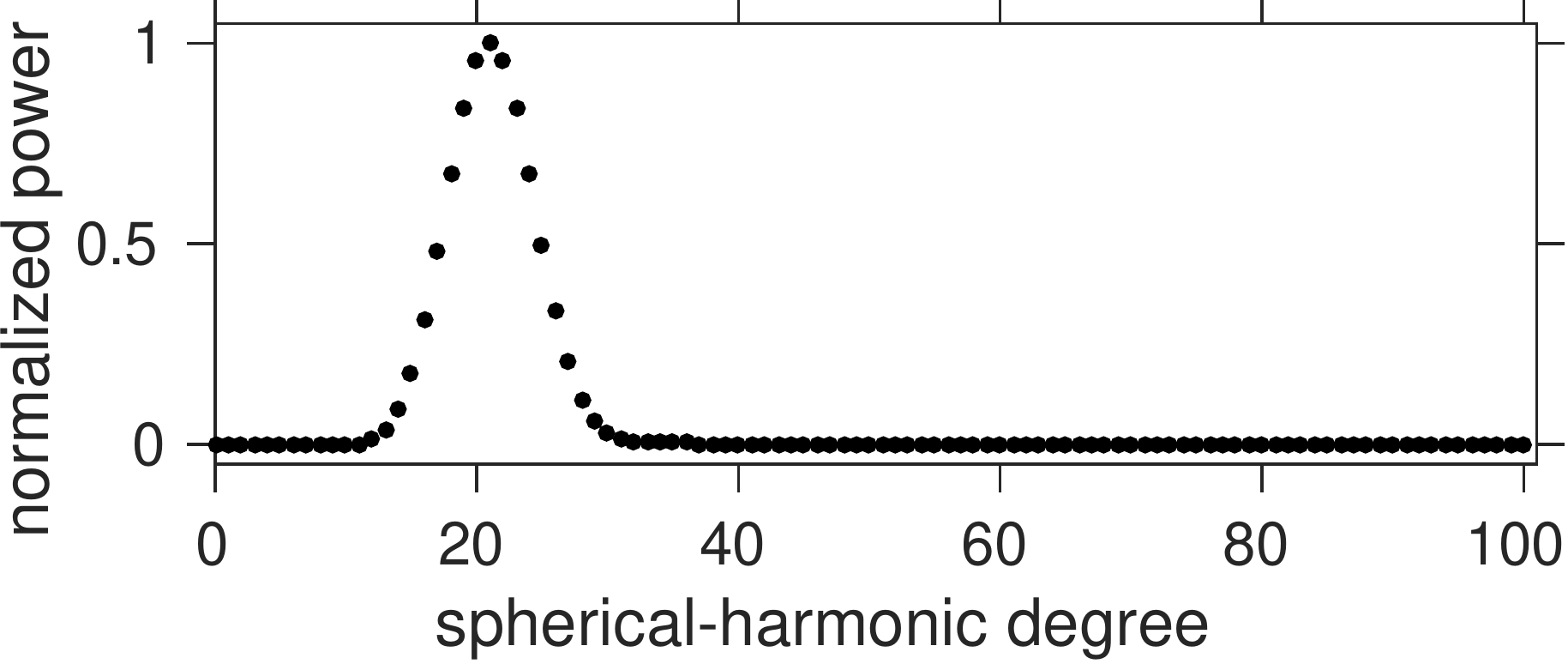}
  \caption{\label{InonlyExamplefunctions1}The best-concentrated
    (highest-eigenvalue) internal-field altitude-cognizant \GVSF{}
    (AC-\aGVSF{}) $\Gfunup_1$, as from eq.~(\ref{intonlyvecslepfun}),
    for region $\region = $ North America, $\Earthrad=6371$~km,
    $\satalt=6671$~km (i.e.~at 300~km satellite altitude), and
    bandwidth $\Lin=100$. Top row: vector (radial, colatitudinal, and
    longitudinal) components of the AC-\aGVSF{} evaluated on the
    planetary surface. Bottom panel: the power spectrum~$\gpower_1$, from
    eq.~(\ref{intonlypowerspec}).}
\end{figure}

Fig.~\ref{InonlyExamplefunctions1} shows the radial, colatitudinal,
and longitudinal components of the highest-eigenvalue internal-field
AC-\aGVSF{} for a planetary radius $\Earthrad=6371$~km, a satellite
radius $\satalt=6671$~km, region $\region=$ North America, and maximum
spherical-harmonic degree $\Lin=100$. Shown are the vector components
$\Gfunup_1\cdot \rvec$, $\Gfunup_1\cdot\thvec$, and $\Gfunup_1\cdot
\phvec$, with their power spectrum~$\gpower_1$. Even though the function
$\Gfunup_1$ has a bandwidth $\Lin=100$, the spatial pattern in
Fig.~\ref{InonlyExamplefunctions1} is consistent with an effective
bandwidth that is much lower, as also seen in the power
spectrum. Herein lies the difference of the AC-\aGVSF{}, first
developed by \cite{Plattner+2015a}, with the `classical' \GVSF
functions (CL-\aGVSF{}) of \cite{Plattner+2015c}, which did not
incorporate the upward-continuation operators $\InUpelm_{lm,l'm'}$ of
eq.~(\ref{InUpelements}) into the optimization solution of
eq.~(\ref{innerlinearproblem}), which now leads to the diagonalization
of the matrix~$\Kmat$ in eq.~(\ref{defKin}). The eigenvector with the
largest eigenvalue in eq.~(\ref{eigenprobleminner}) displays only
small but non-zero coefficients at the highest spherical-harmonic
degrees. Since information at those degrees is more sensitive to noise
amplification under downward-continuation, this was to be expected. It
is furthermore reflected in the power spectrum in
Fig.~\ref{InonlyExamplefunctions1}, where the spherical-harmonic
degrees higher than about $l=40$ are characterized by very low power.

\begin{figure}\centering
   \includegraphics[width=\globewidth,angle=0,trim= 1.0cm 3.6cm 0.2cm 2.5cm,clip]
                   {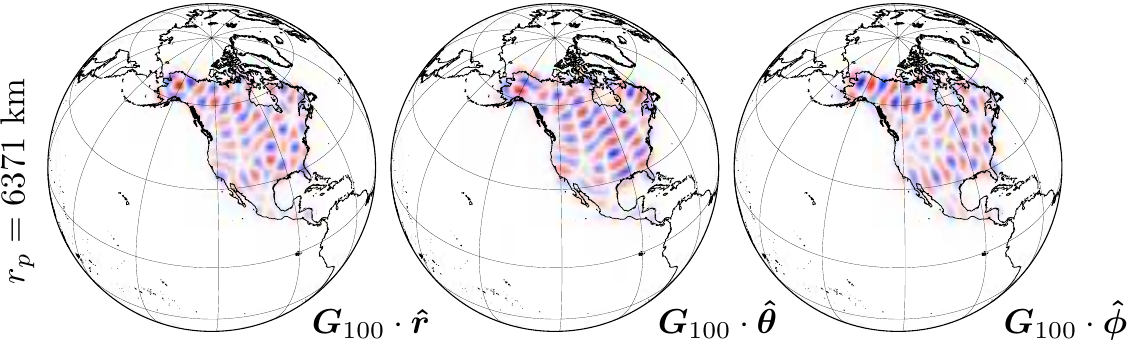}\\[3.5em]
   \includegraphics[width=0.45\textwidth,angle=0,  trim= 0.8cm 9.2cm 2.2cm 10.8cm,clip]
                   {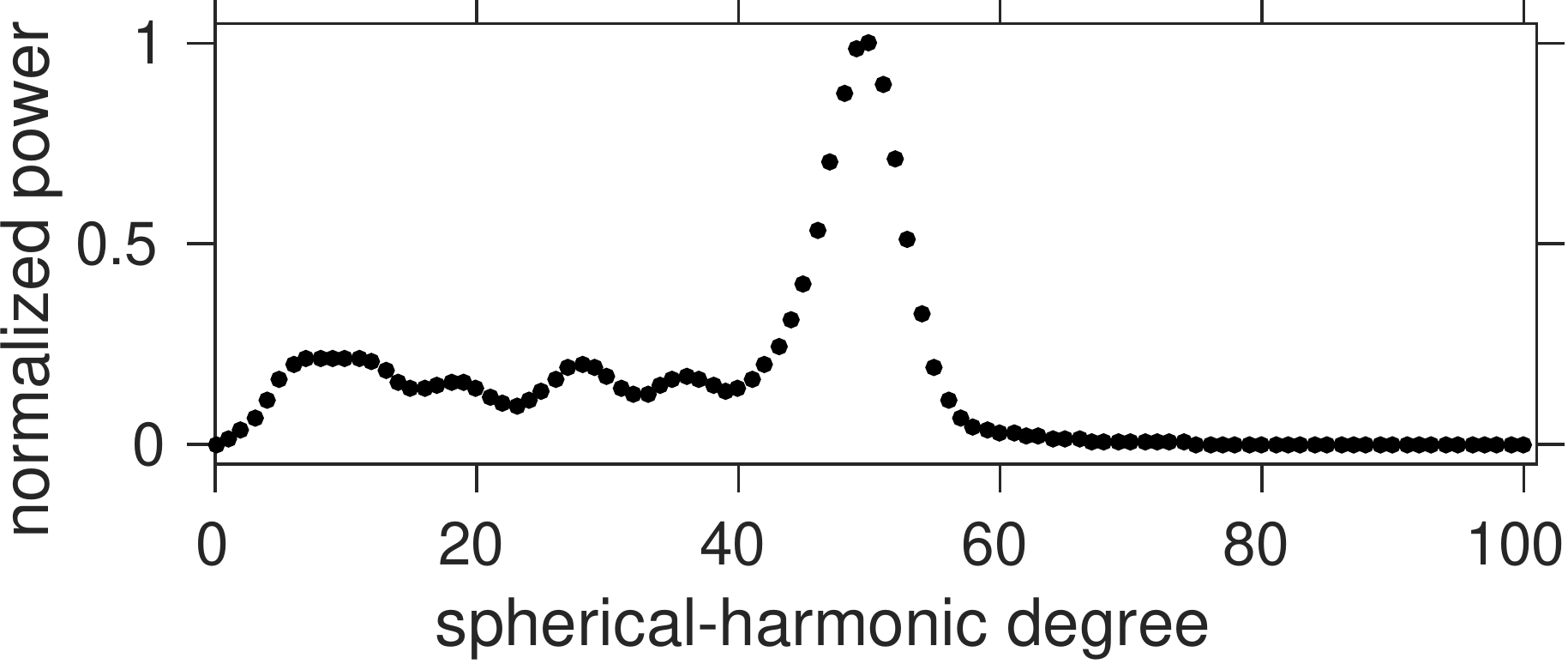}
  \caption{\label{InonlyExamplefunctions2} The 100th-best concentrated
    internal-field AC-\aGVSF{}, $\Gfunup_{100}$, and its power
    spectrum, $\gpower_{100}$, for the same parameters and layout as in
    Fig.~\ref{InonlyExamplefunctions1}.}
\end{figure}

Fig.~\ref{InonlyExamplefunctions2} shows the three components of the
100th best-concentrated internal-field AC-\aGVSF{} for the same
parameters as in Fig.~\ref{InonlyExamplefunctions1}. The function
values of $\Gfunup_{100}$ reveal much finer structures than those of
$\Gfunup_{1}$ in Fig.~\ref{InonlyExamplefunctions1}, and the spectrum
$\gpower_{100}$ manifests more power at the higher spherical-harmonic
degrees than did $\gpower_{1}$ in
Fig.~\ref{InonlyExamplefunctions1}. The increased power at the higher
spherical-harmonic degrees of the lower-eigenvalue AC-\aGVSF{} is in
principle accompanied by the ability to resolve finer spatial
details. However, in the data analysis these functions might also be
more prone than the lower-ranked internal-field AC-\aGVSF{} to fitting
noise rather than signal.

\subsection{\label{sec:innersolutioncont}Continuous solution
by internal-field altitude-cognizant \aGVSF{}}

With the help of eqs~(\ref{defKin})--(\ref{eigenprobleminner}) and the
orthogonality of the matrix~$\Gmat$, the
problem~(\ref{innerlinearproblem}) is rewritten as
\begin{equation}\label{linearsysteminner}
\Kmat\,\estsphcoefEarth=
\Gmat\Lamat\Gmat^\matT\estsphcoefEarth=
\InUpmat\intR\Efunvec_\Lin\cdot\datavec\dOmega. 
\end{equation}
We implement the truncated-SVD approach in using the first
$\nSlepfun$ columns of the matrix~$\Gmat$, hence, using the
formalism of eqs~(\ref{ugh}) and~(\ref{needit}),
\begin{equation}\label{linearsysteminnerregularized}
\Gmat_\nSlepfun\Lamat_\nSlepfun\,\estSlepcoef_{\nSlepfun} =
\InUpmat \intR \Efunvec_\Lin\cdot \datavec \dOmega
,
\end{equation}
where $\Gmat_\nSlepfun$ is as in eq.~(\ref{intslepcoef}), and
$\Lamat_J$ is the diagonal matrix consisting of the $\nSlepfun$
largest eigenvalues $\lambda_1\geq\cdots\geq \lambda_{\nSlepfun} >0$
of $\Kmat$. The new version, eq.~(\ref{linearsysteminnerregularized}),
is not identical to the original problem,
eq.~(\ref{linearsysteminner}), given eq.~(\ref{ugh}), and
$\estSlepcoef_{\nSlepfun}$ is a truncated Slepian-transform
estimator. The value for the regularization parameter $\nSlepfun$
remains to be chosen. If we select $\nSlepfun$ such that all
eigenvalues~$\lambda_\alpha$ are similar in magnitude, the
system~(\ref{linearsysteminnerregularized}) will be well conditioned,
and can be solved by the left-inverse of
$\Gmat_\nSlepfun\Lamat_\nSlepfun$, which is
$\Lamat^{-1}\Gmat_\nSlepfun^\matT$, per eq.~(\ref{ugh}). This defines
a solution, using eq.~(\ref{maria}),
\begin{equation}\label{analyticalsolutioninner}
\estSlepcoef_{\nSlepfun} =  \Lamat_\nSlepfun^{-1} \intR
\Gfunvecup_{\uparrow \nSlepfun}\cdot\datavec \dOmega
.
\end{equation}

The $\nSlepfun$-dimensional vector of coefficients
$\estSlepcoef_{\nSlepfun}$ can be back-projected into the
$\dimin$-dimensional space of internal-field vector
spherical-harmonic coefficients by multiplying it with the
$\dimin\times\nSlepfun$-dimensional matrix $\Gmat_\nSlepfun$. We
can expand the estimated potential field on the planetary surface from
this back-projection, with the help of eq.~(\ref{cox}), to form the
space-domain estimate
\begin{equation}\label{slepianfunctionexpansioninner}
\estIntsignal_\nSlepfun(\Earthrad\rvec)=
\sum_{\alpha=1}^\nSlepfun
\sestSlepcoef_\alpha\,\Gfun_\alpha(\rvec)=
 \Yfunvec_\Lin^\funT \Gmat_\nSlepfun\, \estSlepcoef_{\nSlepfun} 
= \Gfunvec_J^\funT\, \estSlepcoef_{\nSlepfun}
.
\end{equation}
We note emphatically that the estimator in
eq.~(\ref{slepianfunctionexpansioninner}) is different than the one
proposed by \cite{Plattner+2015c}, their eq.~(161), and also briefly
discussed by \cite{Plattner+2015a} in their Section~2.2. It is,
however, the estimator used by \cite{Plattner+2015a} and discussed in
their Section~2.3. The current paper contains the full rationale
behind their doing so. The key to the difference is that we use
eigenfunctions of eq.~(\ref{defKin}), which takes the effects of the
altitude of the observation into account at the optimization stage.

In this section we described the construction of bandlimited
internal-field AC-\aGVSF{}, concentrated in a certain region and
optimized for a representative satellite altitude. In the following
section we expand our method to being able to consider both internal-
and external fields.

\section{Solution by the full-field altitude-cognizant GVSF method}
\label{sec:inoutSlep}

Modeling both the internal and external fields, i.e, the ``full''
field from both $\Intsignal$ and $\Extsignal$ in
eq.~(\ref{superposition}), with a similar spatio-spectrally localized
inversion approach as described in Section~\ref{sec:innersource},
requires Slepian functions that contain both internal-field and
external-field components.

\subsection{Restatement of the inverse problem}

As for eq.~(\ref{dataonlyinregion}) we assume that the data are a
linear combination of a modeled component (signal) and noise, with the
signal now containing both the internally and externally generated vector
fields,
\begin{equation}\label{dataregionintext}
\datavec(\satalt\rvec)=\begin{cases} \vecIntsignal(\satalt\rvec) +
\vecExtsignal(\satalt\rvec) + \vecnoise(\satalt\rvec) &\text{if~} \rvec \in
\region,\\ \text{unknown} &\text{if~} \rvec \in \Omega \setminus
\region.
\end{cases}
\end{equation} 
Complementing the matrix~$\InUpmat$ that serves to upward-continue the
internal-field spherical harmonics, eq.~(\ref{InUpels}), we use the
external-field matrix~$\OutUpmat$ from eq.~(\ref{OutUpels}) to augment
eq.~(\ref{inner-sourceoptimizationproblem}) to take both fields into
account. The internal field is expanded as a linear combination of
upward-continued internal-field vector spherical harmonics
$\InUpmat\Efunvec_\Lin$, with coefficients $\sphcoefEarth$. The
external field is a linear combination of upward-continued
external-field vector spherical harmonics $\OutUpmat\Ffunvec_\Lout$,
with coefficients $\outsphcoefEarth$. The external-field bandwidth,
$\Lout$, can be different from the internal-field
bandwidth,~$\Lin$. We seek a least-squares solution
\begin{equation}\label{inner-outer-sourceoptimizationproblem}
  \begin{pmatrix}
    \displaystyle\estsphcoefEarth\\
\\
    \displaystyle\estoutsphcoefEarth
  \end{pmatrix}
=
\arg\min_{\sphcoefEarth,\outsphcoefEarth}\left\{ \intR
\left(\Efunvec_\Lin^\funT \InUpmat^\matT \sphcoefEarth +
\Ffunvec_\Lout^\funT \OutUpmat^\matT \outsphcoefEarth - \datavec\right)^2 \dOmega
\right\}.
\end{equation}
We solve optimization
problem~(\ref{inner-outer-sourceoptimizationproblem}) by taking its
derivative with respect to the vector $\begin{pmatrix}
  {\sphcoefEarth}^\matT& {\outsphcoefEarth}^\matT \end{pmatrix}^\matT$
and setting it to zero. This yields
\begin{equation}\label{intextanalyticsystem}
  \begin{pmatrix}
    \displaystyle\InUpmat\intR \Efunvec_\Lin
    \cdot \Efunvec_\Lin^\funT
    \dOmega\, \InUpmat^\matT&
    \displaystyle\InUpmat\intR \Efunvec_\Lin
    \cdot \Ffunvec_\Lout^\funT \dOmega\,
    \OutUpmat^\matT\\
    &\\
    \displaystyle \OutUpmat \intR \Ffunvec_\Lout
    \cdot \Efunvec_\Lin^\funT
    \dOmega \,\InUpmat^\matT&
    \displaystyle \OutUpmat \intR \Ffunvec_\Lout
    \cdot \Ffunvec_\Lout^\funT
    \dOmega \,\OutUpmat^\matT
  \end{pmatrix}
  \begin{pmatrix}
    \displaystyle\estsphcoefEarth\\
\\
    \displaystyle\estoutsphcoefEarth
  \end{pmatrix}
=
\begin{pmatrix}
\displaystyle\InUpmat\intR\Efunvec_\Lin\cdot\datavec\dOmega\\
\\
\displaystyle\OutUpmat\intR\Ffunvec_\Lout\cdot\datavec\dOmega
\end{pmatrix}
.
\end{equation}
Substituting $\Ffunvec_\Lout$ for $\Efunvec_\Lin$, the same dot-product notation
as in eqs~(\ref{exto}) and~(\ref{pol}) is used.

\subsection{\label{sec:Slepbasintext}A Slepian approach to the full-field problem} 
As with eq.~(\ref{defKin}) we proceed to regularizing the poorly
conditioned linear system~(\ref{intextanalyticsystem}) by
diagonalization of the combined kernel matrix
\begin{equation}\label{kernel_inout}
  \OutKmat = \begin{pmatrix}
    \displaystyle \InUpmat \intR \Efunvec_\Lin \cdot \Efunvec_\Lin^\funT  \dOmega \,\InUpmat^\matT&
    \displaystyle \InUpmat \intR \Efunvec_\Lin \cdot \Ffunvec_\Lout^\funT  \dOmega\,\OutUpmat^\matT\\
    &\\
    \displaystyle \OutUpmat \intR \Ffunvec_\Lout \cdot \Efunvec_\Lin^\funT \dOmega \,\InUpmat^\matT&
    \displaystyle \OutUpmat \intR \Ffunvec_\Lout \cdot \Ffunvec_\Lout^\funT \dOmega\,\OutUpmat^\matT
  \end{pmatrix},
\end{equation}
which is square, of dimension
$[\dimin+\dimout]\times[\dimin+\dimout]$, and generally fully
populated. In Appendix~\ref{appintext} we show that for symmetric
regions the columns and rows of $\OutKmat$ can be reordered to a
block-diagonal form with the largest block dimension
$(\Lin+\Lout+1)\times(\Lin+\Lout+1)$. In the orthogonal eigenvector
decomposition of this Hermitian positive definite matrix,
\begin{equation}\label{eigenprobleminnerouter}
\OutKmat\OutGmat=\OutGmat\OutLamat,
\where
\OutGmat^\matT\OutGmat=\Imat_{\dimin+\dimout}=\OutGmat\OutGmat^\matT,
\end{equation}
the diagonal matrix of eigenvalues $\Outlambda_1\geq
\Outlambda_2\geq\cdots\geq \Outlambda_{\dimin+\dimout} >0$
is $\OutLamat$, and $\OutGmat$ contains the eigenvectors
$\Outslepfuncoef_\alpha$ in the arrangement
\begin{equation}\label{bo}
\OutGmat=\begin{pmatrix} \Outslepfuncoef_1&\cdots&
\Outslepfuncoef_\alpha&\cdots
&\Outslepfuncoef_{\dimin+\dimout} \end{pmatrix}
=\begin{pmatrix} 
\OutGmat_\nSlepfun\quad\OutGmat_{>\nSlepfun}
\end{pmatrix}
\where
\OutGmat_\nSlepfun=\begin{pmatrix} \Outslepfuncoef_1&\cdots&
\Outslepfuncoef_\alpha&\cdots
&\Outslepfuncoef_\nSlepfun \end{pmatrix}.
\end{equation}
Again it is to be noted that the relations involving the column
restrictions are, for $1\le\nSlepfun\le\dimin+\dimout$ and
any~$n\in\mathbb{N}$,
\begin{equation}\label{ugh2}
\OutKmat\OutGmat_\nSlepfun = \OutGmat_\nSlepfun\OutLamat_\nSlepfun
,\where
\OutGmat_\nSlepfun^\matT\OutGmat_\nSlepfun=\Imat_\nSlepfun
,\but
\left(\OutGmat_\nSlepfun\OutGmat_\nSlepfun^\matT\right)^{n}
=\OutGmat_\nSlepfun\OutGmat_\nSlepfun^\matT\ne\Imat_{\dimin+\dimout}
.
\end{equation}

Each of the column vectors in $\OutGmat$ contains coefficients for the
internal and the external fields, and
$\OutGmat=\OutGmat_{\dimin+\dimout}$,\linebreak to avoid notational
clutter. The first $\dimin$ coefficients of each vector
$\Outslepfuncoef_\alpha$ multiply the internal-field \linebreak vector
spherical harmonics~$\Efunvec_\Lin$ while the last $\dimout$
coefficients expand the external-field \linebreak vector
harmonics~$\Ffunvec_\Lout$. Thus, each vector~$\Outslepfuncoef_\alpha$
decomposes into internal and external parts\linebreak
$\Outslepfuncoef_{\alpha} = \begin{pmatrix}
  \sOutslepfuncoef_{i\,00,\alpha} & \cdots &
  \sOutslepfuncoef_{i\,lm,\alpha} & \cdots &
  \sOutslepfuncoef_{i\,\Lin\Lin,\alpha} &
  \sOutslepfuncoef_{o\,1\,-1,\alpha} & \cdots &
  \sOutslepfuncoef_{o\,lm,\alpha} & \cdots &
  \sOutslepfuncoef_{o\,\Lout\Lout,\alpha} \end{pmatrix}\Tit
= \begin{pmatrix} \Outslepfuncoef_{i\alpha}^\matT &
  \Outslepfuncoef_{o\alpha}^\matT \end{pmatrix}^\matT$. We
baptize~$\OutGmat_i$ the matrix with the first $\dimin$,
and~$\OutGmat_o$ the matrix with the last $\dimout$ rows
of~$\OutGmat$, possibly restricted to their first~$\nSlepfun$ columns,
as follows,
\begin{equation}\label{OutGmatstructure}
\OutGmat=\begin{pmatrix}\OutGmat_i\\
\OutGmat_o\end{pmatrix}
\orelse
\OutGmat_\nSlepfun=\begin{pmatrix}\OutGmat_{i\nSlepfun}\\
\OutGmat_{o\nSlepfun}\end{pmatrix}
,\with
1\le\nSlepfun\le\dimin+\dimout
.
\end{equation}
As a consequence of eqs~(\ref{ugh2})
and~(\ref{OutGmatstructure}), we can see that for all
$1\le\nSlepfun\le\dimin+\dimout$,
\begin{equation}\label{klux}
\OutGmat_{i\nSlepfun}^\matT \OutGmat_{i\nSlepfun} + \OutGmat_{o\nSlepfun}^\matT \OutGmat_{o\nSlepfun} =
\Imat_\nSlepfun
,
\end{equation}
where the individual terms that sum to the identity are rank-deficient
and singular, and we have the pairwise orthogonality relationships
\begin{equation}\label{GinGouttrans}
\OutGmat_i\OutGmat_i^\matT = \Imat_{\dimin} 
,\quad 
\OutGmat_o\OutGmat_o^\matT = \Imat_{\dimout}
,\quad
\OutGmat_i \OutGmat_o^\matT = \Omat 
,\also
\OutGmat_o \OutGmat_i^\matT = \Omat 
,
\end{equation}
none of which, again as in~(\ref{ugh}), apply to their truncated
brethren, and where the latter two matrices have dimensions \linebreak
$\dimin\times[\dimout]$ and $[\dimout]\times\dimin$, respectively.
The power spectra are constructed as in eq.~(\ref{intonlypowerspec}),
for the appropriate field terms.

As with the internal-field transform, we use the building blocks of
the full-field Slepian transform in different ways, but there are four.

[1]~We define the scalar internal- and external altitude-cognizant
Slepian functions on the unit sphere as in eq.~(\ref{intonlypotslepfun}),
\begin{align}
\label{jo}
\OutGfun_{i\alpha}(\rvec)&=
\sumlmLin\sOutslepfuncoef_{i\,lm,\alpha}\Yfun_{lm}(\rvec),\\
\label{ju}
\OutGfun_{o\alpha}(\rvec)&=
\sumlmLoutone\sOutslepfuncoef_{o\,lm,\alpha}\Yfun_{lm}(\rvec).
\end{align}
As in eq.~(\ref{cox}) we vectorize the sets of internal- and
external-field altitude-cognizant scalar functions
$\OutGfun_{i\alpha}= \Yfunvec_\Lin^\funT\Outslepfuncoef_{i\alpha}$
\linebreak and $\OutGfun_{o\alpha}=
\Yfunvec^\funT_\Lout\Outslepfuncoef_{o\alpha}$,
\begin{align}\label{blo}
\OutGfunvec_{i\nSlepfun}& = \begin{pmatrix} \OutGfun_{i1} & \cdots &
  \OutGfun_{i\nSlepfun} \end{pmatrix}^T= \OutGmat_{i
  \nSlepfun}^\matT\Yfunvec_\Lin
,\with
1\le\nSlepfun\le\dimin+\dimout
,\\\label{bli}
 \OutGfunvec_{o\nSlepfun}
&= \begin{pmatrix} \OutGfun_{o1} & \cdots &
  \OutGfun_{o\nSlepfun} \end{pmatrix}^T = \OutGmat_{o\nSlepfun}^\matT
\Yfunvec_\Lout 
,\with
1\le\nSlepfun\le\dimin+\dimout
,
\end{align}
under the partitions
$\OutGfunvec_i=\OutGfunvec_{i\,\dimin+\dimout}=(\OutGfunvec^\funT_{i\nSlepfun}\quad
\OutGfunvec^\funT_{i>\nSlepfun})^\funT$ and
$\OutGfunvec_o=\OutGfunvec_{o\,\dimin+\dimout}=(\OutGfunvec^\funT_{o\nSlepfun}\quad
\OutGfunvec^\funT_{o>\nSlepfun})^\funT$. 

[2]~To evaluate gradients of potential fields on the planetary surface
of radius~$\Earthrad$ or on the outer sphere of radius~$\Outrad$, we
multiply the coefficients with the appropriate continuation factors
and the corresponding vector spherical harmonics, as in
eq.~(\ref{intonlyvecslepfun}),
\begin{align}\label{blablablabla}  
  \OutGfunup_{i\alpha}(\Earthrad\rvec)&=\sumlmLin 
  \InUpelm_l(\Earthrad)\,\sOutslepfuncoef_{i\,lm,\alpha}\,\Efun_{lm}(\rvec) 
,\\ \label{blablablabla2}
  \OutGfunup_{o\alpha}(\Outrad\rvec)&=\sumlmLoutone 
  \OutUpelm_l(\Outrad)\,\sOutslepfuncoef_{o\,lm,\alpha}\,\Ffun_{lm}(\rvec) 
.
\end{align}
As in eq.~(\ref{cix}) we write the vectors containing the functions 
$\OutGfunup_{i\alpha}(\Earthrad\rvec)=\Efunvec_\Lin^\funT\InUpmat(\Earthrad)\,\Outslepfuncoef_{i\alpha}$ 
and
$\OutGfunup_{o\alpha}(\Outrad\rvec)=\Ffunvec_\Lin^\funT\OutUpmat(\Outrad)\,\Outslepfuncoef_{o\alpha}$,
\begin{align}\label{cixout}
\OutGfunvecup_{i\nSlepfun}&=
\begin{pmatrix}
\OutGfunup_{i1}&\cdots&\OutGfunup_{i\alpha}&\cdots&\OutGfunup_{i\nSlepfun} 
\end{pmatrix}\Tit
=
\OutGmat^\matT_{i\nSlepfun}\InUpmat(\Earthrad)\,\Efunvec_\Lin
,\with
1\le\nSlepfun\le\dimin+\dimout
,\\\label{cixout2}
\Gfunvecup_{o\nSlepfun}&=
\begin{pmatrix}
\OutGfunup_{o1}&\cdots&\OutGfunup_{o\alpha}&\cdots&\OutGfunup_{o\nSlepfun} 
\end{pmatrix}\Tit
=
\OutGmat^\matT_{o\nSlepfun}\InUpmat(\Outrad)\,\Ffunvec_\Lin
,\with
1\le\nSlepfun\le\dimin+\dimout
,
\end{align}
again partitioned as
$\OutGfunvecup_i=\OutGfunvecup_{i\,\dimin+\dimout}=(\OutGfunvecup^\funT_{i\nSlepfun}\quad
\OutGfunvecup^\funT_{i>\nSlepfun})^\funT$ and 
$\OutGfunvecup_o=\OutGfunvecup_{o\,\dimin+\dimout}=(\OutGfunvecup^\funT_{o\nSlepfun}\quad
\OutGfunvecup^\funT_{o>\nSlepfun})^\funT$.

[3]~As in eq.~(\ref{maria}), at the common satellite altitude~$\satalt$, the
vectors of AC-\aGVSF{}
$\OutGfunup_{i\uparrow\alpha}(\satalt\rvec)=
\Efunvec_\Lin^\funT\InUpmat\,\Outslepfuncoef_{i\alpha}$ \linebreak
and
$\OutGfunup_{o\uparrow\alpha}(\satalt\rvec)=\Ffunvec_\Lout^\funT\OutUpmat\,\Outslepfuncoef_{o\alpha}$,
\begin{align}\label{chit}
\OutGfunvecup_{i\uparrow\nSlepfun}&=
\begin{pmatrix}
\OutGfunup_{i\uparrow1}&\cdots&\OutGfunup_{i\uparrow\alpha}&\cdots&\OutGfunup_{i\uparrow\nSlepfun} 
\end{pmatrix}\Tit
=\OutGmat_{i\nSlepfun}^\matT\InUpmat\,\Efunvec_\Lin
,\with
1\le\nSlepfun\le\dimin+\dimout\\\label{chat}
\OutGfunvecup_{o\uparrow\nSlepfun}&=
\begin{pmatrix}
\OutGfunup_{o\uparrow1}&\cdots&\OutGfunup_{o\uparrow\alpha}&\cdots&\OutGfunup_{o\uparrow\nSlepfun} 
\end{pmatrix}\Tit
=\OutGmat_{o\nSlepfun}^\matT\OutUpmat\,\Ffunvec_\Lout
,\with
1\le\nSlepfun\le\dimin+\dimout
,
\end{align}
as usual partitioned as
$\OutGfunvecup_{i\uparrow}=\OutGfunvecup_{i\uparrow\dimin+\dimout}=(\OutGfunvecup^\funT_{i\uparrow\nSlepfun}\quad
\OutGfunvecup^\funT_{i\uparrow>\nSlepfun})^\funT$ and 
$\OutGfunvecup_{o\uparrow}=\OutGfunvecup_{o\uparrow\dimin+\dimout}=(\OutGfunvecup^\funT_{o\uparrow\nSlepfun}\quad
\OutGfunvecup^\funT_{o\uparrow>\nSlepfun})^\funT$.

[4]~We finally combine the internal and external AC-\aGVSF{} at
$\satalt$, as $\OutGfunup_{\uparrow\alpha}(\satalt\rvec)=
\OutGfunup_{i\uparrow\alpha}+\OutGfunup_{o\uparrow\alpha}=
\begin{pmatrix}\Efunvec^\funT_\Lin\InUpmat & \Ffunvec^\funT_\Lout\OutUpmat\end{pmatrix}
\begin{pmatrix}\Outslepfuncoef_{i\alpha}^\matT & \Outslepfuncoef_{o\alpha}^\matT\end{pmatrix}^\matT
$,
\begin{equation}\label{jbo}
\OutGfunvecup_{\uparrow\nSlepfun} =
 \begin{pmatrix} 
\OutGfunup_{\uparrow 1}&\cdots& \OutGfunup_{\uparrow \alpha} & \cdots
&\OutGfunup_{\uparrow\nSlepfun}
\end{pmatrix}\Tit
=
\OutGmat_\nSlepfun^\matT
\begin{pmatrix}
\InUpmat\Efunvec_\Lin\\
\OutUpmat\Ffunvec_\Lout\\
\end{pmatrix}  
=\begin{pmatrix}
    \OutGmat_{i\nSlepfun}^\matT&\OutGmat_{o\nSlepfun}^\matT
  \end{pmatrix}
  \begin{pmatrix}
    \InUpmat
    \Efunvec_\Lin\\
    \OutUpmat
    \Ffunvec_\Lout
  \end{pmatrix}
  =  \OutGfunvecup_{i\uparrow \nSlepfun}
  +  \OutGfunvecup_{o\uparrow \nSlepfun}
.
\end{equation}

The properties of the
transformation~(\ref{eigenprobleminnerouter})--(\ref{bli}) cause the
$\dimin+\dimout$-dimensional sets of internal and external scalar
Slepian functions to constitute frames for bandlimited spherical
functions. On $\Earthrad$ and $\Outrad$, respectively,
\begin{align}\label{transfo2}
\Yfunvec_\Lin^\funT\sphcoefEarth=
\Yfunvec_\Lin^\funT\left(\OutGmat_i\OutGmat_i^\matT\right)\sphcoefEarth
&=
\left(\OutGmat_i^\matT\Yfunvec_\Lin\right)^\matT
\left(\OutGmat_i^\matT\sphcoefEarth\right)
=
\OutGfunvec_i^\funT\outSlepcoef_i
\\ \label{transfo2p}
\Yfunvec_\Lout^\funT\outsphcoefEarth=
\Yfunvec_\Lout^\funT\left(\OutGmat_o\OutGmat_o^\matT\right)\outsphcoefEarth
&=
\left(\OutGmat_o^\matT\Yfunvec_\Lout\right)^\matT
\left(\OutGmat_o^\matT\outsphcoefEarth\right)
=
\OutGfunvec_o^\funT\outSlepcoef_o
.
\end{align}

The coefficients $\ssphcoefEarth_{lm}$ and $\soutsphcoefEarth_{lm}$
that expand bandlimited potential fields in the scalar
spherical-harmonic basis $\Yfun_{lm}(\rvec)$ at $\Earthrad$ and
$\Outrad$ transform to the coefficients of any scalar Slepian basis
pair $\OutGfun_{i\alpha}(\rvec)$ and $\OutGfun_{o\alpha}(\rvec)$ designed
with the appropriate bandwidths~$\Lin$ and $\Lout$, as
\begin{align}\label{alltheJoutinstuff0}
\outs_{i\alpha}&=
\intO\Intsignal(\Earthrad\rvec)\,\OutGfun_{i\alpha}(\rvec)\dOmega=
\sumlmLin\sOutslepfuncoef_{i\,lm,\alpha}\,\ssphcoefEarth_{lm}
\quad\text{or, in vector form,}\\ \label{alltheJoutinstuff1}
\outSlepcoef_{i\nSlepfun}&=
\intO\Intsignal(\Earthrad\rvec)\,\OutGfunvec_{i\nSlepfun}(\rvec)\dOmega=
\OutGmat_{i\nSlepfun}^\matT\sphcoefEarth
,\with
1\le\nSlepfun\le\dimin+\dimout,\\  \label{alltheJoutinstuff2}
\outs_{o\alpha}&=
\intO\Extsignal(\Outrad\rvec)\,\OutGfun_{o\alpha}(\rvec) \dOmega=
\sumlmLoutone\sOutslepfuncoef_{o\,lm,\alpha}\,\soutsphcoefEarth_{lm}
\quad\text{or, in vector form,}\\\label{needit2}
\outSlepcoef_{o\nSlepfun}&=
\intO\Extsignal(\Outrad\rvec)\,\OutGfunvec_{o\nSlepfun}(\rvec)\dOmega=
\OutGmat_{o\nSlepfun}^\matT\outsphcoefEarth
,\with
1\le\nSlepfun\le\dimin+\dimout
,
\end{align}
formally mimicking what we wrote down as eq.~(\ref{needit}). To conclude
we also define
\begin{equation}\label{needitalso}
t_\alpha=\outs_{i\alpha}+\outs_{o\alpha}
\quad\text{or, in vector form,}\quad
\OutSlepcoefEarth_\nSlepfun
=
\outSlepcoef_{i\nSlepfun}+\outSlepcoef_{o\nSlepfun}
=
\OutGmat^\matT_\nSlepfun 
\begin{pmatrix}
\displaystyle\sphcoefEarth\\
\displaystyle\outsphcoefEarth
\end{pmatrix}=
\begin{pmatrix}
    \OutGmat_{i\nSlepfun}^\matT&\OutGmat_{o\nSlepfun}^\matT
  \end{pmatrix}
\begin{pmatrix}
\displaystyle\sphcoefEarth\\
\displaystyle\outsphcoefEarth
\end{pmatrix}
,
\end{equation}
as the combined vector of expansion coefficients for bandlimited
potential gradients of the outer and inner types in relation to their
original spherical-harmonic expansion coefficients. Here too, the
arrangement is purely formal, as planetary and outer radii are being
mixed.

\begin{figure}[ht]\centering
\includegraphics[width=\globewidth,angle=0,trim=1.0cm 1.3cm 0.1cm 0.9cm,clip]
               {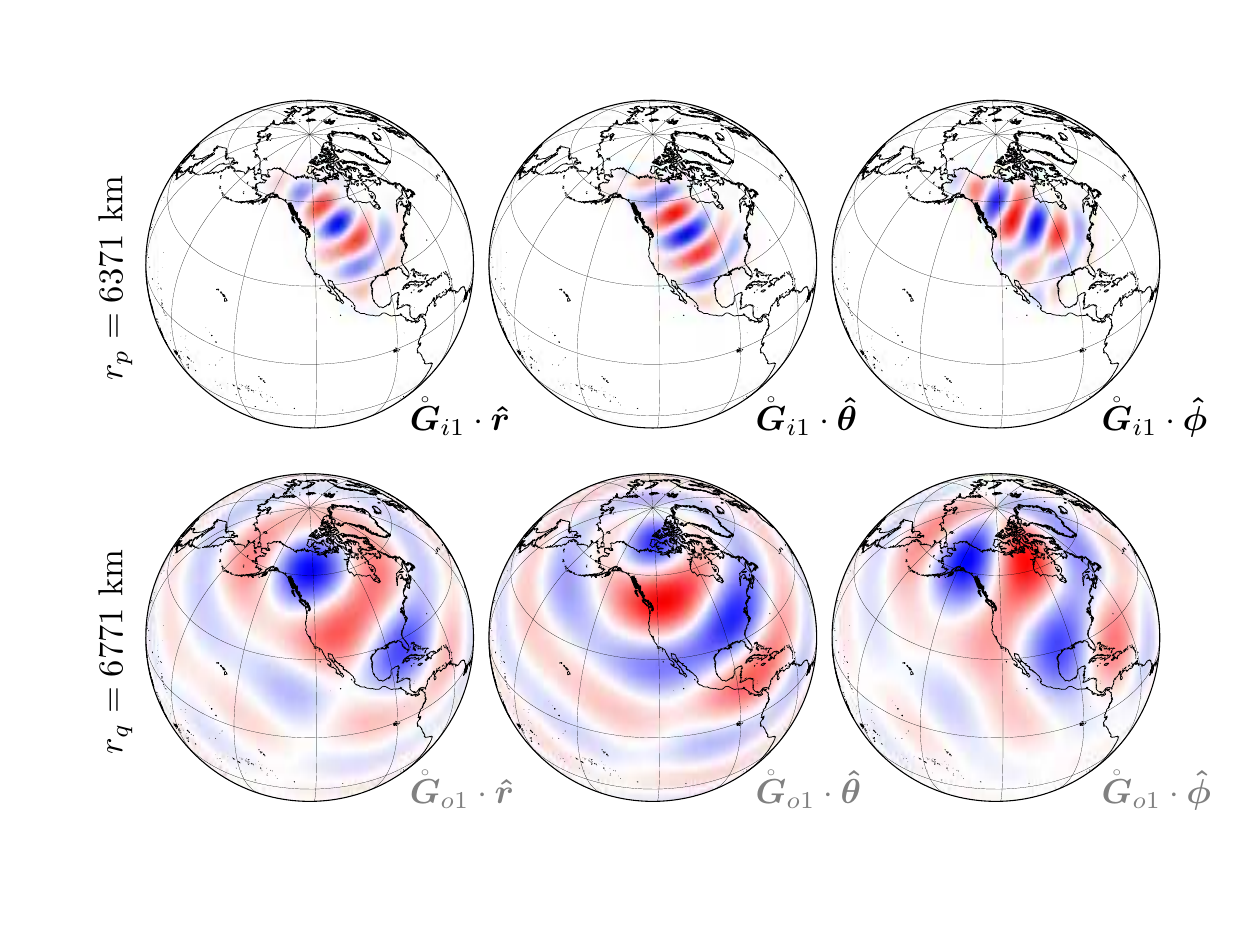}\\[3.5em]
\includegraphics[width=\specwidth,angle=0,trim= 0.8cm 9.2cm 2.2cm 10.8cm,clip]
                {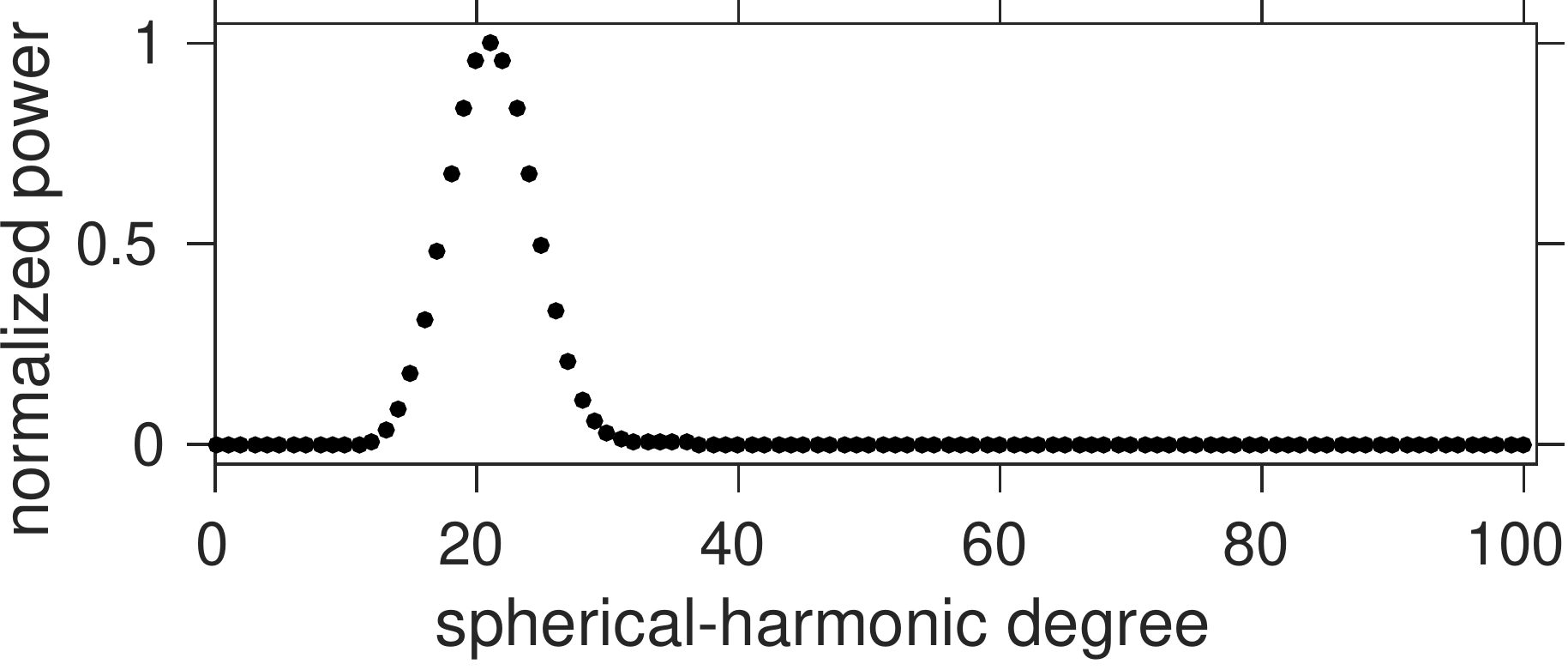}
\hspace{0.75cm}
\includegraphics[width=\specwidth,angle=0,trim= 0.8cm 9.2cm 2.2cm 10.8cm,clip]
                {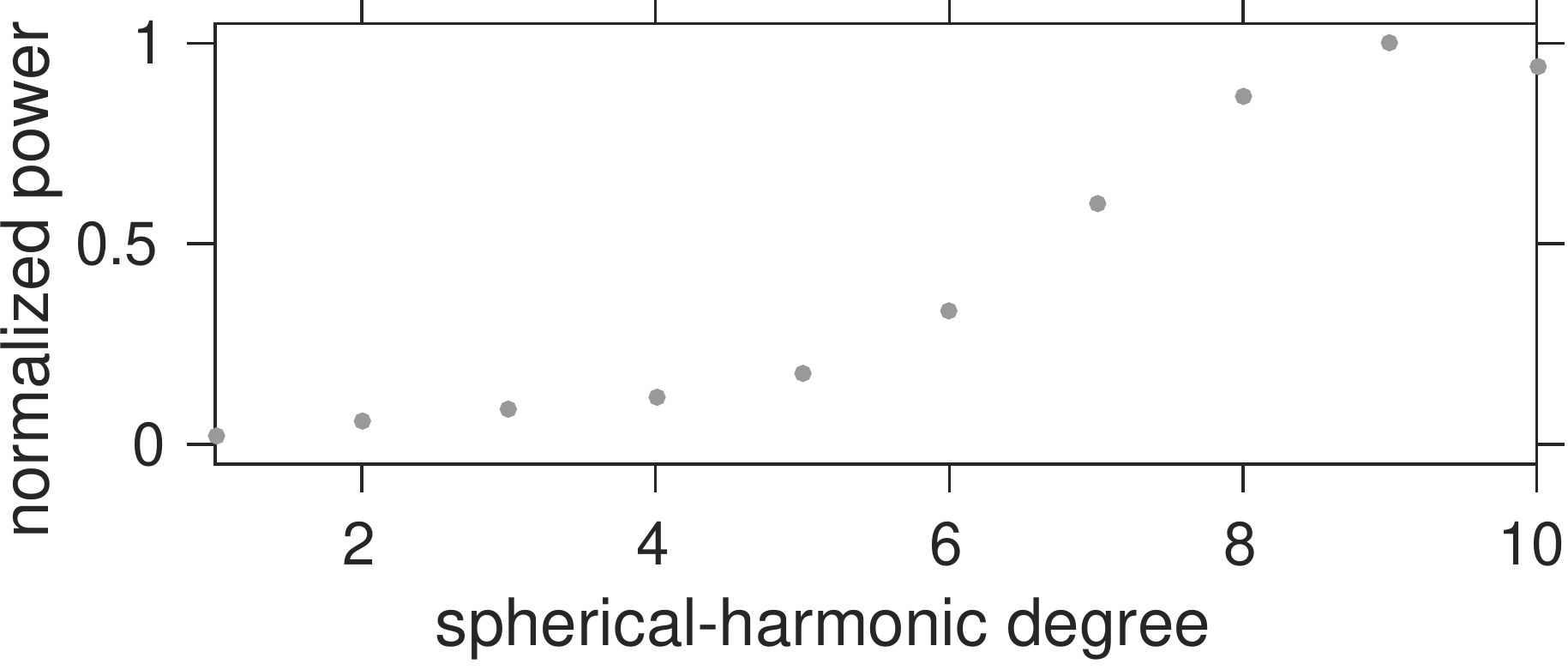}
\caption{\label{Examplefunctions1}The highest-eigenvalue full-field
  altitude-cognizant \GVSF{}, internal field $\OutGfunup_{i1}(\rvec)$
  and external field $\OutGfunup_{o1}(\rvec)$ for region $\region = $
  North America and $\Earthrad=6371$~km, $\Outrad=6771$~km,
  $\satalt=6671$~km, $\Lin=100, \Lout=10$. Upper panels:
  Internal-field part of the function evaluated on the planetary
  surface defined by $\Earthrad$. From left to right: radial
  component, colatitudinal component, longitudinal component. Middle
  panels: External-field part of the function evaluated at the outer
  radius $\Outrad$. From left to right: radial component,
  colatitudinal component, longitudinal component. Bottom panels:
  Power spectra of the internal-field component (bottom
  left) and external-field component (bottom right).}
\end{figure}

\begin{figure}\centering
\includegraphics[width=\globewidth,angle=0,trim= 1.0cm 1.3cm 0.1cm 0.9cm,clip]
                {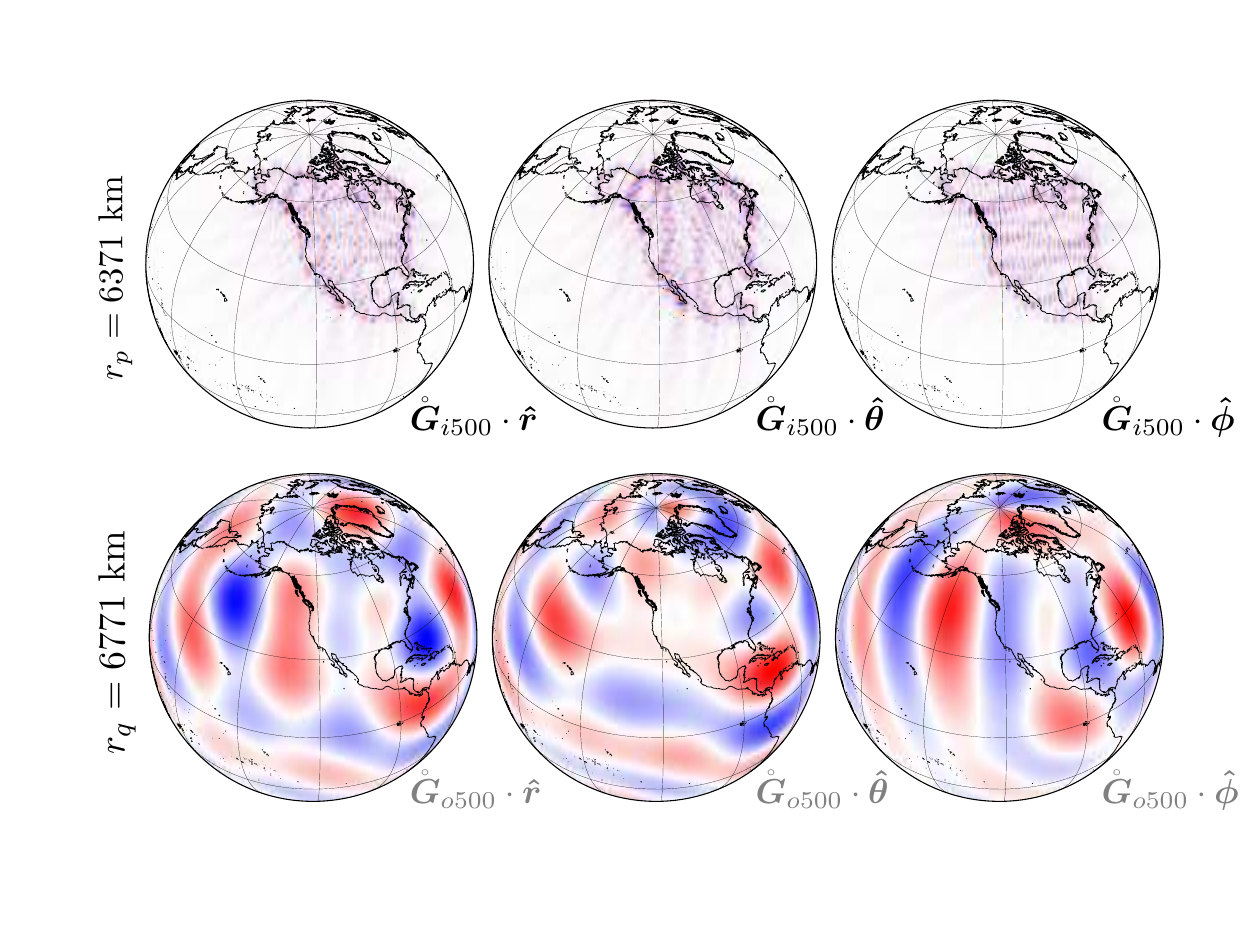}\\[3.5em]
\includegraphics[width=\specwidth,angle=0,trim= 0.8cm 9.2cm 2.2cm 10.8cm,clip]
                {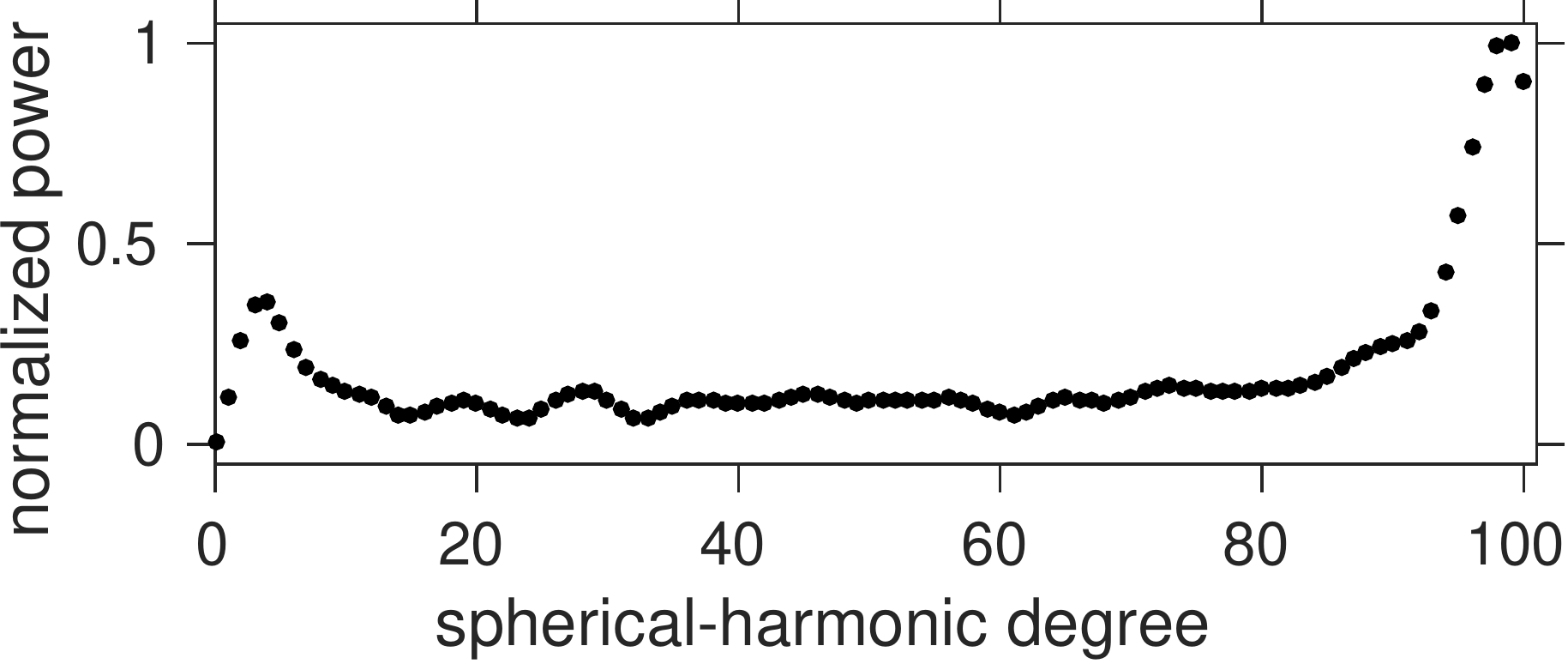}
\hspace{0.75cm}
\includegraphics[width=\specwidth,angle=0,trim= 0.8cm 9.2cm 2.2cm 10.8cm,clip]
                {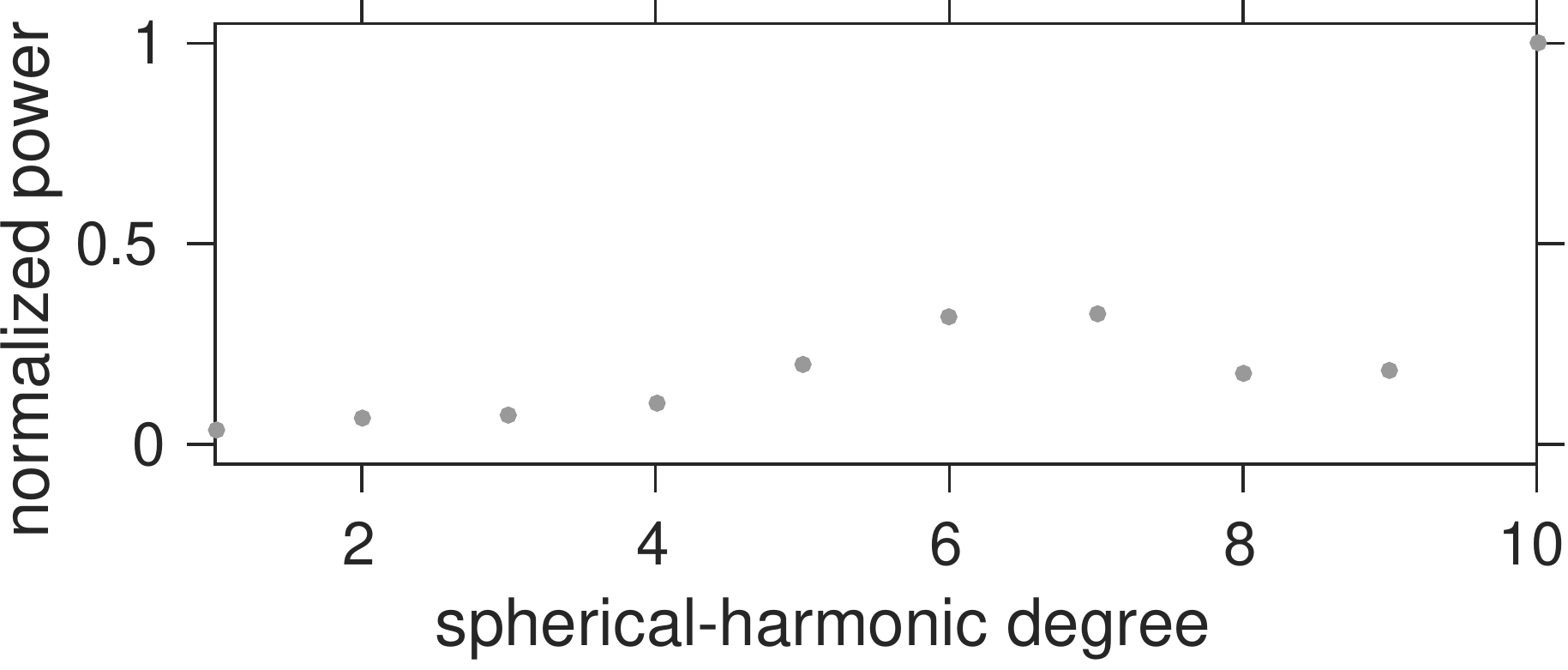}
\caption{\label{Examplefunctions2}Same layout as
  Fig.~\ref{Examplefunctions1} but showing the 500th best-suited
  internal- and external-field altitude-cognizant \GVSF{}, internal
  field $\OutGfunup_{i500}(\rvec)$ and external field
  $\OutGfunup_{o500}(\rvec)$ for region $\region = $ North America and
  $\Earthrad=6371$~km, $\Outrad=6771$~km, $\satalt=6671$~km,
  $\Lin=100, \Lout=10$.}
\end{figure}

Figs.~\ref{Examplefunctions1} and~\ref{Examplefunctions2} show the
best, and the 500th best internal- and external-field AC-\aGVSF{} for
North America, $\OutGfunup_{i1}$, $\OutGfunup_{o1}$,
$\OutGfunup_{i500}$, and $\OutGfunup_{o500}$, with
$\Earthrad=6371$~km, $\satalt=6671$~km, and $\Lin=100$ identical to
the values in Section~\ref{sec:inneraltslepconstruction}, for
comparison with Figs.~\ref{InonlyExamplefunctions1}
and~\ref{InonlyExamplefunctions2}. We set the outer sphere radius
$\Outrad = 6771$~km and the external-field bandwidth $\Lout=10$. In
both figures the internal-field functions are better concentrated
within the region than the external-field functions, a direct
consequence of $\Lin\gg\Lout$. The spatial patterns of the
highest-eigenvalue internal-field AC-\aGVSFs{} shown in
Fig.~\ref{Examplefunctions1} are more consistent with an effective
bandwidth of about~30 than the nominal bandwidth $L=100$ would
suggest. As discussed in Section~\ref{sec:inneraltslepconstruction},
herein lies the difference with the `classical' \GVSFs{} (CL-\aGVSF{})
of \cite{Plattner+2015c}, which were only focused on concentrating
their energy within the region, but disregarded the radial distance
over which they ultimately have to be downward-continued. Since the
higher-degree coefficients are most sensitive to noise amplification
under downward-continuation, the resulting best-suited function has
its energy concentrated over the degrees as shown in the bottom left
panel of Fig.~\ref{Examplefunctions1}. The second-best and third-best,
functions, and so on, contain an increasing proportion of their energy
at higher degrees. The bottom left panel of
Fig.~\ref{Examplefunctions2} shows the power spectrum of the 500th
best internal-field AC-\aGVSF, which has most power toward the tail of
the bandwidth.

\begin{figure}\centering
\includegraphics[width=\eigwidthone,angle=0,trim= 1.7cm 9.7cm 2cm 10.4cm,clip]
{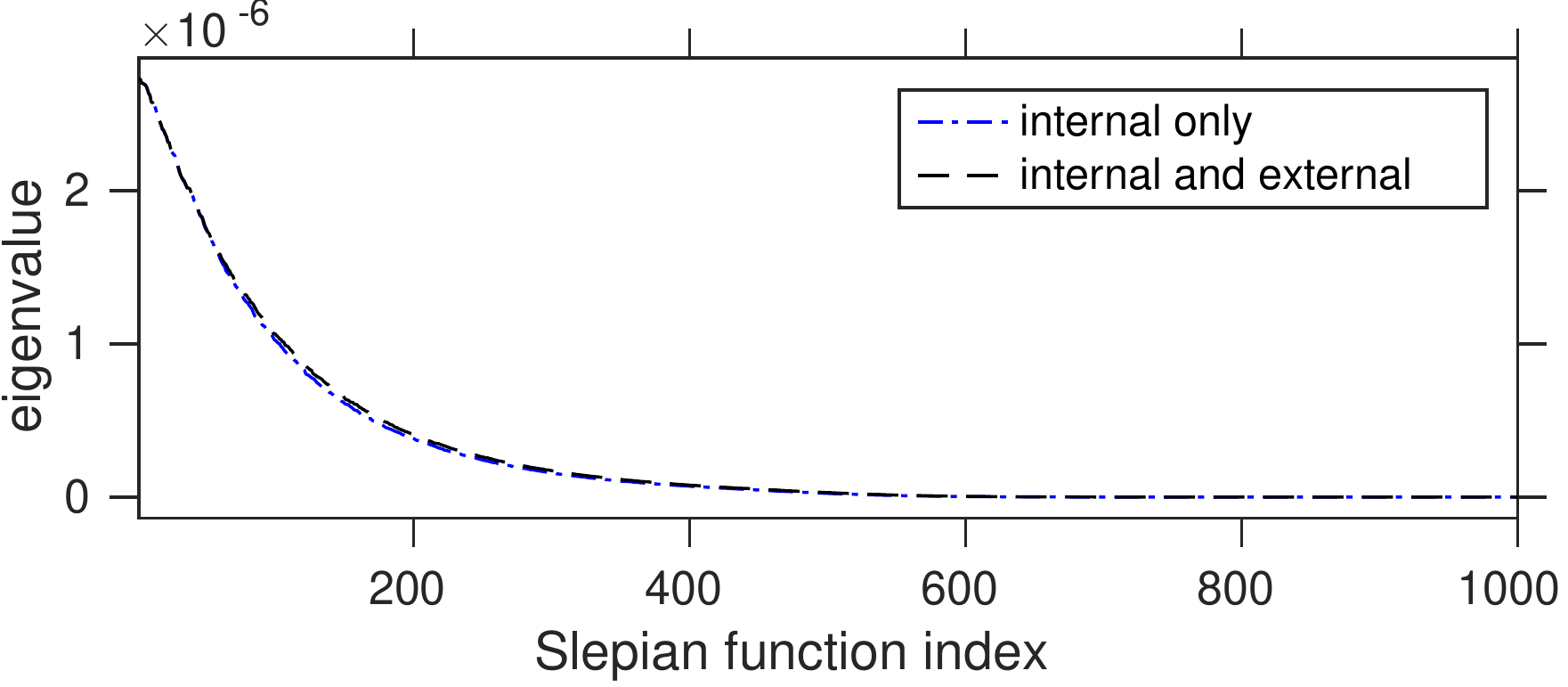}
\caption{\label{EigenvalsFig}The first 1000 eigenvalues
  $\lambda_\alpha$ for the internal-field only altitude-cognizant
  \GVSFs{} $\Gfun_{\alpha}$, and the first 1000
  eigenvalues~$\Outlambda_\alpha$ for the internal- and external-field
  altitude-cognizant \GVSFs{} $\OutGfun_{\alpha}$.}
\end{figure}

In Fig.~\ref{EigenvalsFig} we compare the eigenvalues $\Lamat$ and
$\OutLamat$, obtained from eqs~(\ref{eigenprobleminner})
and~(\ref{eigenprobleminnerouter}), for this parameter set. The two
eigenvalue spectra are similar in character, with few relatively large
eigenvalues and most eigenvalues close to zero. The eigenvalue spectra
for all the different types of classical Slepian functions presented
by \cite{Simons+2006a}, \cite{Plattner+2014a}, and
\cite{Plattner+2015c} typically contained few eigenvalues close to 1
and most eigenvalues close to 0 (the precise numbers depending on the
area of the concentration region). Here, even the largest eigenvalues
drop below $10^{-5}$. This is because the eigenvalues for the
AC-\aGVSF{} also include the effects of harmonic continuation: the
vector spherical harmonics in eqs~(\ref{defKin})
and~(\ref{kernel_inout}) are multiplied by some very small numbers
--- see eqs~(\ref{InUpelements})--(\ref{OutUpelements}).

\subsection{Continuous solution by full-field altitude-cognizant
  \GVSFs{} (AC-\aGVSF{})}

Using eqs~(\ref{kernel_inout})--(\ref{eigenprobleminnerouter}), we
apply the truncated singular value approach to solving
eq.~(\ref{inner-outer-sourceoptimizationproblem}), as we did in
Section~\ref{sec:innersolutioncont}, and obtain
\begin{equation}\label{woot}
\OutKmat
\begin{pmatrix}
\displaystyle\sphcoefEarth\\
\\
\displaystyle\outsphcoefEarth
\end{pmatrix}
=
\OutGmat\OutLamat\OutGmat^\matT
\begin{pmatrix}
\displaystyle\sphcoefEarth\\
\\
\displaystyle\outsphcoefEarth
\end{pmatrix}
=  
\begin{pmatrix}
\displaystyle\InUpmat\intR\Efunvec_\Lin\cdot\datavec\dOmega\\
\\
\displaystyle\OutUpmat\intR\Ffunvec_\Lout\cdot\datavec\dOmega
\end{pmatrix}
.
\end{equation}
Since again we only aim to solve
eq.~(\ref{inner-outer-sourceoptimizationproblem}) for the $\nSlepfun$
best-suited full-field AC-\aGVSF{}, we use eqs~(\ref{ugh2})
and~(\ref{needitalso}) to write
\begin{equation}\label{linearsysteminoutregularized}
\OutGmat_{\nSlepfun}\OutLamat_{\nSlepfun}\,
\estOutSlepcoef_\nSlepfun
=  
\begin{pmatrix}
\displaystyle\InUpmat\intR\Efunvec_\Lin\cdot\datavec\dOmega\\
\\
\displaystyle\OutUpmat\intR\Ffunvec_\Lout\cdot\datavec\dOmega
\end{pmatrix}
,
\end{equation}
with the tilde again distinguishing the truncated solution of
eq.~(\ref{linearsysteminoutregularized}) from the original
statement~(\ref{woot}). The matrix $\OutGmat_{\nSlepfun}$ is as in
eq.~(\ref{OutGmatstructure}), and $\OutLamat_{\nSlepfun}$ is the
$\nSlepfun$-dimensional version of $\OutLamat$. As previously in
eq.~(\ref{analyticalsolutioninner}), a regularized solution of
eq.~(\ref{linearsysteminoutregularized}) then follows in a form that
uses eq.~(\ref{jbo}),
\begin{equation}\label{analyticalsolutioninout}
 \estOutSlepcoef_\nSlepfun = \OutLamat_\nSlepfun^{-1}\intR
\OutGfunvecup_{\uparrow\nSlepfun}\cdot \datavec\dOmega
.
\end{equation}

From the back-projected solution coefficients
$\OutGmat_\nSlepfun\estOutSlepcoef_\nSlepfun$ we can now
individually formulate estimates for the internal potential field at
radius~$\Earthrad$, and for the external potential field at
radius~$\Outrad$, via eqs~(\ref{blo}) and~(\ref{bli}),
\begin{align}\label{slepianfunctionexpansionin}
\estIntsignal_\nSlepfun(\Earthrad\rvec)&=
\sum_{\alpha=1}^\nSlepfun \sestOutSlepcoef_{\alpha}\,\OutGfun_{i\alpha}(\rvec)=
  \Yfunvec_\Lin^\funT \OutGmat_{i\nSlepfun}\, \estOutSlepcoef_{\nSlepfun}
= \OutGfunvec_{i\nSlepfun}^\funT \,\estOutSlepcoef_{\nSlepfun},
\\\label{slepianfunctionexpansionout}
\estExtsignal_\nSlepfun(\Outrad\rvec)&=
\sum_{\alpha=1}^\nSlepfun \sestOutSlepcoef_{\alpha}\,\OutGfun_{o\alpha}(\rvec)=
  \Yfunvec^\funT_\Lout \OutGmat_{o\nSlepfun}\, \estOutSlepcoef_{\nSlepfun}
= \OutGfunvec_{o\nSlepfun}^\funT \,\estOutSlepcoef_{\nSlepfun}
.
\end{align}

\section{Implementation of the internal-field AC-GVSF method}
\label{sec:innersolution}

In Sections~\ref{sec:innersource} and~\ref{sec:inoutSlep} we
constructed altitude-aware spatio-spectrally concentrated `Slepian'
function bases that can be used as alternatives to vector spherical
harmonics to parameterize and solve for potential fields from
continuous data. Of course, instrumental data are always collected at
a discrete set of points. Thus, in this section we describe how to use
the internal-field altitude-cognizant \GVSFs{} (AC-\aGVSF{}) of
Section~\ref{sec:innersource} to solve for an internal potential field
from discrete data at varying satellite altitude. In
Section~\ref{sec:intextest} we will use the full-field AC-\aGVSF{} of
Section~\ref{sec:inoutSlep} to solve for internal- and external
potential fields simultaneously.

The continuous problem stated in
eqs~(\ref{dataonlyinregion})--(\ref{inner-sourceoptimizationproblem}),
after Slepian basis transformation to~(\ref{linearsysteminner}), was
solved approximately, analytically, in the truncated form of
eq.~(\ref{analyticalsolutioninner}). In the present paper, so far, we
have not offered any guidance on how to choose the
parameter~$\nSlepfun$, nor have we shown that solutions of the general
type are actually... any good. Reassuringly, \cite{Plattner+2015a}
showed that they are, and a statistical analysis confirms this in
Section~\ref{sec:statsint}. Hence, in this section, we simply furnish
the details of a method suitable for practical use.

Let the data be a discrete set of vector-valued measurements obtained
at $\npoints$ satellite locations $r_1\rvec_1,\ldots,
r_\npoints\rvec_\npoints$ in the manner of
eq.~(\ref{dataonlyinregion}), densely distributed within a
subregion~$\region$ of the unit sphere~$\Omega$, at the radial
positions $\Earthrad<r_i<\Outrad$ clustered about a representative
average~$\satalt$. We evaluate the vector spherical-harmonics
$\Efun_{lm}$ at the data locations $\rvec_1,\ldots,\rvec_\npoints$ on
the unit sphere, multiply them by the corresponding
upward-continuation terms $\InUpelm_l(r_i)$, and collect the results
in a $\dimin\times3\npoints$-dimensional matrix~$\EpointsUP$. Using
the generic index~$c$ for~$r$,~$\theta$ or~$\phi$ for the radial,
colatitudinal, and longitudinal vector components, and $\bch$ for
$\brh$, $\bthetah$ or $\bphih$ for the unit vectors, we assemble the
pieces
\begin{equation}\label{elmic}
[\Epoints_{\Uparrow lm,i}]_c=
\InUpelm_l(r_i)\Efun_{lm}(\rvec_i)\cdot\bch
\also
[\EpointsUP]_c=\begin{pmatrix} 
[\Epoints_{\Uparrow 00,1}]_c&\cdots&[\Epoints_{\Uparrow 00,k}]_c\\
\vdots&&\vdots\\
[\Epoints_{\Uparrow \Lin\Lin,1}]_c&\cdots&[\Epoints_{\Uparrow \Lin\Lin,k}]_c
\end{pmatrix}
\into
\EpointsUP=
\begin{pmatrix} [\EpointsUP]_r&[\EpointsUP]_\theta&[\EpointsUP]_\phi\end{pmatrix}
.
\end{equation}
Using the Slepian transformation for the region~$\region$ and the
average satellite altitude~$\satalt$, we construct the matrix of
$\nSlepfun$ internal-field AC-\aGVSF{} (compare with eq.~\ref{cix})
evaluated at the actual satellite altitudes~$r_i$, by multiplying
$\EpointsUP$ with the truncated eigenvector matrix~$\Gmat_\nSlepfun$
of eq.~(\ref{intslepcoef}),
\begin{equation}\label{yup}
\GpointsUPJ=\Gmat_\nSlepfun^\matT\EpointsUP.
\end{equation}
Broadly speaking, the success of truncation in the Slepian basis as an
effective means of regularization owes to the eigenvectors of the
`normal' or `Gramian' matrix~(\ref{defKin}) having relatively easily
computable numerical properties and an attractive eigenvalue
structure. Thus, rather than discretizing
eq.~(\ref{innerlinearproblem}) and constructing a truncated-SVD
solution for what would amount to a discretized equivalent of
eqs~(\ref{defKin})--(\ref{eigenprobleminner}), we rely on the data
sampling the region of interest~$\region$ relatively densely, around a
relatively stable altitude~$\satalt$, and therefore, eq.~(\ref{yup})
uses the same, continuously derived, eigenfunctions as in
eq.~(\ref{maria}), except that they are evaluated at the exact,
individual, data altitudes~$r_i$.

The $3\npoints$-dimensional vector of vector-valued data will be
$\dpoints$. Note that we have now introduced a sans-serif
($\dpoints,\Epoints,\Gpoints$) to our lineup of fonts. Using an
$\ell_2$-norm notation (compare with
eq.~\ref{inner-sourceoptimizationproblem}), we restate the inverse
problem in the discrete truncated AC-\aGVSF{} basis~(\ref{yup}) in
terms of its unknown expansion coefficients $\sestSlepcoefi_\alpha$,
$\alpha = 1,\ldots,\nSlepfun$, collecting them in the
vector~$\estSlepcoefi_\nSlepfun$ (notationally distinct from
eqs~\ref{needit} and~\ref{analyticalsolutioninner}), as
\begin{equation}\label{inlinsyspoints}
\arg\min_{\estSlepcoefi_\nSlepfun} \left\lVert \GpointsUPJ^\pointT
\estSlepcoefi_\nSlepfun -\dpoints \right\rVert^2
\orq
\left(\GpointsUPJ\GpointsUPJ^\pointT\right) \estSlepcoefi_\nSlepfun =
\GpointsUPJ\dpoints
.
\end{equation}
Eq.~(\ref{inlinsyspoints}) defines a symmetric positive-definite
$\nSlepfun \times \nSlepfun$-dimensional system of equations whose
condition number depends on the choice of number of Slepian functions
$\nSlepfun$ in virtually the same fashion, given our assumptions, as
the conditioning of eq.~(\ref{analyticalsolutioninner}) depended on
the inverse eigenvalues of the continuous problem. Using the evaluated
continuous-problem eigenfunctions is computationally efficient for the
inversion, and understanding the behavior of the solutions is promoted
through the analysis of their eigenvalues.

However eq.~(\ref{inlinsyspoints}) is solved in numerical practice, we
advocate following up with the iteratively reweighted residual
approach of \cite{Farquharson+1998}. We define the
$3\npoints$-dimensional vector of residuals~$\respoints$ as
\begin{equation}\label{respoints1}
\respoints = \GpointsUPJ^\pointT\,\estSlepcoefi_\nSlepfun - \dpoints
,
\end{equation}
and the $3\npoints\times3\npoints$-dimensional matrix~$\resmat$ as the
diagonal matrix initialized by the identity but in subsequent
iterations populated with the absolute values of the
entries~$\respoints$ on the diagonal, or a threshold value to avoid
division by small numbers for well-fitted data points. Then, we solve
the updated linear problem, repeatedly until convergence, by
satisfying
\begin{equation}\label{initerativeReweight}
\left(\GpointsUPJ \resmat^{-2} \GpointsUPJ^\pointT\right)
\estSlepcoefi_\nSlepfun = \GpointsUPJ\resmat^{-2}\dpoints
.
\end{equation}


From the $\estSlepcoefi_\nSlepfun$ in eq.~(\ref{inlinsyspoints})
or~(\ref{initerativeReweight}) we obtain the estimate~$\estIntsignali_\nSlepfun$
for the potential field on the planetary surface using
eq.~(\ref{intonlypotslepfun}) as in
eq.~(\ref{slepianfunctionexpansioninner}),
\begin{equation}\label{inestpotfield}
\estIntsignali_\nSlepfun(\Earthrad\rvec)=\sum_{\alpha=1}^\nSlepfun
\sestSlepcoefi_\alpha\Gfun_\alpha(\rvec)=\Yfunvec_\Lin^\funT
\Gmat_\nSlepfun\, \estSlepcoefi_{\nSlepfun} =\Gfunvec_\nSlepfun^\funT
\estSlepcoefi_\nSlepfun.
\end{equation}
To evaluate the vector field for the internal potential at a radius
within the shell $\Earthrad \leq r \leq \Outrad$, we expand the
coefficients~$\estSlepcoefi_{\nSlepfun}$ in the Slepian basis
evaluated at a different altitude, using eq.~(\ref{InUpelements}), and
with, as compared to eq.~(\ref{maria}), $\Gfunvecup_{\uparrow
  \nSlepfun}(r\rvec) = \Gmat_\nSlepfun^\matT \InUpmat(r) \,
\Efunvec_\Lin(\rvec)$,
\begin{equation}\label{intuprad}
  \vecestIntsignali_\nSlepfun(r\rvec)= \Gfunvecup^\funT_{\uparrow
    \nSlepfun}(r\rvec)\, \estSlepcoefi_{\nSlepfun} .
\end{equation}

\section{Implementation of the full-field AC-GVSF method}
\label{sec:intextest}

In this section we start from data collected in the manner of
eq.~(\ref{dataregionintext}) and use the full-field AC-\aGVSF{} of
Section~\ref{sec:inoutSlep} to solve for the internal
field~$\estIntsignal$ on the planetary surface~$\Earthrad$, and an
external field $\estExtsignal$ on the outer sphere of
radius~$\Outrad$. Adding to the material developed in
Section~\ref{sec:innersolution} we now need to build the
$[\dimout]\times3\npoints$-dimensional matrix~$\FpointsUP$ of
external-field gradient vector spherical harmonics~$\Ffun_{lm}$
evaluated at the individual data locations $r_1\rvec_1,\ldots,
r_\npoints\rvec_\npoints$. The matrix entries are defined analogously
to eq.~(\ref{elmic}), namely
\begin{equation}\label{flmic}
[\Fpoints_{\Uparrow lm,i}]_c=\OutUpelm_l(r_i)\,
\Ffun_{lm}(\rvec_i)\cdot\bch ,\qquad [\FpointsUP]_c=\begin{pmatrix}
     [\Fpoints_{\Uparrow 1\,-1,1}]_c&\cdots&[\Fpoints_{\Uparrow
         1\,-1,k}]_c\\ \vdots&&\vdots\\ [\Fpoints_{\Uparrow
         \Lout\Lout,1}]_c&\cdots&[\Fpoints_{\Uparrow \Lout\Lout,k}]_c
\end{pmatrix}
\also
\FpointsUP=
\begin{pmatrix} [\FpointsUP]_r&[\FpointsUP]_\theta&[\FpointsUP]_\phi\end{pmatrix}.
\end{equation}

Since the truncated full-field AC-\aGVSF{} coefficient matrix
$\OutGmat_\nSlepfun$ in eq.~(\ref{bo}) contains coefficients
pertaining to both internal- and external-field gradient vector
spherical harmonics, we can assemble the matrix of $\nSlepfun$
full-field altitude cognizant \GVSFs{} evaluated at the varying
satellite locations by multiplying the combined matrices $\EpointsUP$
and $\FpointsUP$ with the Slepian coefficient
matrix~$\OutGmat_\nSlepfun$, 
\begin{equation}\label{biz}
\OutGpointsUPJ=\OutGmat_\nSlepfun^\matT\begin{pmatrix} \EpointsUP
\\ \FpointsUP \end{pmatrix}.
\end{equation}
Compared to eq.~(\ref{inlinsyspoints}), the least-squares formulation
for the full-field problem in the discrete basis~(\ref{biz}) is now in
terms of the unknown $\estOutSlepcoefi_{\nSlepfun}$, 
\begin{equation}\label{bizbiz}
\arg \min_{\estOutSlepcoef_{\nSlepfun}} \left\lVert \OutGpointsUPJ^\pointT
\estOutSlepcoefi_{\nSlepfun} - \dpoints\right\rVert^2
\orq
\left(\OutGpointsUPJ\OutGpointsUPJ^\pointT\right) \estOutSlepcoefi_\nSlepfun =
\OutGpointsUPJ\dpoints
.
\end{equation}

As for the purely internal-field solution, we utilize an iteratively
reweighted residual approach \cite[][]{Farquharson+1998}, with
\begin{equation}\label{reweightedresidual2}
\left( \OutGpointsUPJ \mathring{\resmat}^{-2} \OutGpointsUPJ^\pointT
\right)\estOutSlepcoefi_{\nSlepfun} = \OutGpointsUPJ \mathring{\resmat}^{-2}
\dpoints,
\end{equation}
and where, in the first iteration, the diagonal weighting matrix
$\resmat$ is the identity and, in later iterations, has on its
diagonal the absolute values of~$\mathring{\respoints}$, or a
threshold value for small entries of $\mathring{\respoints}$, where
\begin{equation}\label{theres}
\mathring{\respoints}=\OutGpointsUPJ^\pointT
\estOutSlepcoefi_{\nSlepfun} - \dpoints,
\end{equation}
the $3\npoints$-dimensional vector of residuals at the individual data
locations.


From the obtained coefficient vector $\estOutSlepcoef_{\nSlepfun}$ we
construct estimates of the internal potential
$\estIntsignali_\nSlepfun(\Earthrad\rvec)$ and the external potential
$\estExtsignali_\nSlepfun(\Outrad\rvec)$ at~$\Earthrad$ and~$\Outrad$,
respectively, using eqs~(\ref{jo})--(\ref{ju}) as in
eqs~(\ref{slepianfunctionexpansionin})--(\ref{slepianfunctionexpansionout}),
which leads to
\begin{align}\label{intfieldexpansion}
\estIntsignali_\nSlepfun(\Earthrad\rvec)&= 
\sum_{\alpha=1}^\nSlepfun \sestOutSlepcoefi_{\alpha}\OutGfun_{i\alpha}(\rvec)
=\Yfunvec_\Lin^\funT \OutGmat_{i\nSlepfun}\,\estOutSlepcoefi_{\nSlepfun}
=\OutGfunvec_{i\nSlepfun}^\funT\estOutSlepcoefi_{\nSlepfun},
\\\label{extfieldexpansion}
\estExtsignali_\nSlepfun(\Outrad\rvec)&= 
\sum_{\alpha=1}^\nSlepfun \sestOutSlepcoefi_{\alpha}\OutGfun_{o\alpha}(\rvec)
=\Yfunvec^\funT_\Lout \OutGmat_{o\nSlepfun}\,\estOutSlepcoefi_{\nSlepfun}
=\OutGfunvec_{o\nSlepfun}^\funT\estOutSlepcoefi_{\nSlepfun}
.
\end{align}

To obtain estimates of the internal and external vector fields at
$\Earthrad<r<\Outrad$, we expand~$\estOutSlepcoef_{\nSlepfun}$ in the
upward-continued internal-field or external-field basis. Using
eq.~(\ref{OutUpelements}) and, instead of
eqs~(\ref{chit})--(\ref{chat}), $\OutGfunvecup_{i\uparrow
  \nSlepfun}(r\rvec) = \OutGmat^\matT_{i \nSlepfun} \InUpmat(r) \,
\Efunvec_\Lin(\rvec)$ and $ \OutGfunvecup_{o\uparrow
  \nSlepfun}(r\rvec) = \OutGmat^\matT_{o \nSlepfun} \OutUpmat(r) \,
\Ffunvec_\Lout(\rvec)$,
\begin{align}\label{Intupcontintext}
  \vecestIntsignali_\nSlepfun(r\rvec)&=
  \OutGfunvecup^\funT_{i\uparrow \nSlepfun}(r\rvec)\,\estOutSlepcoefi_{\nSlepfun}
,\\
\label{Extupcontintext}
  \vecestExtsignali_\nSlepfun(r\rvec)&= \OutGfunvecup^\funT_{o\uparrow
    \nSlepfun}(r\rvec)\,\estOutSlepcoefi_{\nSlepfun}
,
\end{align}
or indeed, the complete estimate of the full field in
eq.~(\ref{superposition}), 
\begin{equation}\label{betterbb}
\hat{\Bfun}_\nSlepfun(r\rvec)=\OutGfunvecup^\funT_{\uparrow\nSlepfun}(r\rvec)\,
\estOutSlepcoefi_{\nSlepfun}
,
\end{equation}
where the equivalent to eq.~(\ref{jbo}) now takes the form
\begin{equation}
  \OutGfunvecup_{\uparrow \nSlepfun}(r\rvec)=
  \begin{pmatrix}
    \OutGmat_{i\nSlepfun}^\matT&\OutGmat_{o\nSlepfun}^\matT
  \end{pmatrix}
  \begin{pmatrix}
    \InUpmat(r)\,
    \Efunvec_\Lin(\rvec)\\
    \OutUpmat(r)\,
    \Ffunvec_\Lout(\rvec)
  \end{pmatrix}
  =  \OutGfunvecup_{i\uparrow \nSlepfun}(r\rvec)
  +  \OutGfunvecup_{o\uparrow \nSlepfun}(r\rvec).
\end{equation}

\section{Example: Crustal magnetic field reconstruction} 
\label{numexsect}

We test the internal-field method of Section~\ref{sec:innersolution}
and the full-field method of Section~\ref{sec:intextest} on a
synthetic data set generated as the sum of the internal-field model
NGDC-720~V3 \cite[]{Maus2010}, truncated at spherical-harmonic degree
$\Lin=100$, and an external field simulated from a flat power spectrum
with maximum spherical-harmonic degree $\Lout=10$. We normalize the
external-field coefficients such that the resulting field has 10\% of
the average absolute value of the internal field at the average
satellite radial position $\satalt=6671$~km. Such values seem to be
within a realistic range, see for example \cite{Langel+85} and
\cite{Olsen+2010a}. We evaluate both fields at 15000 uniformly
distributed data locations within North America at a set of satellite
altitudes uniformly distributed between 250~km and 350~km above the
planetary radius $\Earthrad=6371$~km. We add zero-mean uncorrelated
Gaussian noise with standard deviation of 1\% of the mean absolute
value of the combined internal and external fields to the data.

\begin{figure}\centering
\includegraphics[width=\globeswidth,angle=0,trim= 1cm 1.2cm 4.1cm
  0.8cm,clip] {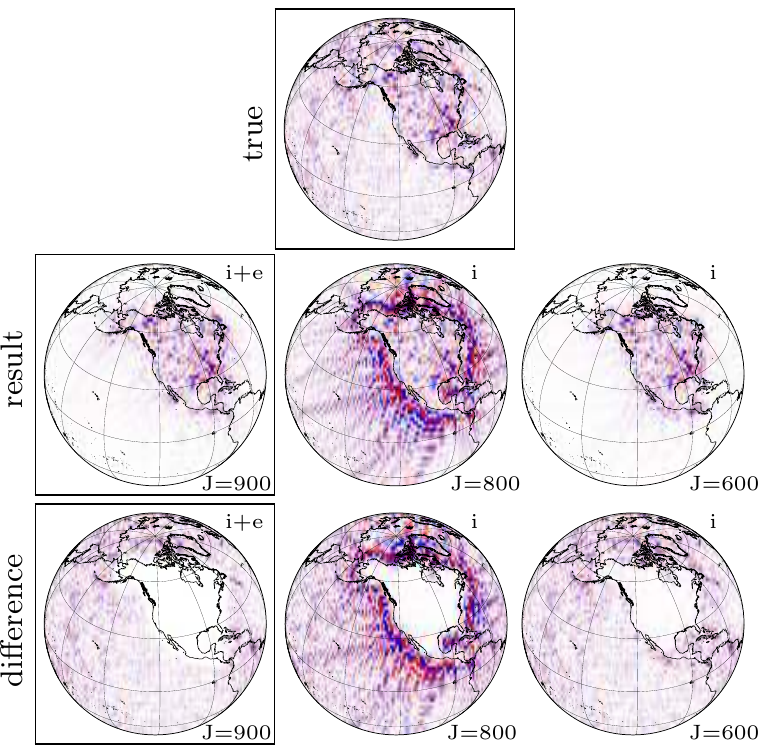}
  \vspace{0.2cm} \includegraphics[width=0.7\textwidth,angle=0,trim=
    3cm 0.5cm 11.5cm 19.2cm,clip] {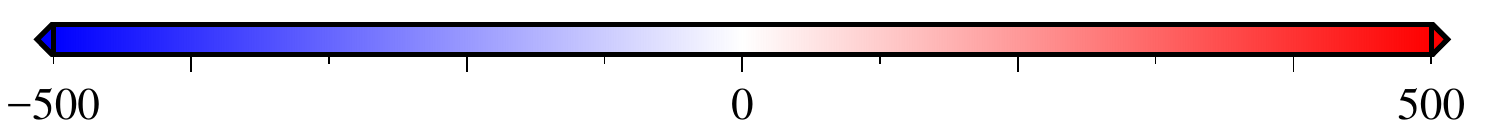}
\Large{radial field [nT]}
\caption{\label{innerouterinvfig}Performance of the methods developed
  in this paper, carried out on a synthetic data set. All panels show
  the radial component of magnetic fields. For the inversion results,
  $\nSlepfun$ identifies the number of Slepian functions used. Top:
  the `true' field, composed of internally and externally generated
  sources, with noise added. Middle left: inversion result using
  full-field altitude-cognizant \GVSFs{} (AC-\aGVSF{}). Middle:
  inversion result using internal-field AC-\aGVSF{}, with
  $\nSlepfun=800$. Middle right: inversion result using internal-field
  AC-\aGVSF{}, with $\nSlepfun=600$. Bottom: difference between the
  true field and the inversion results. The boxed panels are reference
  cases that reappear in subsequent figures.}
\end{figure}

To determine the optimal number $\nSlepfun_\mathrm{opt}$ of Slepian
functions we use the procedure described by \cite{Plattner+2015c}. We
invert for the crustal magnetic field for a series of values
$\nSlepfun$ of Slepian functions and compare each result to the
original field. We select $\nSlepfun_\mathrm{opt}$ as the number of
functions that leads to the smallest mean-squared error within the
region $\region = $ North America. Such a procedure would not be
applicable without knowing the original field. In that case we need to
resort to an indirect strategy, as for example the approach described
by \cite{Plattner+2015a}, or via subsampling methods
\cite[e.g.][]{Davison+97}. For the full-field AC-\aGVSF{} method, our
best number of Slepian functions was $\nSlepfun_\mathrm{opt}=900$. For
the internal-field AC-\aGVSF{}, $\nSlepfun_\mathrm{opt}=800$, though
we note that we achieved a similarly low mean-squared error over North
America when $\nSlepfun=600$.

\begin{figure}\centering
\includegraphics[width=\recswidth,angle=0,trim= 0.2cm 3.7cm 0.2cm
  2.8cm,clip] {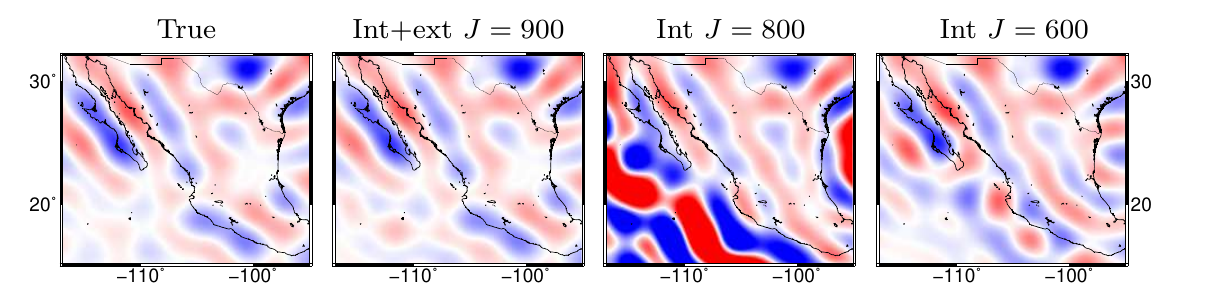}
\includegraphics[width=0.7\textwidth,angle=0,trim= 3cm 0.5cm 11.5cm
  19.2cm,clip] {figures/colbar_e_annot2_500} \Large{radial field [nT]}
\caption{\label{innerouterinvfig2}Performance of the methods developed
  in this paper. Close-up over Mexico of the results shown in
  Fig.~\ref{innerouterinvfig}, identifiable by their labels. While the
  solutions generally look very similar there are differences, in
  particular around longitude $-102^\circ$ and latitude $20^\circ$. A
  seemingly continuous `stripe' in both internal-field solutions,
  $\nSlepfun=600$ and $\nSlepfun=800$, is not continuous in the true
  field nor in the `best' solution that uses $\nSlepfun=900$
  full-field altitude-cognizant \GVSFs{}.}
\end{figure}

\begin{figure}\centering
\includegraphics[width=\globeswidth,angle=0,trim= 0.3cm 1cm 0.3cm 0.8cm,clip]
{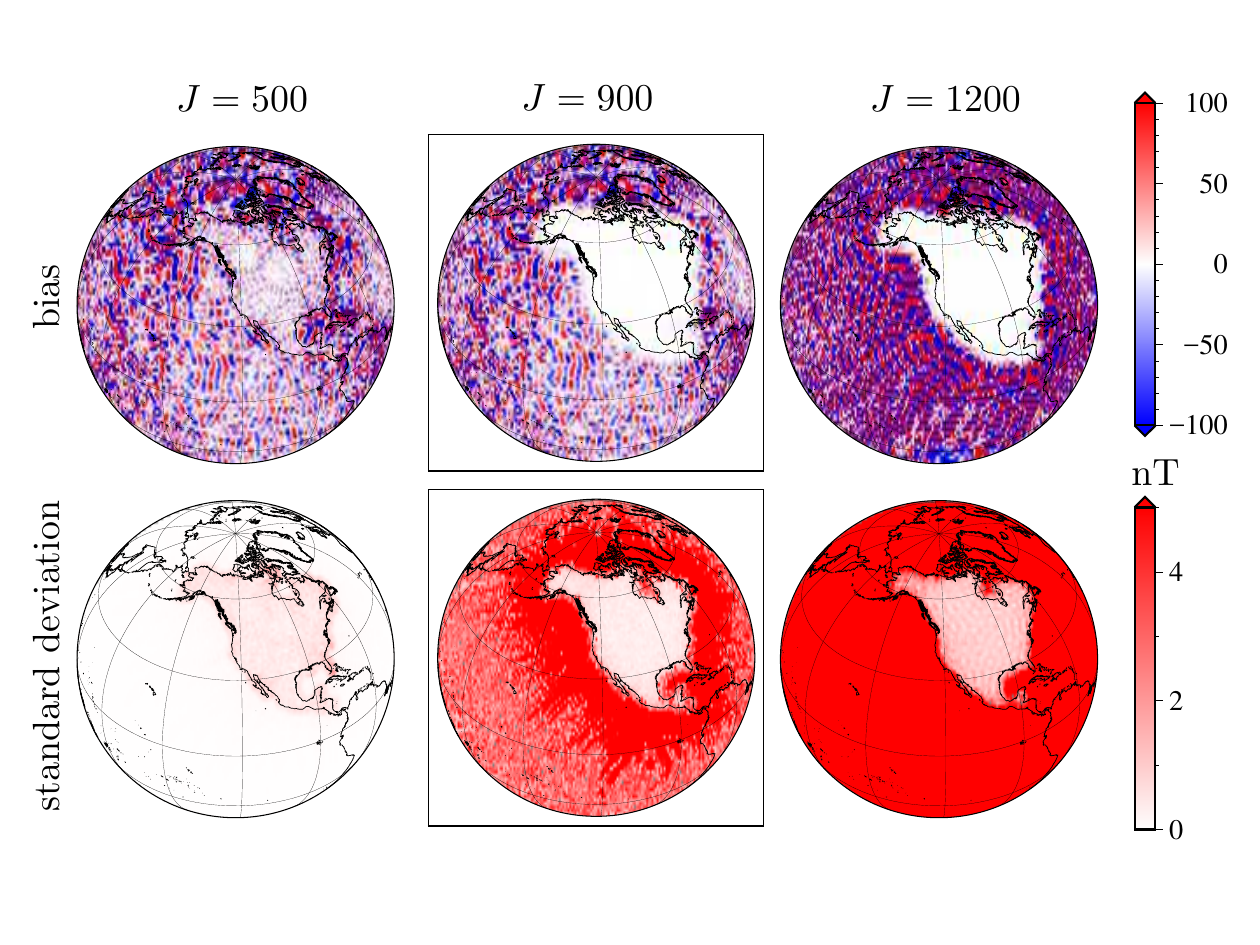}
 \caption{\label{testingvariances}Performance of the methods developed
   in this paper. Bias and standard deviation of the inversion results
   for a sequence of $\nummods=10$ realizations of the data set as
   shown in Fig.~\ref{innerouterinvfig}, conducted using various
   $\nSlepfun$ full-field altitude-cognizant \GVSFs{}. From these
   experiments we determined the bias (top row) and the standard
   deviation (bottom row). Left-hand column: $\nSlepfun = 500$:
   significant bias within the North American target region but low
   variance. Right-hand column: $\nSlepfun = 1200$: low bias, high
   variance. Middle column: $\nSlepfun_\mathrm{opt} = 900$: low bias
   and low variance within the region.}
\end{figure}

Fig.~\ref{innerouterinvfig} summarizes the results. We show the radial
component of the original NGDC-720~V3 internal field on the planetary
surface (top, labeled `true'), together with the models resulting from
our full-field approach using $\nSlepfun_\mathrm{opt} = 900$ (middle,
labeled `i+e'), and the result from the internal-field approach using
$\nSlepfun_\mathrm{opt} = 800$ and $\nSlepfun=600$ (middle, labeled
`i'). The bottom row shows the differences between the true model and
the inversion results. The results for the full-field method for
$\nSlepfun_\mathrm{opt} = 900$ and for the internal-field method when
$\nSlepfun = 600$ show little model strength outside of the North
American region, whereas the result from the internal-field method
when $\nSlepfun = 800$ shows significant ringing off the coast of
North America. This ringing results from the increased model variance
caused by functions that have significant energy outside North
America, where they are unconstrained by data. For larger numbers
$\nSlepfun$ of Slepian functions this variance will also affect the
model within the region of North America. In
Fig.~\ref{innerouterinvfig} we framed the original model, the panels
representing our `optimal' solution, and the difference between the
two. The panels with the original field and the best model will
reappear for comparison later.

The full-field AC-\aGVSF{} solution shown Fig.~\ref{innerouterinvfig}
(middle left) faithfully represents the original model (top), a
finding that is substantiated by their difference (bottom left). At
first glance, the purely internal-field AC-\aGVSF{} solutions (panels
labeled `i') also do appear representative within North America. Upon
closer inspection, shown in Fig.~\ref{innerouterinvfig2}, both
$\nSlepfun=800$ and $\nSlepfun=600$ internal-field solutions contain
similar artifacts which may lead to misinterpretation, in particular
since they persist for different numbers of Slepian functions, and
because they are of smaller length scale than supported by the
bandwidth $\Lout = 10$ of the external field for which we did not
account in those inversions.

When selecting the optimal number $\nSlepfun_\mathrm{opt}$ of Slepian
functions we aim to minimize the mean-squared error of the resulting
model. With increasing $\nSlepfun$ the model bias decreases, whereas
its variance increases. While we postpone a more formal statistical
analysis for the internal-field method to Section~\ref{sec:statsint},
and for the full-field method to Section~\ref{sec:statsintext}, this
behavior can be understood on the basis of elementary considerations
\cite[]{Simons+2006b,Plattner+2015c,Freeden+2016}. Up to a point,
modeling using more Slepian functions implies that less `signal' is
being missed over the target region, but also that more `noise' is
being captured. We calculate the spatial manifestation of the model
bias (the difference between the known truth and the average estimated
model), and its variance (the average of the squared difference
between the estimated models and their average) in our numerical
examples for three different numbers $\nSlepfun$ of Slepian functions,
based on individual inversions for each of $\nummods=10$ realizations
of our synthetic data set, which differ only in the realizations of
the added noise. Bias and standard deviation for the full-field
approach are shown in Fig.~\ref{testingvariances}, for the cases
$\nSlepfun=500$ (too few Slepian functions),
$\nSlepfun_\mathrm{opt}=900$ (our selected solution), and
$\nSlepfun=1200$ (too many Slepian functions). The top row of
Fig.~\ref{testingvariances} shows the bias, the bottom row shows the
corresponding standard deviation.

\begin{figure}\centering
\includegraphics[width=\fglobeswidth,angle=0,trim= 1cm 3.6cm 0.1cm
  3.2cm,clip] {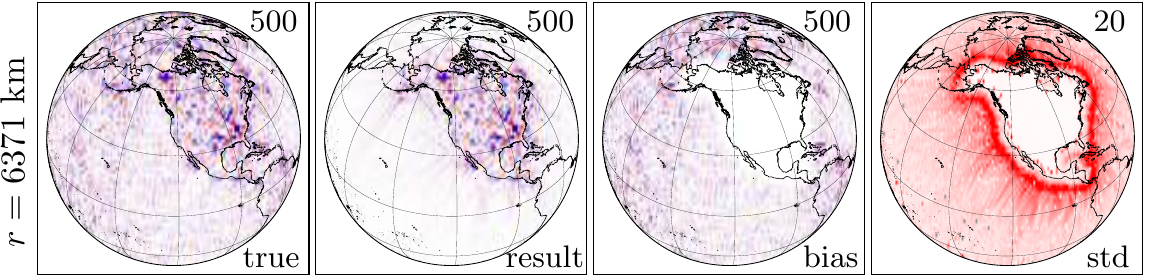}
\includegraphics[width=\fglobeswidth,angle=0,trim= 1cm 3.6cm 0.1cm
  3.2cm,clip] {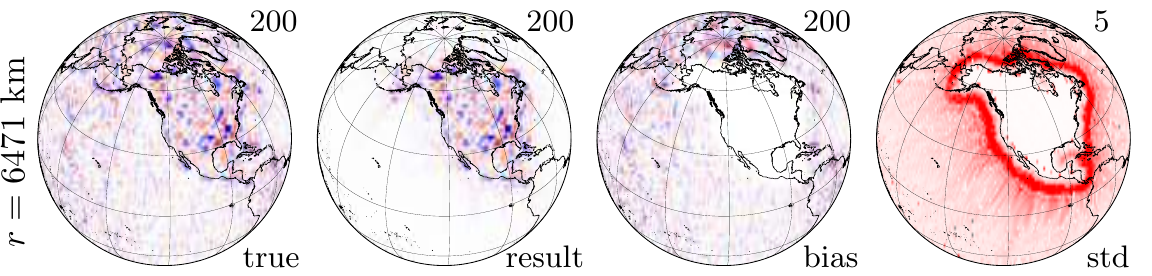}
\includegraphics[width=\fglobeswidth,angle=0,trim= 1cm 3.6cm 0.1cm
  3.2cm,clip] {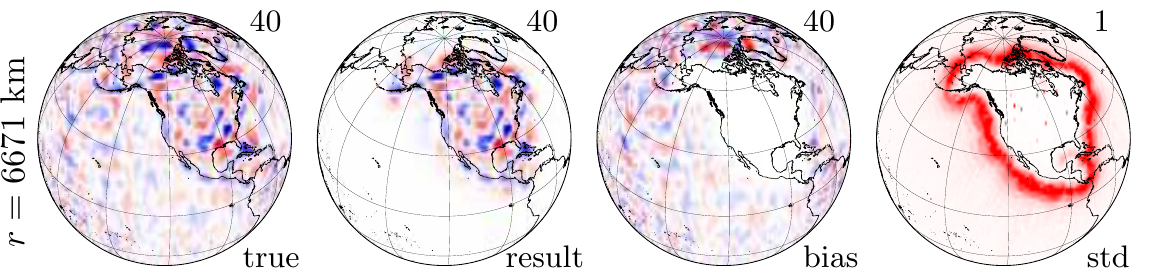}
\includegraphics[width=\fglobeswidth,angle=0,trim= 1cm 3.6cm 0.1cm
  3.2cm,clip] {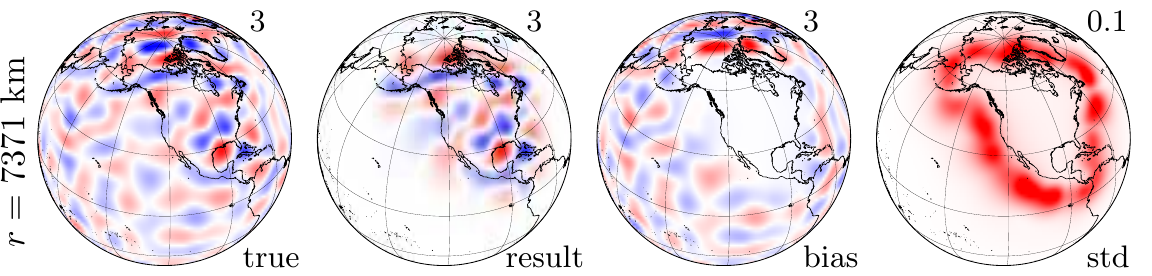}
\includegraphics[width=0.7\textwidth,angle=0,trim= 3.2cm 0.5cm 11.5cm
  19.2cm,clip] {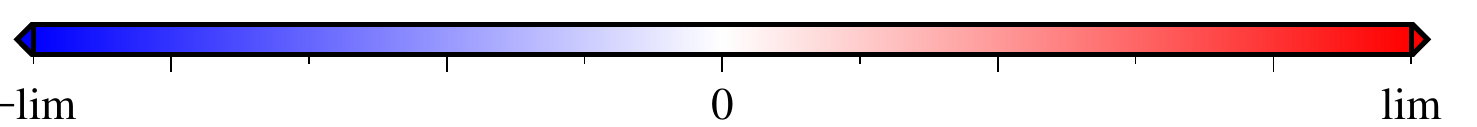}
\caption{\label{inversionfigure}The effect of upward continuation of
  the estimated model. The known model (`true'), internal-field
  results of a full-field inversion (`result'), difference between the
  true model and the average of the inversion results (`bias'), and
  standard deviation of the model results (`std'), over $\nummods=10$
  inversions. All models are evaluated on the planetary surface (first
  row), between the planetary surface and the satellite (at 100~km
  altitude, second row), at average satellite altitude 300~km (third
  row) and at 1000~km above the planet (fourth row). The $\pm$limits
  of the color bar (where we mark them collectively as `$\pm$lim') are
  indicated by the numbers in the top right of each panel.}
\end{figure}

With increasing number $\nSlepfun$ of functions the bias decreases,
but the standard deviation increases. Selecting too few
($\nSlepfun=500$) functions (left-hand column), a low standard
deviation is achieved at the expense of a large bias within the North
American target region $\region$. If, on the other hand, we use too
many ($\nSlepfun=1200$) functions (right-hand column), a small bias
within the region comes at the expense of a large standard
deviation. Only the right number ($\nSlepfun_\mathrm{opt}=900$) of
functions (middle column) has both low bias and a small standard
deviation within the region. The framed panels correspond to the
reference solutions shown in Fig.~\ref{innerouterinvfig}.

To illustrate the effects of upward continuing the estimated internal
field, as described by eq.~(\ref{Intupcontintext}),
Fig.~\ref{inversionfigure} shows the original model (leftmost column),
the full-field inversion result (second from left), bias (third from
left), and standard deviation (rightmost column) for different
evaluation altitudes. The framed panels in the top row show the
original field, resulting model, bias, and variance on the planetary
surface, exactly as they appeared in Figs.~\ref{innerouterinvfig}
and~\ref{testingvariances}. The second through fourth rows of
Fig.~\ref{inversionfigure} show the original field and results
reevaluated at different altitudes, together with their bias and
standard deviation. A single color bar serves all panels, with the
color limits listed in the upper right corner of each panel. For
radial position $r=6471$~km, $100$~km altitude, and for radial
position $r=6671$~km, corresponding to the average satellite altitude
of the simulated data, both bias and standard deviation are low within
the North American target region.

Upward continuation beyond the satellite altitude, up to 1000~km above
the planetary surface, inflates the standard deviation within the
region, where it reaches up to 3\% of the maximum values of the
field. Outside of the target region the model is not well
resolved. This leads to a strong dependence on noise which is greatest
close to the region where the selected Slepian functions still have
relatively high energy but are not well constrained by the
data. Further away from the region, the Slepian functions have less
power, and the resulting model power and, with it, the standard
deviation, are weaker.

\section{Analysis of the internal-field AC-GVSF method}
\label{sec:statsint}

In the previous sections we discussed how our methods work, while the
results of \cite{Plattner+2015a} showed that, indeed, they do work. In
this section, we show why. To this end we introduce some more
notation. We take inspiration from eq.~(\ref{maria}) to define the
vectors of (truncated) downward-continued AC-\aGVSF{}, namely
\begin{equation}\label{pola}
\Gfunvecup_{\downarrow} = \Gmat^\matT \InUpmat^{-1}\Efunvec_\Lin
\also
\Gfunvecup_{\downarrow \nSlepfun} = \Gmat_\nSlepfun^\matT \InUpmat^{-1}\Efunvec_\Lin,
\end{equation}
as well as the complement $\Gfunvecup_{\uparrow >\nSlepfun}$, the
vector with the remaining $\dimin-\nSlepfun$ altitude-cognizant
\GVSFs. With these expressions we state the important relationships
\begin{equation}\label{inupdown}
\Gfunvecup_{\uparrow}^\funT\Gfunvecup_{\downarrow} 
=\Gfunvecup_{\downarrow}^\funT\Gfunvecup_{\uparrow} 
= \Efunvec_\Lin^\funT\InUpmat^{-\matT}\Gmat\Gmat^\matT \InUpmat\,\Efunvec_\Lin 
= \Efunvec_\Lin^\funT \Efunvec_\Lin.
\end{equation}
With the above we can now rewrite the infinitely wideband
eq.~(\ref{gradv}), at satellite altitude, in the following equivalent
forms,
\begin{align}\label{convon}
\vecIntsignal
&=
  \Efunvec_\Lin^\funT\intO \Efunvec_\Lin\cdot
  \vecIntsignal \dOmega
 + \Efunvec_{>\Lin}^\funT\intO \Efunvec_{>\Lin}\cdot
  \vecIntsignal \dOmega\\\label{wes}
&=
  \Gfunvecup^\funT_{\uparrow}\intO \Gfunvecup_{\downarrow}\cdot
  \vecIntsignal \dOmega
 + \Efunvec_{>\Lin}^\funT\intO \Efunvec_{>\Lin}\cdot
  \vecIntsignal \dOmega\\\label{wts}
&=\Efunvec_\Lin^\funT\InUpmat\sphcoefEarth
 + \Efunvec_{>\Lin}^\funT\intO \Efunvec_{>\Lin}\cdot
  \vecIntsignal \dOmega
.
\end{align}

\subsection{Relationship to classical spherical Slepian functions} 
\label{sec:relint1}

The optimization problem in
eq.~(\ref{inner-sourceoptimizationproblem}) led to the diagonalization
of the matrix~$\Kmat$ in eq.~(\ref{defKin}) via
eq.~(\ref{eigenprobleminner}). The coefficients~$\Gmat$ in
eq.~(\ref{intslepcoef}) also solve an energy concentration
maximization problem in the space of bandlimited upward-continued
vector spherical-harmonic functions, as we can see through the
formalism in eqs~(\ref{cox}) and~(\ref{maria}), given the equivalency
\begin{equation}\label{optimizationprobleminner}
\lambda
= \frac{\slepfuncoef^\matT \Kmat\,
  \slepfuncoef}{\slepfuncoef^\matT \slepfuncoef}
=  \frac{\displaystyle \intR
 \left(\slepfuncoef^\matT \InUpmat\,\Efunvec_\Lin\right) \cdot
\left( \Efunvec_\Lin^\funT   \InUpmat^\matT \slepfuncoef \right)d\Omega}
{\displaystyle \intO
\left(  \slepfuncoef^\matT \Yfunvec_\Lin\right)\left(\Yfunvec_\Lin^\funT  \slepfuncoef\right)d\Omega}
=  \frac{\displaystyle \intR
\Gfunup_\uparrow^2 \dOmega}{\displaystyle \intO
  G^2\dOmega}
=\mathrm{maximum}
.
\end{equation}
Appearing without the factor~$\InUpmat$ of eq.~(\ref{InUpels}) in the
numerator, eq.~(\ref{optimizationprobleminner}) is a `classical'
(gradient-)vector spherical-harmonic concentration problem in the
style of \cite{Maniar+2005}, \cite{Plattner+2014a,Plattner+2015c}, and
\cite{Jahn+2014}, much as these authors generalized the `classical'
scalar problem of \cite{Albertella+99}, \cite{Simons+2006a},
\cite{Simons+2006b}, and others --- see also \cite{Eshagh2009a}.
Among all bandlimited upward-continued gradient-vector functions that
are linear combinations of the basis set~$\InUpmat\,\Efunvec_\Lin$,
the first altitude-cognizant \GVSF{}, $\Gfunup_{\uparrow 1}$, is the
best-concentrated in the sense
of~(\ref{optimizationprobleminner}). The concentration factor
$\lambda_1$ is the first eigenvalue associated with the
diagonalization problem~(\ref{eigenprobleminner}). The second-best
concentrated AC-\aGVSF{}, $\Gfunup_{\uparrow 2}$, and its
corresponding $\lambda_2$, is the next best function in the
sense~(\ref{optimizationprobleminner}) that is orthogonal to
$\Gfunup_{\uparrow 1}$, and so on.

Evaluated at satellite altitude, the internal-field AC-\aGVSF{}
$\Gfunvecup_{\uparrow\alpha}$ of eq.~(\ref{maria}) are mutually
orthogonal over the region~$\region$ but not over the
sphere~$\Omega$. On the planetary surface, the corresponding scalar
functions $\Gfunvec_{\alpha}$ of eq.~(\ref{cox}) are
orthogonal over the entire sphere but not over~$\region$. With
$\Lamat$ the eigenvalue matrix as in eq.~(\ref{eigenprobleminner})
and~$\Imat$ the identity, it is straightforward to verify that
\begin{align}
\label{intregionaltinner}
\intR \Gfunvecup_\uparrow\cdot\Gfunvecup_\uparrow^\funT \dOmega 
&=\Lamat\also    
\intO \Gfunvecup_\uparrow\cdot\Gfunvecup_\uparrow^\funT \dOmega 
= \Gmat^\matT \InUpmat \InUpmat^\matT \Gmat,\\
\intO \Gfunvec\,\Gfunvec^\funT \dOmega 
&=\Imat,\also\label{intregionsurfaceinner}
\intR \Gfunvec\,\Gfunvec^\funT \dOmega =
\Gmat^\matT  \left(\intR \Yfunvec_\Lin \Yfunvec_\Lin^\funT d\Omega\right)\Gmat. 
\end{align}
For truncated Slepian bases we also have the corresponding projective
relationships
\begin{equation}\label{nece}
\intR
\Gfunvecup_{\uparrow\nSlepfun}\cdot\Gfunvecup_{\uparrow}^\funT
\dOmega=\begin{pmatrix}\Lamat_{\nSlepfun}&\Omat\end{pmatrix}
\quad \text{ and }\quad
\intO
\Gfunvec_{\nSlepfun}\,\Gfunvec^\funT \dOmega =
\begin{pmatrix}\Imat_{\nSlepfun}&\Omat\end{pmatrix}.
\end{equation}
The `localization' matrix $\intR \Yfunvec_\Lin
\Yfunvec_\Lin^\funT d\Omega$ in eq.~(\ref{intregionsurfaceinner}) is
one that appears in the construction of the classical scalar Slepian
functions \cite[]{Simons+2006a,Simons+2006b}. Its eigenvectors lead to
spherical functions that are orthogonal over the region~$\region$, but
also over the entire sphere~$\Omega$. Incorporating the
upward-continuation into the construction of the Slepian functions has
induced a loss of orthogonality over the region~$\region$ on the
planetary surface (eq.~\ref{intregionsurfaceinner}) --- but we gained
orthogonality within~$\region$ at satellite altitude
(eq.~\ref{intregionaltinner}).

\subsection{Spatially restricted, spectrally concentrated internal-field Slepian functions}
\label{sexo}

As shown by \cite{Simons+2006a}, bandlimited spatially concentrated
Slepian functions have broadband relatives that are spacelimited but
spectrally concentrated. As shown by \cite{Simons+2006b} such
functions play an important role in the analysis of inversion problems
like the one that we are treating in this paper. Spacelimited vector
Slepian functions were introduced by \cite{Plattner+2014a}, and
spacelimited gradient-vector Slepian functions by
\cite{Plattner+2015c}. As to the spacelimited
altitude-cognizant-\GVSFs{} that we will be needing here, we define
the vector with the the first $\nSlepfun$ of the $\Hfuno_{\uparrow
  \alpha,>\Lin}= \vslepfuncoef^\matT_{\uparrow \alpha,
  >\Lin}\Efunvec_{>\Lin}$,
\begin{equation}\label{InUpTruncRelation}
  \Hfunvec_{\uparrow J,>\Lin}=
\begin{pmatrix}\Hfuno_{\uparrow 1, >\Lin}&\cdots& 
\Hfuno_{\uparrow \alpha, >\Lin}&\cdots& 
\Hfuno_{\uparrow \nSlepfun, >\Lin}\end{pmatrix}\Tit
=
 \left( \intR
  \Gfunvecup_{\uparrow \nSlepfun} \cdot \Efunvec_{>\Lin}^\funT
  \dOmega\right) \Efunvec_{>\Lin}
,
\end{equation}
where the infinite-dimensional vector $\vslepfuncoef_{\uparrow
  \alpha}$ that contains the expansion coefficients in the full basis
set $\Efun_{lm}$, $0<l<\infty$, $-l<m<l$, of the bandlimited
AC-\aGVSF{} $\Gfunup_{\uparrow \alpha}$ after spatial truncation to
the region~$R$, and the infinite
vector~$\vslepfuncoef_{\uparrow\alpha, >\Lin}$ containing only those
components at the degrees $l>\Lin$, respectively, for each
$\alpha=1,\ldots,\dimin$, are, from eq.~(\ref{maria}),
\begin{equation}\label{inuptrunc}
\vslepfuncoef_{\uparrow \alpha} = \intR \Efunvec \cdot
\Gfunup_{\uparrow \alpha}\dOmega
=\left(\intR \Efunvec\cdot\Efunvec_\Lin^\funT d\Omega\right)
\InUpmat^\matT \slepfuncoef_\alpha
\also
\vslepfuncoef_{\uparrow\alpha, >\Lin} = \intR
\Efunvec_{>\Lin} \cdot \Gfunup_{\uparrow \alpha}\dOmega
=\left(\intR \Efunvec_{>\Lin}\cdot\Efunvec_\Lin^\funT d\Omega\right)
\InUpmat^\matT \slepfuncoef_\alpha
.
\end{equation}
The coefficient sets $\vslepfuncoef_{\uparrow \alpha}$ and
$\vslepfuncoef_{\uparrow\alpha, >\Lin}$ relate to the set of
coefficients~$\InUpmat^\matT\slepfuncoef_{\alpha}$ of the bandlimited
functions~$\Gfunup_{\uparrow\alpha}$ via broadband extensions of the
localization matrix $\intR
\Efunvec_\Lin\cdot\Efunvec_\Lin^\funT d\Omega$ in
eq.~(\ref{innerlinearproblem}), in the same manner as did their
equivalents in the scalar and vector cases discussed above.

\subsection{Statistical analysis of the internal-field method}
\label{sec:intanalysis}

We now return to the issue we mentioned in Section~\ref{sec:problem},
namely, how spherical-harmonic model bandlimitation affects the
estimate made from data that have, per eq.~(\ref{gradv}), in
principle, infinite bandwidth. What are we missing? And, what is the
effect of truncation of the Slepian basis?

We begin by rewriting the bandlimited portion of the internal
potential~(\ref{Vrr}) in terms of the internal-field
altitude-cognizant scalar Slepian functions of eq.~(\ref{cox}), which,
owing to their orthogonality~(\ref{intregionsurfaceinner}), remain a
complete basis for bandlimited functions on the sphere, see
eqs~(\ref{transfo})--(\ref{needit}). To the Slepian expansion we add
the broadband components,
\begin{equation}\label{hoopla}
\Intsignal(\Earthrad \rvec) = 
 \Gfunvec_\nSlepfun^\funT \intO  \Gfunvec_\nSlepfun
 \Intsignal(\Earthrad \rvec) \dOmega  
+ \Gfunvec_{>\nSlepfun}^\funT \intO  \Gfunvec_{>\nSlepfun}
\Intsignal(\Earthrad \rvec)\dOmega
+\Yfunvec_{>\Lin}^\funT \intO  \Yfunvec_{>\Lin}
\Intsignal(\Earthrad \rvec)\dOmega
.
\end{equation}
Next, we rewrite eq.~(\ref{dataonlyinregion}) in terms of the
AC-\aGVSF{} with the help of eq.~(\ref{wes}). The continuous data
representation is then given by
\begin{equation}\label{indataexpansion}
  \datavec =\Gfunvecup_{\uparrow}^\funT \intO \Gfunvecup_{\downarrow}
  \cdot \vecIntsignal \dOmega
  + \Efunvec_{>\Lin}^\funT\intO \Efunvec_{>\Lin}\cdot
  \vecIntsignal \dOmega + \vecnoise.
\end{equation}
Finally, we return to the form of the bandlimited truncated internal-field
AC-\aGVSF{} estimator in
eqs~(\ref{analyticalsolutioninner})--(\ref{slepianfunctionexpansioninner}),
restated as
\begin{equation}\label{inestpotfieldanl}
\estIntsignal_\nSlepfun(\Earthrad\rvec) =
\Gfunvec_\nSlepfun^\funT\Lamat^{-1}_\nSlepfun \intR
\Gfunvecup_{\uparrow \nSlepfun}\cdot\datavec\dOmega.
\end{equation}
As noted before, the estimator~(\ref{inestpotfieldanl}) is the one
used by \cite{Plattner+2015a}, which is rather radically different
from its counterpart discussed by \cite{Plattner+2015c}. However, and
issues of notation cast aside, the derivations below retain the full
character of the material presented by \cite{Plattner+2015c} --- or,
mutatis mutandis, by \cite{Simons+2006b} --- hence our abridged
treatment here.

Inserting eq.~(\ref{indataexpansion}) into
eq.~(\ref{inestpotfieldanl}), we use eqs~(\ref{nece})
and~(\ref{InUpTruncRelation}), and further using eqs~(\ref{wts}),
(\ref{pola}), the orthonormality~(\ref{ortho}) of the $\Efunvec$, and, at last,
eqs~(\ref{transfo})--(\ref{needit}), we obtain the expression
\begin{equation}\label{intestexpanded}
\estIntsignal_\nSlepfun(\Earthrad\rvec)
=\Gfunvec_\nSlepfun^\funT \intO \Gfunvec_\nSlepfun
\Intsignal(\Earthrad \rvec)\dOmega 
+ \Gfunvec_\nSlepfun^\funT\Lamat^{-1}_\nSlepfun \left(
 \intO \Hfunvec_{\uparrow J,>\Lin } \cdot \vecIntsignal \dOmega 
+
 \intR \Gfunvecup_{\uparrow\nSlepfun}\cdot \vecnoise\dOmega
 \right)
.
\end{equation}
Spherical-harmonic bandlimitation, Slepian-function truncation, and
noise are the three ingredients necessary to understand the quality of
our estimates. A direct comparison between the estimate in
eq.~(\ref{intestexpanded}) and the unknown truth in eq.~(\ref{hoopla})
reveals that the bandlimited estimate~$\estsignal$ of the internal
potential field~$\signal$ does not only depend on the bandlimited part
of the true but also on its broadband portion, and the
noise. Bandlimitation introduces a direct bias term, whether the
estimate uses a truncated Slepian basis or not. The latter two terms
in eq.~(\ref{intestexpanded}) are amplified by the inverse
eigenvalues, which typically become large for increasing~$\nSlepfun$,
which is the primary reason for truncating the Slepian expansions. The
more Slepian functions we use to estimate the internal potential
field, the more leakage contributions we pick up from the neglected
broadband components and from the noise. On the other hand, if we use
too few Slepian functions, then we cannot solve for enough details of
the bandlimited internal field.

If we now make the defensible assumptions that the noise term has zero
mean, and that the noise is uncorrelated with the signal, we can
obtain palatable expressions for the bias, variance, and mean-squared
error of our estimates in terms of the power-spectral densities of
both signal and noise. Again, we follow the recipes outlined by
\cite{Simons+2006b} or \cite{Plattner+2015c}, with the modifications
appropriate to our case at hand. The estimation bias, the difference
between the expected value of the estimator~(\ref{intestexpanded})
and the truth~(\ref{hoopla}), is 
\begin{equation}\label{knot}
b=-\Gfunvec_{>\nSlepfun}^\funT \intO  \Gfunvec_{>\nSlepfun}
\Intsignal(\Earthrad \rvec)\dOmega
-\Yfunvec_{>\Lin}^\funT \intO  \Yfunvec_{>\Lin}
\Intsignal(\Earthrad \rvec)\dOmega
+\Gfunvec_\nSlepfun^\funT\Lamat^{-1}_\nSlepfun 
  \intO \Hfunvec_{\uparrow J,>\Lin } \cdot \vecIntsignal \dOmega 
.
\end{equation}
It is clear from eq.~(\ref{knot}) that `bias' is caused by `missing'
and `mismapped' signal, from the combination of spherical-harmonic
bandlimitation and Slepian truncation. A key feature of the Slepian
function apparatus, however, is that because the AC-\aGVSF{} are
spatially concentrated within the target region~$\region$, with the
concentration measured by the usually rapidly diminishing ranked
eigenvalues, the bias from neglecting low-ranked eigenfunctions will
mostly affect the regions that are not of interest or where no data
were collected. Certainly, care should be taken to define the optimal
truncation level~$\nSlepfun_\mathrm{opt}$, but in
Section~\ref{numexsect} we outlined a procedure precisely for doing
so.

To avoid complicating matters from now on, we drop broadband terms
(the second term in eq.~\ref{intestexpanded}, and the last two terms
in eq.~\ref{knot}), as without a priori modeling of what we truly do
not know: the signal at the unmodeled spherical-harmonic degrees, we
are in no position to remediate the broadband bias nor its leakage
into the bandlimited estimate. We now consider both the true signal
and the noise to be realizations of a time-independent random process
that can be characterized by a power-spectral density, or,
equivalently, a certain covariance function, in the scalar and vector
forms
\begin{align}\label{pipi}
\sigpower(\Earthrad\rvec,\Earthrad\rvec')&=\left\langle
\Intsignal(\Earthrad\rvec)\,\Intsignal(\Earthrad\rvec')\right\rangle
,\\\label{popo} \noisepower(\satalt\rvec,\satalt\rvec')&=\left\langle
\vecnoise(\satalt\rvec)\cdot\vecnoise(\satalt\rvec')\right\rangle
.
\end{align}
Upon doing so, the mean-squared estimation error will be given by the
expression
\begin{equation}\label{mse}
\mathrm{mse}=  \Gfunvec_\nSlepfun^\funT \Lamat^{-1}_\nSlepfun
\left( \intR  \Gfunvecup_{\uparrow\nSlepfun} \cdot
\noisepower(\satalt\rvec,\satalt\rvec')
\cdot
 \Gfunvecup^\funT_{\uparrow\nSlepfun} \dOmega \right) \Lamat^{-1}_\nSlepfun
 \Gfunvec_\nSlepfun
+
\Gfunvec_{>\nSlepfun}^\funT
\left(\intO\intO \Gfunvec_{>\nSlepfun}
\sigpower(\Earthrad\rvec,\Earthrad\rvec') \,
\Gfunvec_{>\nSlepfun}^\funT \dOmega\dOmega'\right)
\Gfunvec_{>\nSlepfun}
,
\end{equation}
where the first term is the estimation variance, readily computed from
the difference between the expectation of the square of
eq.~(\ref{intestexpanded}) and its squared expectation, and the second
the expected value of the squared estimation bias in eq.~(\ref{knot}),
see \cite{Cox+74}.

It is a most welcome feature of the Slepian framework that truncation
of the basis neatly separates the effects of variance and bias, by
projection onto different basis functions altogether. But of course it
remains a feature common to all inverse problems that diminishing
variance comes at the price of increasing bias, and that solutions
that minimize the mean-squared estimation errors are accessible only
after experimentation and iteration, as determined by the
signal-to-noise ratio of what, ultimately, should turn out to be
signal, and what, noise. Disregarding the ultimate complexity of what,
practically, needs to be achieved to perform statistically efficient
internal-field estimation, eq.~(\ref{mse}) shows the `knobs' of the
system: a region~$\region$, a bandwidth~$\Lin$, an accompanying
satellite-altitude-cognizant Slepian basis that is driving the
mean-squared error through its eigenvalue structure~$\Lamat$, and a
truncation level~$\nSlepfun$ that remains to be judiciously chosen.

\subsection{Case study~I: Bandlimited and spectrally flat signal and noise}
\label{casestudyinonly}

Under the admittedly unrealistic if not mathematically impossible
scenario where both the signal and the noise should be bandlimited (to
the same degree~$\Lin$ as the Slepian functions used) and `white'
(uncorrelated between any two different space points), with the signal
and the noise completely uncorrelated, and with signal
power~$\sigpower$ and noise power~$\noisepower$, respectively, the
expression for the mean-squared error would take a simple form derived
from the identities in eq.~(\ref{intregionaltinner}), namely
\begin{equation}\label{poop}
\mathrm{mse}=
\noisepower\,\Gfunvec_\nSlepfun^\funT \Lamat^{-1}_\nSlepfun \Gfunvec_\nSlepfun
+
\sigpower\,\Gfunvec_{>\nSlepfun}^\funT\,\Gfunvec_{>\nSlepfun}
.
\end{equation}
Eq.~(\ref{poop}) is decidedly more palatable when contrasted with the
equivalent result~(174) of \cite{Plattner+2015c}, and thus illustrates
the benefits of using altitude-cognizant functions as advocated
here. Moreover, we find again eq.~(146) from \cite{Simons+2006b},
where it only applied to the zero-altitude case.

The mean-squared error~(\ref{poop}) is a function of all of
space~$\Omega$. To obtain the relative mean-squared error over the
target region~$\region$ we calculate its integral over the region and
normalize it by the integral over~$\region$ of the signal
power~$\sigpower$, for an example similar to the one described in
Section~\ref{numexsect}.

\begin{figure}\centering
\includegraphics[width=\eigwidthtwo,angle=0,trim= 1.5cm 10cm 1.8cm 10.5cm,clip]
{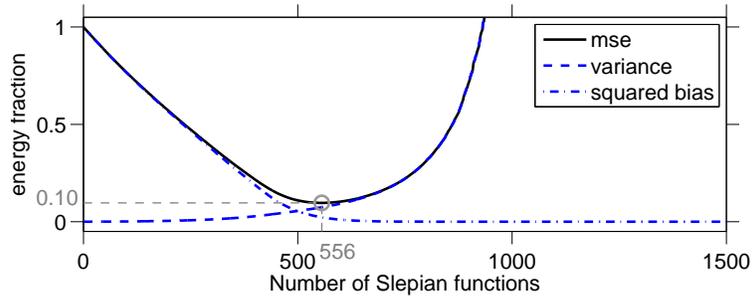}
\caption{\label{relativeMSEfig}Statistical performance of the
  truncated Slepian-function solution to the internal-field inversion
  problem, as computed via eq.~(\ref{poop}). Mean-squared error and
  its constituents, estimation variance and squared bias, all relative
  to the untruncated values. The optimal number of Slepian functions
  $\nSlepfun_\mathrm{opt}=556$ and the corresponding relative
  mean-squared error (0.1).}
\end{figure}

We set $\Lin=100, \Earthrad=6371$~km, $\satalt = 6671$~km, and
$\region = $ North America. As per eqs~(\ref{pipi})--(\ref{popo}) the
signal power~$\sigpower=\sigpower(\Earthrad)$ is given on the
planetary surface, while the noise
power~$\noisepower=\noisepower(\satalt)$ is at satellite altitude. To
calculate the relative mean-squared error for a realistic
signal-to-noise level, we need to calculate the signal power on the
planetary surface, $\sigpower(\Earthrad)$, as a function of that at
satellite altitude, $\sigpower(\satalt)$. Because the signal is white
on the surface we distribute the power evenly over the degrees such
that each degree contributes $\sigpower(\Earthrad)/(\Lin+1)$. We then
upward-continue to the satellite altitude by multiplying the values at
each degree power by the corresponding factor $\InUpelm_{lm,l'm'}^2$
in eq.~(\ref{InUpelements}), and obtain the signal power at satellite
altitude by summing those to obtain $\sigpower(\satalt)$. For the
values for $\Lin$, $\satalt$, and $\Earthrad$ that we chose, we obtain
$\sigpower(\Earthrad) \approx 1.1\times 10^{6}\,\sigpower(\satalt)$.
Fig.~\ref{relativeMSEfig} shows the region-average relative
mean-squared error as it depends on the number $\nSlepfun$ of
internal-field AC-\aGVSF{} used, together with the relative squared
bias and variance. In this example we set $\sigpower(\satalt)=1$ and
$\noisepower(\satalt)=0.01$. The optimal number of Slepian functions
$\nSlepfun_\mathrm{opt}=556$, with a relative mean-squared error
of~0.1.

\section{Analysis of the full-field AC-GVSF method}
\label{sec:statsintext}

To understand the effect of bandlimitation and Slepian truncation on
the full-field solution, we begin by defining the vectors of
(truncated) downward-continued AC-\aGVSF{} inspired by
eqs~(\ref{chit})--(\ref{chat}), namely
\begin{align}
\label{dobi}
\OutGfunvecup_{i\downarrow}&=\OutGmat_i^\matT\InUpmat^{-1}\Efunvec_\Lin
\also
\OutGfunvecup_{i\downarrow\nSlepfun}=\OutGmat_{i\nSlepfun}^\matT\,\InUpmat^{-1}\Efunvec_\Lin,\\
\label{doba}
\OutGfunvecup_{o\downarrow}&=
\OutGmat_o^\matT\OutUpmat^{-1}\Ffunvec_\Lout
\also
\OutGfunvecup_{o\downarrow\nSlepfun}=
\OutGmat_{o\nSlepfun}^\matT\,\OutUpmat^{-1}\Ffunvec_\Lout
,
\end{align}
for $1\le\nSlepfun\le\dimin+\dimout$, and their complements,
$\OutGfunvecup_{i\downarrow>\nSlepfun}$ and
$\OutGfunvecup_{o\downarrow>\nSlepfun}$. From the above definitions
and together with eq.~(\ref{jbo}) and eq.~(\ref{GinGouttrans}), we
obtain relationships similar to the ones we have for the purely
internal-field AC-\aGVSF{} in eq.~(\ref{inupdown}),
\begin{align}\label{axu}
\OutGfunvecup_{\uparrow}^\funT \OutGfunvecup_{i\downarrow} 
&=\OutGfunvecup_{i\uparrow}^\funT \OutGfunvecup_{i\downarrow} 
=\OutGfunvecup_{i\downarrow}^\funT \OutGfunvecup_{i\uparrow} 
=\Efunvec_\Lin^\funT \InUpmat^{-\matT} \OutGmat_i\OutGmat_i^\matT \InUpmat\,
\Efunvec_\Lin = \Efunvec_\Lin^\funT\Efunvec_\Lin,\\\label{axo}
\OutGfunvecup_{\uparrow}^\funT \OutGfunvecup_{o\downarrow} 
&=\OutGfunvecup_{o\uparrow}^\funT \OutGfunvecup_{o\downarrow} 
=\OutGfunvecup_{o\downarrow}^\funT \OutGfunvecup_{o\uparrow} 
=\Ffunvec_\Lout^\funT \OutUpmat^{-\matT} \OutGmat_o\OutGmat_o^\matT \OutUpmat\,
\Ffunvec_\Lout= \Ffunvec_\Lout^\funT \Ffunvec_\Lout.
\end{align}
With these, we rewrite the wideband eqs~(\ref{gradv})--(\ref{gradw})
in the following equivalent forms,
\begin{align}
\label{wesa}
\vecIntsignal
&=\Gfunvecup^\funT_{i\uparrow}\intO \Gfunvecup_{i\downarrow}\cdot
  \vecIntsignal \dOmega
 + \Efunvec_{>\Lin}^\funT\intO \Efunvec_{>\Lin}\cdot
  \vecIntsignal \dOmega\\
\label{wtsa}
&=\Efunvec_\Lin^\funT\InUpmat\sphcoefEarth
 + \Efunvec_{>\Lin}^\funT\intO \Efunvec_{>\Lin}\cdot
  \vecIntsignal \dOmega\\
\vecExtsignal
\label{wesb}
&=\Gfunvecup^\funT_{o\uparrow}\intO \Gfunvecup_{o\downarrow}\cdot
  \vecExtsignal \dOmega
 + \Ffunvec_{>\Lout}^\funT\intO \Ffunvec_{>\Lout}\cdot
  \vecExtsignal \dOmega\\
\label{wtsb}
&=\Ffunvec_\Lout^\funT\InUpmat\outsphcoefEarth
 + \Ffunvec_{>\Lout}^\funT\intO \Ffunvec_{>\Lout}\cdot
  \vecExtsignal \dOmega
.
\end{align}

\subsection{Relationship to classical spherical Slepian functions}
\label{sec:relint2}

We obtained the full-field altitude-cognizant \GVSFs{} from solving
misfit-minimization problem
eq~(\ref{inner-outer-sourceoptimizationproblem}) and diagonalizing the
matrix~$\OutKmat$ in eq.~(\ref{kernel_inout}) via
eq.~(\ref{eigenprobleminnerouter}). The coefficients~$\OutGmat$ in
eq.~(\ref{bo}) can alternatively be obtained by solving an energy
maximization problem, as were, for example, the classical vector
Slepian functions of \cite{Plattner+2014a}. In
Section~\ref{sec:relint1}, we maximized the energy over the target
region of the upward-continued function relative to its scalar
incarnation on the planetary surface. Here, we have the additional
complication that we have two scalar fields inhabiting two different
radial positions, $\Earthrad$ and $\Outrad$. Using
eqs~(\ref{OutGmatstructure}), (\ref{blo})--(\ref{bli})
and~(\ref{jbo}), we write
\begin{align}\label{optimizationprobleminout}
  \lambda 
&=
  \frac{\displaystyle\Outslepfuncoef^\matT \OutKmat\,\Outslepfuncoef}{\displaystyle \Outslepfuncoef^\matT
    \Outslepfuncoef}
=
  \frac{\displaystyle
\begin{pmatrix}\Outslepfuncoef_i^\matT & \Outslepfuncoef_o^\matT\end{pmatrix}
 \OutKmat\, 
\begin{pmatrix}\Outslepfuncoef_i^\matT & \Outslepfuncoef_o^\matT\end{pmatrix}^\matT
}{\displaystyle\Outslepfuncoef_i^\matT \Outslepfuncoef_i +
    \Outslepfuncoef_o^\matT\Outslepfuncoef_o} =\frac{ \displaystyle
    \intR
    \left(\Outslepfuncoef_i^\matT\InUpmat\,\Efunvec_\Lin
    +\Outslepfuncoef_o^\matT\OutUpmat\Ffunvec_\Lout\right) 
    \cdot \left(\Efunvec_\Lin^\funT\InUpmat^\matT \Outslepfuncoef_i+
    \Ffunvec_\Lout^\funT \OutUpmat^\matT \Outslepfuncoef_o\right)
    d\Omega} {\displaystyle \intO \left(\Outslepfuncoef_i^\matT
    \Yfunvec_\Lin\right)\left(\Yfunvec_\Lin^\funT \Outslepfuncoef_i\right)
    \dOmega + \intO \left(\Outslepfuncoef_o^\matT
    \Yfunvec_\Lout\right) \left(\Yfunvec_\Lout^\funT
    \Outslepfuncoef_o\right) \dOmega}\\ &= \frac{\displaystyle
    \intR \OutGfunup_\uparrow^2 \dOmega}{\displaystyle
    \intO \OutGfun_i^2\dOmega + \intO \OutGfun_o^2\dOmega}
  =\mathrm{maximum}.
\end{align} 
Among all bandlimited upward-continued gradient-vector functions that
are linear combinations of the basis sets~$\InUpmat\,\Efunvec_\Lin$
and $\OutUpmat\Ffunvec_\Lout$, the first AC-\aGVSF{},
$\OutGfunup_{\uparrow 1}$, is the best-concentrated in the
sense~(\ref{optimizationprobleminner}). The concentration factor
$\lambda_1$ is the first eigenvalue associated with the
diagonalization problem~(\ref{eigenprobleminner}). The second-best
concentrated AC-\aGVSF{}, $\OutGfunup_{\uparrow 2}$, and its
corresponding $\lambda_2$, is the next best function in the
sense~(\ref{optimizationprobleminner}) that is orthogonal to
$\OutGfunup_{\uparrow 1}$, and so on.

The full-field AC-\aGVSF{} $\OutGfunvecup_{\uparrow\alpha}$ of
eq.~(\ref{jbo}) obey the same orthogonality relations as their purely
internal-field siblings described in Section~\ref{sec:relint1}. Their
gradient vector incarnations at average satellite altitude are
orthogonal over the region $\region$ but not over the entire
sphere~$\Omega$, and the full-field scalar functions of
eqs~(\ref{blo})--(\ref{bli}) are orthogonal over~$\Omega$ but not
over~$\region$ at the construction altitudes. With $\OutLamat$ the
eigenvalue matrix as in eq.~(\ref{eigenprobleminnerouter}) and~$\Imat$
the identity, we have relations equivalent to
eqs~(\ref{intregionaltinner})--(\ref{intregionsurfaceinner}), namely
\begin{align}
  \label{intregionaltinout}
  \intR \OutGfunvecup_\uparrow\cdot\OutGfunvecup_\uparrow^\funT
  d\Omega &= \OutLamat
\also
  \intO \OutGfunvecup_\uparrow\cdot\OutGfunvecup_\uparrow^\funT
 d\Omega = \OutGmat^\matT
  \begin{pmatrix}
    \InUpmat\InUpmat^\matT&\zerovec\\
 \zerovec&\OutUpmat\OutUpmat^\matT
  \end{pmatrix}
  \OutGmat,\\
  \label{intallsurfaceinout}
  \intO \left(\OutGfunvec_i\OutGfunvec_i^\funT  +
  \OutGfunvec_o\OutGfunvec_o^\funT\right) d\Omega &=\Imat
\also
  \intR \left(\OutGfunvec_i\OutGfunvec_i^\funT  +
 \OutGfunvec_o\OutGfunvec_o^\funT\right) d\Omega=
  \OutGmat_i^\matT \left(\intR \Yfunvec_\Lin
  \Yfunvec_\Lin^\funT d\Omega\right)\OutGmat_i +\OutGmat_o^\matT 
\left(\intR \Yfunvec_\Lout \Yfunvec_\Lout^\funT d\Omega\right)\OutGmat_o.
\end{align}
Note that in eq.~(\ref{intallsurfaceinout}), as we recall from
eq.~(\ref{OutGmatstructure}), the matrix $\OutGmat_i$ is of dimension
$\dimin\times[\dimin+\dimout]$, whereas the matrix $\OutGmat_o$ is of
size $[\dimout]\times[\dimin+\dimout]$. Both
$\OutGmat_i^\matT\OutGmat_i$ and $\OutGmat_o^\matT\OutGmat_o$\linebreak are
$[\dimin+\dimout]\times[\dimin+\dimout]$, and they are both
singular. Their sum is a unit matrix, see eq.~(\ref{klux}).

As it did in eq.~(\ref{nece}), basis truncation to the
first~$\nSlepfun$ vectors leads to the appropriately resized
subscripted relations
\begin{equation}
  \label{hohoho}
  \intR
  \OutGfunvecup_{\uparrow\nSlepfun}\cdot\OutGfunvecup_{\uparrow}^\funT
  \dOmega=\begin{pmatrix}\OutLamat_\nSlepfun&\Omat\end{pmatrix} \also
  \intO \left(
  \OutGfunvec_{i\nSlepfun}\OutGfunvec_i^\funT +
  \OutGfunvec_{o\nSlepfun}\OutGfunvec_o^\funT\right) d\Omega
  =\begin{pmatrix}\Imat_\nSlepfun&\Omat\end{pmatrix} .
\end{equation}

\subsection{Spatially restricted, spectrally concentrated full-field Slepian functions}

As we did in eq.~(\ref{InUpTruncRelation}) we define specific sets of
exactly spacelimited, broadband full-field altitude-cognizant \GVSFs{}
obtained, respectively, as $\Hfunor^{\Efunvec}_{\uparrow
  \alpha,>\Lin}= \big(\vslepfuncoef^{\Efunvec}_{\uparrow \alpha,
  >\Lin}\big)^\matT\Efunvec_{>\Lin}$ and $\Hfunor^{\Ffunvec}_{\uparrow
  \alpha,>\Lout}= \big(\vslepfuncoef^{\Ffunvec}_{\uparrow \alpha,
  >\Lout}\big)^\matT\Ffunvec_{>\Lout}$, in the truncated vectors
\begin{align}\label{Hrelationsint}
\OutHfunvec^{\Efunvec}_{\uparrow\nSlepfun,>\Lin}
&=\begin{pmatrix}
\Hfunor^{\Efunvec}_{\uparrow 1,>\Lin}&\cdots& 
\Hfunor^{\Efunvec}_{\uparrow\alpha,>\Lin}&\cdots& 
\Hfunor^{\Efunvec}_{\uparrow\nSlepfun,>\Lin}\end{pmatrix}\Tit
= \left(\intR \OutGfunvecup_{\uparrow\nSlepfun}\cdot
\Efunvec_{>\Lin}^\funT \dOmega\right) \Efunvec_{>\Lin},\\ 
\label{Hrelationsout}
\OutHfunvec^{\Ffunvec}_{\uparrow\nSlepfun,>\Lout}
&=\begin{pmatrix}
\Hfunor^{\Ffunvec}_{\uparrow 1,>\Lout}&\cdots& 
\Hfunor^{\Ffunvec}_{\uparrow\alpha,>\Lout}&\cdots& 
\Hfunor^{\Ffunvec}_{\uparrow\nSlepfun,>\Lout}\end{pmatrix}\Tit
= \left(\intR \OutGfunvecup_{\uparrow\nSlepfun}\cdot
\Ffunvec_{>\Lout}^\funT \dOmega\right) \Ffunvec_{>\Lout}. 
\end{align}

The infinite vector of coefficients
$\Outvslepfuncoef^{\Efunvec}_{\uparrow\alpha,>\Lin}$ contains the
$\Efun_{lm}$-components at the spherical-harmonic degrees exceeding
$\Lin$, and the infinite vector
$\Outvslepfuncoef^{\Ffunvec}_{\uparrow\alpha,>\Lout}$ contains the
$\Ffun_{lm}$-components at the degrees above $\Lout$, of the hard
spatial truncation of the bandlimited AC-\aGVSF{}
$\OutGfunup_{\uparrow\alpha}$ to the region~$\region$. As for the
internal-field case in eq.~(\ref{inuptrunc}), we can obtain these
coefficients directly from the full-field AC-\aGVSF{} coefficients
$\Outslepfuncoef_{\alpha}$ by multiplication with the appropriate
broadband localization kernel extensions, for each
$\alpha=1,\dots,\dimin+\dimout$, respectively,
\begin{align}\label{blipE}
\Outvslepfuncoef^{\Efunvec}_{\uparrow\alpha,>\Lin}
&=\intR\Efunvec_{>\Lin}\cdot\OutGfunup_{\uparrow\alpha}\dOmega
=\left(\intR\Efunvec_{>\Lin}\cdot\Efunvec_\Lin^\funT
d\Omega\right)\InUpmat^\matT\Outslepfuncoef_{i\alpha}
+\left(\intR\Efunvec_{>\Lin}\cdot\Ffunvec_\Lout^\funT
d\Omega\right)\OutUpmat^\matT\Outslepfuncoef_{o\alpha},\\ \label{blipF}
\Outvslepfuncoef^{\Ffunvec}_{\uparrow\alpha,>\Lout}
&=\intR\Ffunvec_{>\Lout}\cdot\OutGfunup_{\uparrow \alpha}\dOmega
=\left(\intR\Ffunvec_{>\Lout}\cdot\Efunvec_\Lin^\funT
d\Omega\right)\InUpmat^\matT\Outslepfuncoef_{i\alpha}
+\left(\intR\Ffunvec_{>\Lout}\cdot\Ffunvec_\Lout^\funT
d\Omega\right)\OutUpmat^\matT\Outslepfuncoef_{o\alpha}.
\end{align}
The above relations are easily derived from eq.~(\ref{jbo}). 

\subsection{Statistical analysis of the full-field method}
\label{sec:fullanalysis}

As in Section~\ref{sec:intanalysis} we recall the infinitely broadband
target field as composed of the pieces in
eqs~(\ref{gradv})--(\ref{gradw}). We decompose the internal-field and
external-field potentials (\ref{Vrr}) and~(\ref{Wrr}) in terms of the
full-field altitude-cognizant scalar Slepian functions of
eqs~(\ref{blo})--(\ref{bli}) into a bandlimited part, for which we use
eqs~(\ref{transfo2})--(\ref{needit2}), and a broadband complement,
\begin{align}\label{hooplo}
\Intsignal(\Earthrad \rvec)&= \OutGfunvec_{i\nSlepfun}^\funT \intO
\OutGfunvec_{i\nSlepfun}\Intsignal(\Earthrad \rvec) \dOmega +
\OutGfunvec_{i>\nSlepfun}^\funT \intO \OutGfunvec_{i>\nSlepfun}
\Intsignal(\Earthrad \rvec)\dOmega +\Yfunvec_{>\Lin}^\funT
\intO \Yfunvec_{>\Lin} \Intsignal(\Earthrad \rvec)\dOmega,\\
\label{ergo}
\Extsignal(\Outrad \rvec)&= \OutGfunvec_{o\nSlepfun}^\funT \intO
\OutGfunvec_{o\nSlepfun}\Extsignal(\Outrad \rvec) \dOmega +
\OutGfunvec_{o>\nSlepfun}^\funT \intO \OutGfunvec_{o>\nSlepfun}
\Extsignal(\Outrad \rvec)\dOmega +\Yfunvec_{>\Lout}^\funT
\intO \Yfunvec_{>\Lout} \Extsignal(\Outrad \rvec)\dOmega
.
\end{align}
The data in eq.~(\ref{dataregionintext}) are broken down in terms of
the contributions by the AC-\aGVSF{} with the help of
eqs~(\ref{axu})--(\ref{axo}), (\ref{wesa}) and~(\ref{wesb}),
\begin{equation}\label{dintextexpanded}
\datavec=
\OutGfunvecup_{\uparrow}^\funT\intO\OutGfunvecup_{i\downarrow}\cdot\vecIntsignal\dOmega+
\OutGfunvecup_{\uparrow}^\funT\intO\OutGfunvecup_{o\downarrow}\cdot\vecExtsignal\dOmega+
\Efunvec_{>\Lin}^\funT\intO\Efunvec_{>\Lin}\cdot\vecIntsignal\dOmega+
\Ffunvec_{>\Lout}^\funT\intO\Ffunvec_{>\Lout}\cdot\vecExtsignal\dOmega+\vecnoise.
\end{equation}
The bandlimited truncated full-field AC-\aGVSF{} estimator described
in eqs~(\ref{analyticalsolutioninout})
through~(\ref{slepianfunctionexpansionout}) is the sum of two terms,
\begin{align}\label{intanalyticsol}
\estIntsignal_\nSlepfun(\Earthrad\rvec)&=
\OutGfunvec_{i\nSlepfun}^\funT\,\OutLamat_J^{-1}\intR
\OutGfunvecup_{\uparrow\nSlepfun}\cdot \datavec\dOmega,\\
\label{extanalyticsol}
\estExtsignal_\nSlepfun(\Outrad\rvec)&=
\OutGfunvec_{o\nSlepfun}^\funT\,\OutLamat_J^{-1}\intR
\OutGfunvecup_{\uparrow\nSlepfun}\cdot \datavec\dOmega
.
\end{align}

Inserting eq.~(\ref{dintextexpanded}) into
eqs~(\ref{intanalyticsol})--(\ref{extanalyticsol}) and using
eqs~(\ref{hohoho}) and~(\ref{Hrelationsint})--(\ref{Hrelationsout}),
and then eqs~(\ref{wtsa}), (\ref{wtsb})
and~(\ref{dobi})--(\ref{doba}), the orthonormality~(\ref{ortho2}) of
the $\Efunvec$ and the $\Ffunvec$, and, finally,
eqs~(\ref{transfo2})--(\ref{needit2}), we obtain
\begin{align}\nonumber
\estIntsignal_\nSlepfun(\Earthrad\rvec)
\nonumber
&= \OutGfunvec_{i\nSlepfun}^\funT
  \intO
 \OutGfunvec_{i\nSlepfun}\Intsignal(\Earthrad \rvec)
 \dOmega + \OutGfunvec_{i\nSlepfun}^\funT \intO
  \OutGfunvec_{o\nSlepfun}\Extsignal(\Outrad \rvec)
 \dOmega\\
\label{estinternaltotal}
&\qquad+\OutGfunvec_{i\nSlepfun}^\funT \OutLamat_{\nSlepfun}^{-1}
  \left(\intO 
  \OutHfunvec^{\Efunvec}_{\uparrow \nSlepfun, >\Lin} \cdot \vecIntsignal
  \dOmega + \intO
  \OutHfunvec^{\Ffunvec}_{\uparrow \nSlepfun, >\Lout} \cdot  \vecExtsignal
  \dOmega +\intR\OutGfunvecup_{\uparrow\nSlepfun}
  \cdot\vecnoise\dOmega \right),\\\nonumber
  \estExtsignal_\nSlepfun(\Outrad\rvec)
&= \OutGfunvec_{o\nSlepfun}^\funT
  \intO 
  \OutGfunvec_{o\nSlepfun} \Extsignal(\Outrad \rvec)
 \dOmega + \OutGfunvec_{o\nSlepfun}^\funT \intO
  \OutGfunvec_{i\nSlepfun} \Intsignal(\Earthrad \rvec)
  \dOmega
  \\ &\qquad+ \OutGfunvec_{o\nSlepfun}^\funT
  \OutLamat_{\nSlepfun}^{-1} \left( \intO
  \OutHfunvec^{\Efunvec}_{\uparrow \nSlepfun, >\Lin} \cdot  \vecIntsignal
 \dOmega + \intO 
\OutHfunvec^{\Ffunvec}_{\uparrow  \nSlepfun, >\Lout} \cdot \vecExtsignal
\dOmega
  +\intR\OutGfunvecup_{\uparrow\nSlepfun} \cdot\vecnoise\dOmega
  \right).\label{hungry}
\end{align}
The first right-hand side term in eq.~(\ref{estinternaltotal})
describes the component of the estimated bandlimited internal
potential field~$\Intsignal$ that stems from the bandlimited internal
vector field $\vecIntsignal$. From the second term in
eq.~(\ref{estinternaltotal}) we learn that our estimation of
$\Intsignal$ includes leakage from the external field also. This
leakage stems from Slepian truncation, decreases with increasing
$\nSlepfun$, and vanishes when $\nSlepfun=\dimin+\dimout$, due to
eq.~(\ref{GinGouttrans}), which does not hold for
$\nSlepfun<\dimin+\dimout$. The next terms describe leakage from the
broadband components of the internal and external vector
fields, and from the noise. These last components are multiplied with
the inverse of the eigenvalues $\Outlambda_\alpha$ from
eq.~(\ref{eigenprobleminnerouter}), which approach zero for
$\nSlepfun$ large. Hence, broadband and noise leakage increases with
increasing $\nSlepfun$. Equivalent considerations apply to the
interpretation of eq.~(\ref{hungry}).

Under the assumption of zero-mean, uncorrelated noise, the difference
between the expected values of
eqs.~(\ref{estinternaltotal})--(\ref{hungry}) and the truths in
eqs~(\ref{hooplo})--(\ref{ergo}) yields the estimation bias terms
\begin{align}\label{bibiint1}
  b_\Intsignal&=-\OutGfunvec_{i>\nSlepfun}^\funT \intO
  \OutGfunvec_{i>\nSlepfun} \Intsignal(\Earthrad\rvec)\dOmega
  -\Yfunvec_{>\Lin}^\funT \intO \Yfunvec_{>\Lin}
  \Intsignal(\Earthrad \rvec)\dOmega + \OutGfunvec_{i\nSlepfun}^\funT
  \intO \OutGfunvec_{o\nSlepfun}
  \Extsignal(\Outrad\rvec)\dOmega,\\ \label{bibiext1}
  b_\Extsignal&=-\OutGfunvec_{o>\nSlepfun}^\funT
  \intO \OutGfunvec_{o>\nSlepfun} \Extsignal(\Outrad\rvec)\dOmega
  -\Yfunvec_{>\Lout}^\funT \intO \Yfunvec_{>\Lout}
  \Extsignal(\Outrad \rvec)\dOmega + \OutGfunvec_{o\nSlepfun}^\funT
  \intO \OutGfunvec_{i\nSlepfun}
  \Intsignal(\Earthrad\rvec)\dOmega,\\ \label{bibiint2}
  b_{\Intsignal\Extsignal}&=\OutGfunvec_{o\nSlepfun}^\funT
  \OutLamat_{\nSlepfun}^{-1} \left( \intO
  \OutHfunvec^{\Efunvec}_{\uparrow \nSlepfun, >\Lin} \cdot
  \vecIntsignal \dOmega + \intO \OutHfunvec^{\Ffunvec}_{\uparrow
    \nSlepfun, >\Lout} \cdot \vecExtsignal \dOmega \right),\\ \label{bibiext2}
  b_{\Extsignal\Intsignal}&=\OutGfunvec_{i\nSlepfun}^\funT
  \OutLamat_{\nSlepfun}^{-1} \left( \intO
  \OutHfunvec^{\Efunvec}_{\uparrow \nSlepfun, >\Lin} \cdot
  \vecIntsignal \dOmega + \intO \OutHfunvec^{\Ffunvec}_{\uparrow
    \nSlepfun, >\Lout} \cdot \vecExtsignal \dOmega \right)
,
\end{align}
where the bias of the internal-field estimate is given by $b_\Intsignal+b_{\Intsignal\Extsignal}$ and the
bias of the external-field estimate is given by $b_\Extsignal+b_{\Extsignal\Intsignal}$. In the absence of
Slepian-function truncation, when $\nSlepfun=\dimin+\dimin$, the terms
that are subscripted $>\nSlepfun$ are `squeezed' to vanish altogether,
as are the terms that then involve
$\OutGfunvec_{i\nSlepfun}^\funT\,\OutGfunvec_{o\nSlepfun}=
\OutGfunvec_{i}^\funT\OutGfunvec_{o}=0$ and
$\OutGfunvec_{o\nSlepfun}^\funT \,\OutGfunvec_{i\nSlepfun}=
\OutGfunvec_{o}^\funT\OutGfunvec_{i}=0$, again by virtue of
eqs~(\ref{blo})--(\ref{bli}) and~(\ref{GinGouttrans}). What is then
left are the bias terms that arise from forming bandlimited estimates
of broadband fields, which we deem unavoidable.

If the internal and external fields, and the noise are thought of as
mutually uncorrelated random processes with two-point covariances
\begin{align}\label{pipi2}
\sigpower(\Earthrad\rvec,\Earthrad\rvec')&=\left\langle
\Intsignal(\Earthrad\rvec)\,\Intsignal(\Earthrad\rvec')\right\rangle
,\\\label{papa2}
\sigpowerw(\Outrad\rvec,\Outrad\rvec')&=\left\langle
\Extsignal(\Outrad\rvec)\,\Extsignal(\Outrad\rvec')\right\rangle
,\\\label{popo2}
\noisepower(\satalt\rvec,\satalt\rvec')&=\left\langle
\vecnoise(\satalt\rvec)\cdot\vecnoise(\satalt\rvec')\right\rangle
,
\end{align}
and removing all the essentially unknowable broadband terms from
consideration, the mean-squared estimation errors will be given by 
\begin{align}\nonumber
\mathrm{mse}_\Intsignal= \OutGfunvec_{i\nSlepfun}^\funT
\OutLamat^{-1}_\nSlepfun \left( \intR
\OutGfunvecup_{\uparrow\nSlepfun} \cdot
\noisepower(\satalt\rvec,\satalt\rvec') \cdot
\OutGfunvecup^\funT_{\uparrow\nSlepfun} \dOmega \right)
\OutLamat^{-1}_\nSlepfun \OutGfunvec_{i\nSlepfun} &+
\OutGfunvec_{i>\nSlepfun}^\funT \left(\intO\intO
\OutGfunvec_{i>\nSlepfun} \sigpower(\Earthrad\rvec,\Earthrad\rvec') \,
\OutGfunvec_{i>\nSlepfun}^\funT \dOmega\dOmega'\right)
\OutGfunvec_{i>\nSlepfun}\\\label{mseV} & +
\OutGfunvec_{i\nSlepfun}^\funT \left(\intO\intO \OutGfunvec_{o}
\sigpowerw(\Earthrad\rvec,\Earthrad\rvec') \, \OutGfunvec_{o}^\funT
\dOmega\dOmega'\right) \OutGfunvec_{i\nSlepfun},\\\nonumber
\mathrm{mse}_\Extsignal= \OutGfunvec_{o\nSlepfun}^\funT
\OutLamat^{-1}_\nSlepfun \left( \intR
\OutGfunvecup_{\uparrow\nSlepfun} \cdot
\noisepower(\satalt\rvec,\satalt\rvec') \cdot
\OutGfunvecup^\funT_{\uparrow\nSlepfun} \dOmega \right)
\OutLamat^{-1}_\nSlepfun \OutGfunvec_{o\nSlepfun} &+
\OutGfunvec_{o>\nSlepfun}^\funT \left(\intO\intO
\OutGfunvec_{o>\nSlepfun} \sigpowerw(\Earthrad\rvec,\Earthrad\rvec') \,
\OutGfunvec_{o>\nSlepfun}^\funT \dOmega\dOmega'\right)
\OutGfunvec_{o>\nSlepfun}\\\label{mseW} & +
\OutGfunvec_{o\nSlepfun}^\funT \left(\intO\intO \OutGfunvec_{i}
\sigpower(\Earthrad\rvec,\Earthrad\rvec') \, \OutGfunvec_{i}^\funT
\dOmega\dOmega'\right) \OutGfunvec_{o\nSlepfun} .
\end{align}
Again, the first terms in eqs~(\ref{mseV}) and~(\ref{mseW}) are due
variance, and the remaining terms due to bias squared
\cite[]{Cox+74}. As to the former, the less we truncate (at high
$\nSlepfun$) the solution, the more we pick up data noise. As to the
latter, the more we truncate (at low $\nSlepfun$), the more signal we
leave unaccounted for. Between the two effects, we recognize the
customary trade-off, as we remarked with eq.~(\ref{mse}).

\begin{figure}\centering
\includegraphics[width=\eigwidthtwo,angle=0,trim= 1.5cm 10cm 1.8cm 10.5cm,clip]
                  {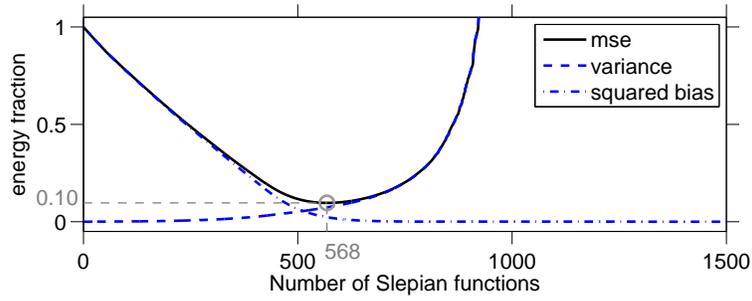}
\caption{\label{relativeMSEintext}Statistical performance of the
  truncated Slepian-function internal-field solution to the full-field
  inversion problem, as computed via eq.~(\ref{paap}). The optimal
  number of Slepian functions $\nSlepfun_\mathrm{opt}=568$ and the
  corresponding relative mean-squared error (0.1) are
  indicated. Layout is as in Fig.~\ref{relativeMSEfig}.}
\end{figure}

\subsection{Case study~II: Bandlimited and spectrally flat signal and noise} 
\label{casestudyintext}

If the internal field is bandlimited with the same bandwidth~$\Lin$ as
the internal-field Slepian functions, with a `whitish' spectrum on the
planetary surface with power $\sigpower$, and if the external field is
bandlimited with the same bandwidth~$\Lout$ as the external-field
Slepian functions, and with a white spectrum on the outer sphere with
power $\sigpowerw$, eqs~(\ref{mseV}) and~(\ref{mseW}) simplify through
eq.~(\ref{intregionaltinout}) to the rather digestible forms
\begin{align}\label{paap}
\text{mse}_\Intsignal&=
\noisepower \OutGfunvec_{i\nSlepfun}^\funT \OutLamat_\nSlepfun^{-1}
  \OutGfunvec_{i\nSlepfun}+
  \sigpower\OutGfunvec_{i>\nSlepfun}^\funT
\left(  \OutGmat_{i>\nSlepfun}^\matT\OutGmat_{i>\nSlepfun} \right)
  \OutGfunvec_{i>\nSlepfun} +
  \sigpowerw\OutGfunvec_{i\nSlepfun}^\funT
\left(  \OutGmat_{o\nSlepfun}^\matT\OutGmat_{o\nSlepfun} \right)
    \OutGfunvec_{i\nSlepfun},\\
\text{mse}_\Extsignal&=
\noisepower \OutGfunvec_{o\nSlepfun}^\funT \OutLamat_\nSlepfun^{-1}
  \OutGfunvec_{o\nSlepfun}+
  \sigpower\OutGfunvec_{o>\nSlepfun}^\funT
\left(  \OutGmat_{o>\nSlepfun}^\matT\OutGmat_{o>\nSlepfun} \right)
  \OutGfunvec_{o>\nSlepfun} +
  \sigpowerw\OutGfunvec_{o\nSlepfun}^\funT
\left(  \OutGmat_{i\nSlepfun}^\matT\OutGmat_{i\nSlepfun} \right)
    \OutGfunvec_{o\nSlepfun}.
\end{align}

To observe the behavior of $\text{mse}_\Intsignal$ depending on the
truncation number of Slepian functions $\nSlepfun$ for a case similar
to the one considered in Section~\ref{numexsect}, we set $\Lin=100$,
$\Lout=10$, $\Earthrad = 6371$ km, $\satalt= 6671$ km, $\Outrad =
6771$ km, and $\region = $ North America. To obtain a realistic
signal-to-noise ratio at satellite altitude we apply the same
principle as earlier to the internal-field power $\sigpower$ and,
mutatis mutandis, to the external-field power $\sigpowerw$. We obtain
$\sigpower(\Earthrad) = 1.1\times 10^{6}\,\sigpower(\satalt)$ and
$\sigpowerw(\Outrad) = 1.5\times 10^{6}\,\sigpowerw(\satalt)$. For the
example presented in Fig.~\ref{relativeMSEintext} we chose
$\sigpower(\satalt) = 1 $, $\sigpowerw(\satalt) =0.1$, and
$\noisepower = 0.01$. As for the internal-field case described in
Section~\ref{casestudyinonly}, we calculate the regional integral of
the mean-squared error and normalize it by the regional integral of
the signal power. Fig.~\ref{relativeMSEintext} shows how the bias term
decreases with increasing number $\nSlepfun$ of Slepian functions. On
the other hand, the variance term increases with
increasing~$\nSlepfun$. Together, the variance and bias lead to an
optimal number~$\nSlepfun_\mathrm{opt} = 568$ that minimizes the
mean-squared error at~0.1.

\section{Conclusions}
\label{sec:conclusions}

We presented two methods to invert for a regional representation of
potential fields on a planetary or lunar surface from discrete,
regionally available, vector data collected at varying radial
positions. The first method only considers internal fields, whereas
the second simultaneously inverts for internal and external
fields. Both methods are based on systems of functions that arise from
solving optimization problems that take the region of data
availability and the ensuing downward continuation of the field into
account. In our numerical tests we observed that under favorable noise
conditions, the estimated internal field faithfully represents the
true field within the region of data availability, but is
unconstrained outside of this region. Our tests also revealed some of
the dangers of not considering external fields when they are present
and are not removed by other means. When large-scale external fields
were left unaccounted for, the solution contained erroneous
small-scale features. \cite{Plattner+2015a} previously applied the
internal-field method described in this paper to map the South Polar
crustal magnetic field of Mars, after subtracting an external-field
model from the data before solving for the internal field. We provided
a detailed statistical analysis of both methods, highlighting, in
particular, the leakage induced by unaccounted-for data
components. For a contrived special case of spectrally flat and
bandlimited data, we derived simple analytic expressions for the bias,
variance, and mean-squared error of the estimates, which allows us to
predict the solution error, which is dominantly controlled by the
number of altitude-cognizant \GVSFs{} used in the truncated model
expansion. These methods are constructed to maximize the numerical
conditioning of the solution under downward continuation from an
average satellite altitude. In principle the harmonic continuation
matrices could be replaced with any other invertible matrix, such as a
noise covariance, and the solution optimized to counteract the
influence of noisy measurements, downward continuation, or both. The
construction of altitude-cognizant \GVSFs{} necessitates solving an
eigenvalue problem which, at large bandwidths, may become
computationally expensive. However, for symmetric regions, such as
spherical caps, belts or rings, the original eigenvalue problem can be
simplified into a set of smaller eigenvalue problems, which can be
solved in parallel, dramatically reducing the computational cost.

\section{Acknowledgments}

This work was sponsored by the National Aeronautics and Space
Administration under grant NNX14AM29G. FJS thanks the Institute for
Advanced Study for a hospitable environment during the academic year
2014--2015, and the K.U.~Leuven for a productive working environment
in the summers of 2015 and 2016. The authors thank Nils Olsen, an
anonymous referee, and the Associate Editor, Kosuke Heki, for
constructive reviews of the submitted manuscript.

\bibliographystyle{natbib}
\bibliography{./bib}

\begin{thebibliography}{}

\bibitem[Albee {\em et~al.}(2001)Albee, Arvidson, Palluconi, and
  Thorpe]{Albee+2001}
Albee, A.~L., Arvidson, R.~E., Palluconi, F., and Thorpe, T. (2001).
\newblock Overview of the {M}ars {G}lobal {S}urveyor mission.
\newblock {\em J.~Geophys.~Res.}, {\bf 106}(E10), 23291--23316, doi:
  10.1029/2000JE001306.

\bibitem[Albertella {\em et~al.}(1999)Albertella, Sans{\`o}, and
  Sneeuw]{Albertella+99}
Albertella, A., Sans{\`o}, F., and Sneeuw, N. (1999).
\newblock Band-limited functions on a bounded spherical domain: the {S}lepian
  problem on the sphere.
\newblock {\em J.~Geod.}, {\bf 73}, 436--447.

\bibitem[Amirbekyan {\em et~al.}(2008)Amirbekyan, Michel, and
  Simons]{Amirbekyan+2008b}
Amirbekyan, A., Michel, V., and Simons, F.~J. (2008).
\newblock Parameterizing surface-wave tomopgraphic models with harmonic
  spherical splines.
\newblock {\em Geophys.~J.~Int.}, {\bf 174}(2), 617--628, doi:
  {10.1111/j.1365--246X.2008.03809.x}.

\bibitem[Aster {\em et~al.}(2013)Aster, Borchers, and Thurber]{Aster+2013}
Aster, R.~C., Borchers, B., and Thurber, C.~H. (2013).
\newblock {\em Parameter Estimation {a}nd Inverse Problems\/}.
\newblock Elsevier Academic Press, San Diego, Calif.

\bibitem[Backus(1986)Backus]{Backus86}
Backus, G. (1986).
\newblock Poloidal and toroidal fields in geomagnetic field modeling.
\newblock {\em Rev.~Geophys.}, {\bf 24}(1), 75--109, doi:
  10.1029/RG024i001p00075.

\bibitem[Backus {\em et~al.}(1996)Backus, Parker, and Constable]{Backus+96}
Backus, G.~E., Parker, R.~L., and Constable, C.~G. (1996).
\newblock {\em Foundations of Geomagnetism\/}.
\newblock Cambridge Univ.~Press, Cambridge, UK.

\bibitem[Bates {\em et~al.}(2017)Bates, Khalid, and Kennedy]{Bates+2017}
Bates, A., Khalid, Z., and Kennedy, R. (2017).
\newblock Efficient computation of {S}lepian functions for arbitrary regions on
  the sphere.
\newblock {\em IEEE Trans.~Signal~Process.}, page under review.

\bibitem[Baur and Sneeuw(2011)Baur and Sneeuw]{Baur+2011}
Baur, O. and Sneeuw, N. (2011).
\newblock Assessing {G}reenland ice mass loss by means of point-mass modeling:
  a viable methodology.
\newblock {\em J.~Geod.}, {\bf 85}(9), 607--615, doi:
  10.1007/s00190--011--0463--1.

\bibitem[Blakely(1995)Blakely]{Blakely95}
Blakely, R.~J. (1995).
\newblock {\em Potential Theory {i}n Gravity {a}nd Magnetic Applications\/}.
\newblock Cambridge Univ.~Press, New York.

\bibitem[B{\"o}lling and Grafarend(2005)B{\"o}lling and
  Grafarend]{Boelling+2005}
B{\"o}lling, K. and Grafarend, E.~W. (2005).
\newblock Ellipsoidal spectral properties of the {E}arth's gravitational
  potential and its first and second derivatives.
\newblock {\em J.~Geod.}, {\bf 79}(6--7), 300--330, doi:
  10.1007/s00190--005--0465--y.

\bibitem[Connerney(2015)Connerney]{Connerney2015}
Connerney, J. E.~P. (2015).
\newblock Planetary magnetism.
\newblock In T.~Spohn, editor, {\em Treatise on Geophysics\/}, volume~10, pages
  195--237, doi: 10.1016/B978--0--444--53802--4.00171--8. Elsevier, Amsterdam,
  Neth., 2 edition.

\bibitem[Cox and Hinkley(1974)Cox and Hinkley]{Cox+74}
Cox, D.~R. and Hinkley, D.~V. (1974).
\newblock {\em Theoretical Statistics\/}.
\newblock Chapman and Hall, London, UK.

\bibitem[Dahlen and Tromp(1998)Dahlen and Tromp]{Dahlen+98}
Dahlen, F.~A. and Tromp, J. (1998).
\newblock {\em Theoretical Global Seismology\/}.
\newblock Princeton Univ.~Press, Princeton, N.J.

\bibitem[Davison and Hinkley(1997)Davison and Hinkley]{Davison+97}
Davison, A.~C. and Hinkley, D.~V. (1997).
\newblock {\em Bootstrap Methods and Their Application\/}.
\newblock Cambridge Univ.~Press, Cambridge, UK.

\bibitem[Eshagh(2009)Eshagh]{Eshagh2009a}
Eshagh, M. (2009).
\newblock Spatially restricted integrals in gradiometric boundary value
  problems.
\newblock {\em Artif.~Sat.}, {\bf 44}(4), 131--148, doi:
  {10.2478/v10018--009--0025--4}.

\bibitem[Farquharson and Oldenburg(1998)Farquharson and
  Oldenburg]{Farquharson+1998}
Farquharson, C.~G. and Oldenburg, D.~W. (1998).
\newblock Non-linear inversion using general measures of data misfit and model
  structure.
\newblock {\em Geophys.~J.~Int.}, {\bf 134}, 213--227, doi:
  10.1046/j.1365--246x.1998.00555.x.

\bibitem[Freeden and Michel(1999)Freeden and Michel]{Freeden+99}
Freeden, W. and Michel, V. (1999).
\newblock Constructive approximation and numerical methods in geodetic research
  today --- an attempt at a categorization based on an uncertainty principle.
\newblock {\em J.~Geod.}, {\bf 73}(9), 452--465.

\bibitem[Freeden and Schreiner(2009)Freeden and Schreiner]{Freeden+2009}
Freeden, W. and Schreiner, M. (2009).
\newblock {\em Spherical Functions of Mathematical Geosciences: {A} Scalar,
  Vectorial, and Tensorial Setup\/}.
\newblock Springer, Berlin, Germany.

\bibitem[Freeden {\em et~al.}(2016)Freeden, Michel, and Simons]{Freeden+2016}
Freeden, W., Michel, V., and Simons, F.~J. (2016).
\newblock Spherical-harmonics based special function systems and constructive
  approximation methods.
\newblock In W.~Freeden, editor, {\em Handbook of Mathematical Geodesy\/}, page
  under review. Springer, Berlin, Germany.

\bibitem[Gauss(1839)Gauss]{Gauss1839}
Gauss, C.~F. (1839).
\newblock Allgemeine {T}heorie des {E}rdmagnetismus.
\newblock In C.~F. Gauss and W.~Weber, editors, {\em {R}esultate aus den
  {B}eobachtungen des magnetischen {V}ereins im {J}ahre 1838\/}, pages 1--57.
  Weidmannsche Buchhandlung, Leipzig, Germany.

\bibitem[Gerhards(2011)Gerhards]{Gerhards2011}
Gerhards, C. (2011).
\newblock Spherical decompositions in a global and local framework: {t}heory
  and an application to geomagnetic modeling.
\newblock {\em Intern.~J.~Geomath.}, {\bf 1}(2), 205--256, doi:
  10.1007/s13137--010--0011--9.

\bibitem[Gerhards(2012)Gerhards]{Gerhards2012}
Gerhards, C. (2012).
\newblock Locally supported wavelets for the separation od spherical vector
  fields with respect to their sources.
\newblock {\em Int.~J.~Wavelets Multiresolut.\ Inf.\ Process.}, {\bf 10}(4),
  1250034, doi: 10.1142/S0219691312500348.

\bibitem[Gerhards(2014)Gerhards]{Gerhards2014a}
Gerhards, C. (2014).
\newblock A combination of downward continuation and local approximation for
  harmonic potentials.
\newblock {\bf 30}, 085004, doi: 10.1088/0266--5611/30/8/08500.

\bibitem[Holschneider {\em et~al.}(2003)Holschneider, Chambodut, and
  Mandea]{Holschneider+2003}
Holschneider, M., Chambodut, A., and Mandea, M. (2003).
\newblock From global to regional analysis of the magnetic field on the sphere
  using wavelet frames.
\newblock {\em Phys.~Earth Planet.~Inter.}, {\bf 135}, 107--124.

\bibitem[Jahn and Bokor(2014)Jahn and Bokor]{Jahn+2014}
Jahn, K. and Bokor, N. (2014).
\newblock Revisiting the concentration problem of vector fields within a
  spherical cap: {A} commuting differential operator solution.
\newblock {\em J.~Fourier Anal.~Appl.}, {\bf 288}(2), 421--451.

\bibitem[Kaula(1967)Kaula]{Kaula67a}
Kaula, W.~M. (1967).
\newblock Theory of statistical analysis of data distributed over a sphere.
\newblock {\em Rev.~Geophys.}, {\bf 5}(1), 83--107, doi:
  10.1029/RG005i001p00083.

\bibitem[Kaula(1968)Kaula]{Kaula68}
Kaula, W.~M. (1968).
\newblock {\em An Introduction to Planetary Physics. The Terrestrial
  Planets\/}.
\newblock Wiley, New York.

\bibitem[Lambeck(1988)Lambeck]{Lambeck88}
Lambeck, K. (1988).
\newblock {\em Geophysical Geodesy\/}.
\newblock Oxford Univ.~Press, New York.

\bibitem[Langel and Estes(1985)Langel and Estes]{Langel+85}
Langel, R.~A. and Estes, R.~H. (1985).
\newblock Large-scale, near-field magnetic fields from external sources and the
  corresponding induced internal field.
\newblock {\em J.~Geophys.~Res.}, {\bf 90}(B3), 2487--2494.

\bibitem[Langel and Hinze(1998)Langel and Hinze]{Langel+98}
Langel, R.~A. and Hinze, W.~J. (1998).
\newblock {\em The Magnetic Field of the {E}arth's Lithosphere: {T}he Satellite
  Perspective\/}.
\newblock Cambridge Univ.~Press, Cambridge, UK.

\bibitem[Langlais {\em et~al.}(2010)Langlais, Lesur, Purucker, Connerney, and
  Mandea]{Langlais+2010}
Langlais, B., Lesur, V., Purucker, M.~E., Connerney, J. E.~P., and Mandea, M.
  (2010).
\newblock Crustal magnetic fields of terrestrial planets.
\newblock {\em Space Sci.~Rev.}, {\bf 152}(1), 223--249.

\bibitem[Lesur(2006)Lesur]{Lesur2006}
Lesur, V. (2006).
\newblock Introducing localized constraints in global geomagnetic field
  modelling.
\newblock {\em Earth Planets Space\/}, {\bf 58}(4), 477--483.

\bibitem[Maniar and Mitra(2005)Maniar and Mitra]{Maniar+2005}
Maniar, H. and Mitra, P.~P. (2005).
\newblock The concentration problem for vector fields.
\newblock {\em Int.~J.~Bioelectromagn.}, {\bf 7}(1), 142--145.

\bibitem[Maus(2010)Maus]{Maus2010}
Maus, S. (2010).
\newblock An ellipsoidal harmonic representation of {E}arth's lithospheric
  magnetic field to degree and order 720.
\newblock {\em Geochem.~Geophys.~Geosys.}, {\bf 11}(6), Q06015, doi:
  {10.1029/2010GC003026}.

\bibitem[Mayer and Maier(2006)Mayer and Maier]{Mayer+2006}
Mayer, C. and Maier, T. (2006).
\newblock Separating inner and outer {E}arth's magnetic field from {CHAMP}
  satellite measurements by means of vector scaling functions and wavelets.
\newblock {\em Geophys.~J.~Int.}, {\bf 167}, 1188--1203, doi:
  {10.1111/j.1365--246X.2006.03199.x}.

\bibitem[Merrill {\em et~al.}(1998)Merrill, McElhinny, and
  Mc{F}adden]{Merrill+98}
Merrill, R.~T., McElhinny, M.~W., and Mc{F}adden, P.~L. (1998).
\newblock {\em The Magnetic Field {o}f {t}he {E}arth\/}.
\newblock Academic Press, San Diego, Calif.

\bibitem[Newman(2016)Newman]{Newman2016}
Newman, W.~I. (2016).
\newblock {\em Mathematical Methods for Geophysics and Space Physics\/}.
\newblock Princeton Univ.~Press, Princeton, N.J.

\bibitem[O'Brien and Parker(1994)O'Brien and Parker]{OBrien+94}
O'Brien, M.~S. and Parker, R.~L. (1994).
\newblock Regularized geomagnetic field modelling using monopoles.
\newblock {\em Geophys.~J.~Int.}, {\bf 118}(3), 566--578, doi:
  10.1111/j.1365--246X.1994.tb03985.x.

\bibitem[Olsen {\em et~al.}(2010a)Olsen, Glassmeier, and Jia]{Olsen+2010a}
Olsen, N., Glassmeier, K.-H., and Jia, X. (2010a).
\newblock Separation of the magnetic field into external and internal parts.
\newblock {\bf 152}, 135--157, doi: 10.1007/s11214--009--9563--0.

\bibitem[Olsen {\em et~al.}(2010b)Olsen, Hulot, and Sabaka]{Olsen+2010b}
Olsen, N., Hulot, G., and Sabaka, T.~J. (2010b).
\newblock Sources of the geomagnetic field and the modern data that enable
  their investigation.
\newblock In W.~Freeden, M.~Z. Nashed, and T.~Sonar, editors, {\em Handbook of
  {G}eomathematics\/}, chapter~5, pages 105--124, doi:
  {10.1007/978--3--642--01546--5\_5}. Springer, Heidelberg, Germany.

\bibitem[Plattner and Simons(2014)Plattner and Simons]{Plattner+2014a}
Plattner, A. and Simons, F.~J. (2014).
\newblock Spatiospectral concentration of vector fields on a sphere.
\newblock {\em Appl.~Comput.~Harmon.~Anal.}, {\bf 36}, 1--22, doi:
  10.1016/j.acha.2012.12.001.

\bibitem[Plattner and Simons(2015a)Plattner and Simons]{Plattner+2015a}
Plattner, A. and Simons, F.~J. (2015a).
\newblock High-resolution local magnetic field models for the {M}artian {S}outh
  {P}ole from {M}ars {G}lobal {S}urveyor data.
\newblock {\em J.~Geophys.~Res.}, {\bf 120}, 1543--1566, doi:
  10.1002/2015JE004869.

\bibitem[Plattner and Simons(2015b)Plattner and Simons]{Plattner+2015c}
Plattner, A. and Simons, F.~J. (2015b).
\newblock Potential-field estimation using scalar and vector {S}lepian
  functions at satellite altitude.
\newblock In W.~Freeden, M.~Z. Nashed, and T.~Sonar, editors, {\em Handbook of
  {G}eomathematics\/}, pages 2003--2055, doi:
  10.1007/978--3--642--54551--1\_64. Springer, Heidelberg, Germany, 2 edition.

\bibitem[Sabaka {\em et~al.}(2015)Sabaka, Hulot, and Olsen]{Sabaka+2015}
Sabaka, T.~J., Hulot, G., and Olsen, N. (2015).
\newblock Mathematical properties relevant to geomagnetic field modeling.
\newblock In W.~Freeden, M.~Z. Nashed, and T.~Sonar, editors, {\em Handbook of
  {G}eomathematics\/}, chapter~17, pages 835--877, doi:
  10.1007/978--3--642--54551--1\_17. Springer, Heidelberg, Germany.

\bibitem[Schachtschneider {\em et~al.}(2012)Schachtschneider, Holschneider, and
  Mandea]{Schachtschneider+2012}
Schachtschneider, R., Holschneider, M., and Mandea, M. (2012).
\newblock Error distribution in regional modelling of the geomagnetic field.
\newblock {\em Geophys.~J.~Int.}, {\bf 191}, 1015--1024, doi:
  {10.1111/j.1365--246X.2012.05675.x}.

\bibitem[Schmidt {\em et~al.}(2007)Schmidt, Fengler, Mayer-G{\"u}rr, Eicker,
  Kusche, S{\'a}nchez, and Han]{Schmidt+2007}
Schmidt, M., Fengler, M., Mayer-G{\"u}rr, T., Eicker, A., Kusche, J.,
  S{\'a}nchez, L., and Han, S.-C. (2007).
\newblock Regional gravity modeling in terms of spherical base functions.
\newblock {\em J.~Geod.}, {\bf 81}(1), 17--38, doi:
  {10.1007/s00190--006--0101--5}.

\bibitem[Shure {\em et~al.}(1982)Shure, Parker, and Backus]{Shure+82}
Shure, L., Parker, R.~L., and Backus, G.~E. (1982).
\newblock Harmonic splines for geomagnetic modeling.
\newblock {\em Phys.~Earth Planet.~Inter.}, {\bf 28}, 215--229.

\bibitem[Simons and Dahlen(2006)Simons and Dahlen]{Simons+2006b}
Simons, F.~J. and Dahlen, F.~A. (2006).
\newblock Spherical {S}lepian functions and the polar gap in geodesy.
\newblock {\em Geophys.~J.~Int.}, {\bf 166}(3), 1039--1061, doi:
  10.1111/j.1365--246X.2006.03065.x.

\bibitem[Simons and Plattner(2015)Simons and Plattner]{Simons+2015}
Simons, F.~J. and Plattner, A. (2015).
\newblock Scalar and vector {S}lepian functions, spherical signal estimation
  and spectral analysis.
\newblock In W.~Freeden, M.~Z. Nashed, and T.~Sonar, editors, {\em Handbook of
  {G}eomathematics\/}, pages 2563--2608, doi:
  10.1007/978--3--642--54551--1\_30. Springer, Heidelberg, Germany, 2 edition.

\bibitem[Simons {\em et~al.}(2006)Simons, Dahlen, and Wieczorek]{Simons+2006a}
Simons, F.~J., Dahlen, F.~A., and Wieczorek, M.~A. (2006).
\newblock Spatiospectral concentration on a sphere.
\newblock {\em SIAM Rev.}, {\bf 48}(3), 504--536, doi:
  10.1137/S0036144504445765.

\bibitem[Slepian(1983)Slepian]{Slepian83}
Slepian, D. (1983).
\newblock Some comments on {F}ourier analysis, uncertainty and modeling.
\newblock {\em SIAM Rev.}, {\bf 25}(3), 379--393.

\bibitem[Slepian and Pollak(1961)Slepian and Pollak]{Slepian+61}
Slepian, D. and Pollak, H.~O. (1961).
\newblock Prolate spheroidal wave functions, {F}ourier analysis and uncertainty
  --- {I}.
\newblock {\em Bell Syst.~Tech.~J.}, {\bf 40}(1), 43--63.

\bibitem[Sneeuw(1994)Sneeuw]{Sneeuw94}
Sneeuw, N. (1994).
\newblock Global spherical harmonic-analysis by least-squares and numerical
  quadrature methods in historical perspective.
\newblock {\em Geophys.~J.~Int.}, {\bf 118}(3), 707--716.

\bibitem[Snieder(2004)Snieder]{Snieder2004}
Snieder, R. (2004).
\newblock {\em A Guided Tour of Mathematical Methods for the Physical
  Sciences\/}.
\newblock Cambridge Univ.~Press, Cambridge, UK, 2 edition.

\bibitem[Solomon {\em et~al.}(2001)Solomon, Jr, Gold, Acu{\~{n}}a, Baker,
  Boynton, Chapman, Cheng, Gloeckler, {III}, Krimigis, McClintock, Murchie,
  Peale, Phillips, Robinson, Slavin, Smith, Strom, Trombka, and
  Zuber]{Solomon+2001}
Solomon, S.~C., Jr, R. L.~M., Gold, R.~E., Acu{\~{n}}a, M.~H., Baker, D.~N.,
  Boynton, W.~V., Chapman, C.~R., Cheng, A.~F., Gloeckler, G., {III}, J. W.~H.,
  Krimigis, S.~M., McClintock, W.~E., Murchie, S.~L., Peale, S.~J., Phillips,
  R.~J., Robinson, M.~S., Slavin, J.~A., Smith, D.~E., Strom, R.~G., Trombka,
  J.~I., and Zuber, M.~T. (2001).
\newblock The {MESSENGER} mission to {M}ercury: {s}cientific objectives and
  implementation.
\newblock {\em Planet.~Space Sci.}, {\bf 49}(14--15), 1445--1465, doi:
  10.1016/S0032--0633(01)00085--X.

\bibitem[Solomon {\em et~al.}(2007)Solomon, Jr., Gold, and
  Domingue]{Solomon+2007}
Solomon, S.~C., Jr., R. L.~M., Gold, R.~E., and Domingue, D.~L. (2007).
\newblock {MESSENGER} mission overview.
\newblock {\bf 131}(1), 3--39, doi: 10.1007/s11214--007--9247--6.

\bibitem[Th{\'e}bault {\em et~al.}(2006)Th{\'e}bault, Schott, and
  Mandea]{Thebault+2006}
Th{\'e}bault, E., Schott, J.~J., and Mandea, M. (2006).
\newblock Revised spherical cap harmonic analysis ({R-SCHA}): Validation and
  properties.
\newblock {\em J.~Geophys.~Res.}, {\bf 111}(B1), B01102, doi:
  {10.1029/2005JB003836}.

\bibitem[Trampert and Snieder(1996)Trampert and Snieder]{Trampert+96b}
Trampert, J. and Snieder, R. (1996).
\newblock Model estimations biased by truncated expansions: {P}ossible
  artifacts in seismic tomography.
\newblock {\em Science\/}, {\bf 271}(5253), 1257--1260, doi:
  10.1126/science.271.5253.1257.

\bibitem[Watkins {\em et~al.}(2015)Watkins, Wiese, Yuan, Boening, and
  Landerer]{Watkins+2015}
Watkins, M.~M., Wiese, D.~N., Yuan, D.-N., Boening, C., and Landerer, F.~W.
  (2015).
\newblock Improved methods for observing {E}arth's time variable mass
  distribution with {GRACE} using spherical cap mascons.
\newblock {\em J.~Geophys.~Res.}, {\bf 120}(4), 2648--2671, doi:
  10.1002/2014JB011547V.

\bibitem[Wieczorek(2015)Wieczorek]{Wieczorek2015}
Wieczorek, M.~A. (2015).
\newblock The gravity and topography of the terrestrial planets.
\newblock In T.~Spohn, editor, {\em Treatise on Geophysics\/}, volume~10, pages
  153--2193, doi: 10.1016/B978--0--444--53802--4.00169--X. Elsevier, Amsterdam,
  Neth., 2 edition.

\bibitem[Xu(1992)Xu]{Xu92b}
Xu, P. (1992).
\newblock The value of minimum norm estimation of geopotential fields.
\newblock {\em Geophys.~J.~Int.}, {\bf 111}, 170--178.

\end{thebibliography}

\appendix

\section{Computational considerations}
\label{app:comp}

In our construction of the altitude-cognizant \GVSFs{} (AC-\aGVSF) in
Sections~\ref{sec:inneraltslepconstruction}
and~\ref{sec:Slepbasintext} we calculated the eigenvectors of the
matrices $\Kmat$ and $\OutKmat$, which are of dimensions
$\dimin\times\dimin$ and $[\dimin+\dimout]\times[\dimin+\dimout]$,
respectively. For large $\Lin$, or large $\Lout$, numerical
computations may become prohibitively expensive. Fortunately, for
symmetric regions such as rings, belts or caps, we can reorder the
matrices $\Kmat$ and $\OutKmat$ into block-diagonal form, with blocks
of maximum size $[\Lin+1] \times [\Lin+1]$, or $[\Lin + \Lout +1]
\times [\Lin + \Lout +1]$. Considering that a full eigenvalue
decomposition of an $n\times n$ matrix has a numerical complexity of
$\mathcal{O}(n^3)$, the block diagonal reordering significantly
reduces the computational costs and allows us to construct
altitude-cognizant \GVSFs{} with otherwise prohibitively high maximum
spherical-harmonic degrees. For example, a fully populated matrix for
internal-field bandlimit $\Lin=500$ and external-field bandlimit
$\Lout=10$ containing double-precision floating point numbers would
require $(501^2+6^2-1)^2\cdot 8$ bytes $\approx$ 470~GB of memory.
Calculating the eigenvalue distribution of such a large matrix would
be prohibitively computationally demanding. On the other hand, the
largest block-matrix for the symmetric region has dimensions $[\Lin +
  \Lout + 1]\times[\Lin + \Lout + 1]$, which would require
$(500+5+1)^2\cdot 8$ bytes $\approx$ 2~MB. Eigenvalue decompositions
for such matrices are feasible even on small computers. The overall
memory required for all block-diagonal matrices is
$\sum_{m=0}^{\max(\Lin,\Lout)} \left[ (\Lin+1-m)_+ +
  (\Lout+1-\max(m,1))_+ \right]^2$, where $(a)_+$ is $a$ for $a\geq 0$
and 0 otherwise. For $\Lin=500$ and $\Lout=5$, this amounts to
$\approx$ 320~MB, less than 0.1\% of the full matrix
requirement. Besides, the eigenvalue decompositions only need to be
performed on the much smaller ($<2$MB) block matrices. Recently,
\cite{Bates+2017} proposed a method to calculate approximate classical
scalar Slepian functions for generic regions by first obtaining a
basis of Slepian functions for a spherical cap surrounding the generic
target region and then, in a second step, calculating the Slepian
functions for the generic region using the Slepian functions for the
spherical cap. Their approach greatly reduces the computational
costs. A similar procedure may be implemented for the AC-\aGVSFs{}.

\begin{figure}[ht]\centering
\includegraphics[width=0.8\textwidth,angle=0,trim= 0.5cm 0.8cm 0.5cm 0.5cm,clip]
{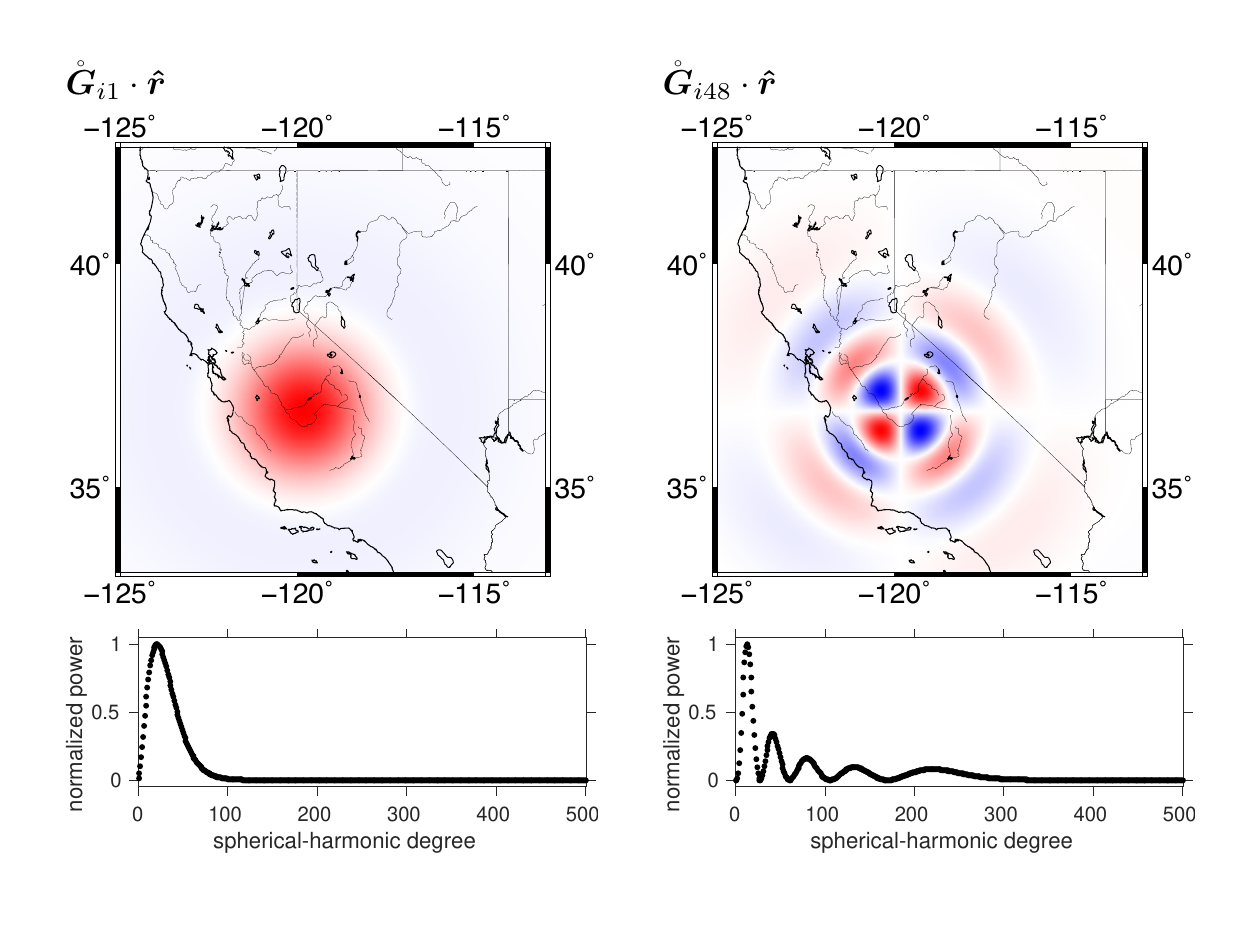}
\caption{\label{ExamplefunctionCap}Full-field altitude-cognizant
  \GVSF{} for a polar cap of opening angle~$1^\circ$ centered over
  Peeve's Public House in Fresno, California with bandwidths
  $\Lin=500$, and $\Lout=10$, $\Earthrad=6371$~km, $\Outrad=6771$~km,
  $\satalt=6671$~km. Top left: radial component of highest-eigenvalue
  internal-field function. Bottom left: Power spectral density of that
  function. Top right: 48th best internal-field and its power spectral
  density, bottom right.}
\end{figure}

In Fig.~\ref{ExamplefunctionCap} we show examples for such
high-maximum-spherical-harmonic internal- and external-field
AC-\aGVSF{} constructed for a spherical cap of opening angle $1^\circ$
centered on the first author's current favorite pub, with
internal-field bandwidth $\Lin=500$, external-field bandwidth
$\Lout=10$, and planetary and satellite radii, and external-field
radius as discussed in the legend. The left column, above, shows the
radial component of the internal field of the best-suited function for
this setting, and its power spectral density, below. The right column
shows the radial component of the internal field of the
48th-best-suited function, above, and, below, its power spectral
density. We constructed these functions on an off-the-shelf desktop
computer. We describe our block-diagonalization approaches in
Sections~\ref{app:compint} and~\ref{appintext}.

\subsection{Internal-field altitude-cognizant \GVSFs{}}  
\label{app:compint}

To reorder the kernel matrix~$\Kmat$ of eq.~(\ref{defKin}) when the
region $\region$ is a spherical cap, we follow the derivations of
\cite{Plattner+2014a}. As they do, we define $\Theta$ as the half
opening angle of the spherical cap. We can describe our internal-field
gradient vector spherical harmonics $\Efun_{lm}$ as linear combination
of the radial and tangential vector spherical harmonics, $\Pfun_{lm}$
and $\Bfun_{lm}$, defined by \cite{Dahlen+98} as
\begin{equation}\label{DahlenVSPH}
  \Pfun_{lm}(\rvec) = \rvec\hsom \Yfun_{lm}(\rvec)
  \also
  \Bfun_{lm}(\rvec) = \frac{ \bnabla_1
  \Yfun_{lm}(\rvec)}{\sqrt{l(l+1)}},
\end{equation}
for~$1\leq l \leq \Lin$ and~$\-l\leq m \leq l$. For $l=m=0$ we set
$\Pfun_{00}(\rvec) = \rvec$. Owing to the pointwise orthogonality
between $\Pfun_{lm}$ and $\Bfun_{lm}$, the regional integrals over
products of $\Efun_{lm}$ are linear combinations of integrals of
products of $\Pfun_{lm}$ and $\Bfun_{lm}$ with the same
spherical-harmonic degrees and orders as the $\Efun_{lm}$,
\begin{equation}\label{PolarcapintegralElm}
  \intR \Efun_{lm}\cdot\Efun_{l'm'} \dOmega
  = 
  \sqrt{\frac{(l+1)(l'+1)}{(2l+1)(2l'+1)}}
  \intR \Pfun_{lm} \cdot \Pfun_{l'm'}\dOmega + 
  \sqrt{\frac{l\hspace{0.1em}l'}{(2l+1)(2l'+1)}}
  \intR \Bfun_{lm} \cdot \Bfun_{l'm'}\dOmega.
\end{equation}
\cite{Plattner+2014a}, in their equations~(70)--(76), derived the
relationships, using a prime to denote differentiation,
\begin{align}
  \label{Plmlmp}
  \intR \Pfun_{lm} \cdot \Pfun_{l'm'} \dOmega
  &=  2\pi\hsom\deltamm\intoth \Xlm\Xlpm \sindth,\\
  \label{Blmlmp}
  \intR \Bfun_{lm} \cdot \Bfun_{l'm'} \dOmega
  &=\frac{\displaystyle 2\pi\hsom\deltamm\intoth 
    \left[ \dXlm \dXlpm + m^2 \divsinsq \Xlm \Xlpm \right]
    \sindth}{\displaystyle \renormprod}
.
\end{align}
Eqs~(\ref{Plmlmp})--(\ref{Blmlmp}) imply that for $m\neq m'$, the
regional integral over $\Bfun_{lm}\cdot\Pfun_{l'm'}$ is zero, which
amounts to the bulk of the entries of matrix $\Kmat$ in
eq.~(\ref{defKin}). We rearrange the non-zero elements using an
orthogonal permutation matrix $\Qmat$ such that, for each $m$, all the
corresponding $l,l'\geq m$ entries of matrix $\Kmat$ are within one
block of size $\Lin-m+1$ and the blocks are collected in a block
diagonal matrix $\Qmat\Kmat\Qmat^\matT$. The eigenvectors of
$\Qmat\Kmat\Qmat^\matT$ are the permutations $\Qmat\slepfuncoef$ of
the eigenvectors $\slepfuncoef$ defined in eq.~(\ref{intslepcoef})
with the same eigenvalues, thanks to
\begin{equation}\label{orthogonaltransformation}
  (\Qmat\Kmat\Qmat^\matT)(\Qmat\Gmat)
  = \Qmat\Kmat\Gmat
  = (\Qmat\Gmat)\Lamat.
\end{equation}
For the case or $\region = $ polar caps, the entries in each block
\begin{equation}
\sKm=\begin{pmatrix}
   K^m_{mm}&\cdots&K^m_{mL}\\
   \vdots&&\vdots\\
   K^m_{Lm}&\cdots&K^m_{LL}
 \end{pmatrix}
\end{equation}
have analytic expressions that can be obtained from
\begin{equation}\label{ElmElmCap}
K^m_{ll'} = \Earthrad^{-2}\enormfactprod
\left(\frac{\satalt}{\Earthrad}\right)^{-l-l'-4} \intR
\Efun_{lm}\cdot\Efun_{l'm} \dOmega 
,
\end{equation} 
together with eqs~(\ref{PolarcapintegralElm})--(\ref{Blmlmp}) and the
derivations by \cite{Simons+2006b} and \cite{Plattner+2014a}. To
calculate internal-field AC-\aGVSF{} for spherical caps not centered
on the North Pole, we can again make use of
relationship~(\ref{orthogonaltransformation}) with~$\Qmat$ defined as
the orthogonal spherical-harmonic rotation matrix using the
appropriate Euler angles \cite[]{Dahlen+98}. We first calculate the
eigenvectors for a north-polar cap region with our chosen angle
$\Theta$ and then rotate these using the matrix~$\Qmat$. To obtain
internal-field AC-\aGVSF{} for polar rings we subtract the kernel
matrices for the inner cap from the kernel matrices for the large cap.

\subsection{Full-field altitude-cognizant \GVSFs{}}
\label{appintext}

To reorder matrix~$\OutKmat$ in eq.~(\ref{kernel_inout}) in a similar
fashion as in Appendix~\ref{app:compint} we need to also consider the
non-vanishing integrals over products of $\Efun_{lm}$ and
$\Ffun_{lm}$. The following equations describe the relationships
between products of $\Ffun_{lm}$ and $\Efun_{m}$, and products of
$\Pfun_{lm}$ and $\Bfun_{lm}$, 
\begin{equation}\label{PolarcapintegralFlm}
\intR \Ffun_{lm}\cdot\Ffun_{l'm'} \dOmega = 
\sqrt{\frac{l\hspace{0.01em}l'}{(2l+1)(2l'+1)}}
\intR \Pfun_{lm}\cdot \Pfun_{l'm'}\dOmega + 
\sqrt{\frac{(l+1)(l'+1)}{(2l+1)(2l'+1)}}
\intR \Bfun_{lm}\cdot \Bfun_{l'm'}\dOmega,
\end{equation}
and
\begin{equation}\label{PolarcapintegralElmFlm}
\intR \Efun_{lm}\cdot\Ffun_{l'm'} \dOmega = 
\sqrt{\frac{(l+1)l'}{(2l+1)(2l'+1)}}
\intR \Pfun_{lm}\cdot \Pfun_{l'm'}\dOmega -
\sqrt{\frac{l(l'+1)}{(2l+1)(2l'+1)}}
\intR \Bfun_{lm}\cdot \Bfun_{l'm'}\dOmega.
\end{equation}
These integrals are zero unless $m=m'$, rendering the matrix
$\OutKmat$ sparse. To make optimal use of this sparsity and save
computational resources in the calculation of its eigenvectors, we
reorder the matrix elements into block-diagonal shape using orthogonal
transformation. The resulting eigenvectors can be translated back into
eigenvectors of the original matrix $\OutKmat$ by reordering the
vector entries as is explained in
eq.~(\ref{orthogonaltransformation}). To define the individual blocks,
we reuse the definition of the matrix entries in eq.~(\ref{ElmElmCap})
and set $K^{im}_{ll'}=K^{m}_{ll'}$. We also define
\begin{equation}\label{FlmFlmCap}
K^{om}_{ll'} = \Outrad^{-2}\fnormfactprod 
\left(\frac{\satalt}{\Outrad}\right)^{l+l'-2} 
 \intR \Ffun_{lm}\cdot\Ffun_{l'm} \dOmega,
\end{equation}
and
\begin{equation}\label{ElmFlmCap}
K^{iom}_{ll'} = (\Earthrad\Outrad)^{-1}\efnormfactprod 
\left(\frac{\satalt}{\Earthrad}\right)^{-l-2}
\left(\frac{\satalt}{\Outrad}\right)^{l'-1} 
 \intR \Efun_{lm}\cdot\Ffun_{l'm} \dOmega
,
\end{equation}
and assemble these elements in the matrices
\begin{equation}
\sKom=\begin{pmatrix}
   K^{om}_{mm}&\cdots&K^{om}_{m\Lout}\\
   \vdots&&\vdots\\
   K^{om}_{\Lout m}&\cdots&K^{om}_{\Lout\Lout}
 \end{pmatrix}
 \also
 \sKiom=\begin{pmatrix}
   K^{iom}_{mm}&\cdots&K^{iom}_{m\Lout}\\
   \vdots&&\vdots\\
   K^{iom}_{\Lin m}&\cdots&K^{iom}_{\Lin\Lout}
 \end{pmatrix}.
\end{equation}
Note that the matrix $\sKom$ is of dimension 
$[\Lout+1 - \max(m,1)]\times[\Lout+1 - \max(m,1)]$ 
while matrix $\sKiom$ is of dimension
$[\Lin+1-m ]\times[\Lout+1 - \max(m,1)]$.
We collect these matrices in the single block
\begin{equation}
\OutsKm=
\begin{pmatrix}
\sKm&\sKiom\\
 (\sKiom)^\matT& \sKom
\end{pmatrix}.
\end{equation}

If we elect to calculate internal- and external-field
altitude-cognizant \GVSFs{} for spherical caps not centered at the
North Pole, we can transform the resulting eigenvectors from the
north-polar caps into eigenvectors of spherical caps centered at any
other point using orthonormal spherical-harmonic rotation matrices. To
obtain internal- and external-field altitude-cognizant \GVSFs{} for
spherical-ring regions, we subtract the matrices $\OutsKm$ for the
polar cap constituting the inner gap of the ring from the matrices
$\OutsKm$ for the larger polar cap. We then transform the resulting
eigenvectors using orthonormal spherical-harmonic rotation matrices.

\newgeometry{textwidth=7.7in,textheight=9.8in}

\section{Table of Symbols}

\begin{center}
\begin{longtable}{c|l|c}
Symbol&Definition, description, usage&eq.\\\hline
\endfirsthead
Symbol&Definition, description, usage&eq.\\\hline
\endhead
\multicolumn{3}{r}{{\textit{Continued on next page}}} \\ 
\endfoot
\hline
\endlastfoot
$\Intsignal$&(scalar) internal-source potential field, a quantity that we attempt to estimate from the data&(\ref{scalarLaplaceequation})\strutt\\
$\Extsignal$&(scalar) external-source potential field, a quantity that we attempt to estimate from the data&(\ref{scalarLaplaceequation})\\
$\Earthrad$&outer boundary radius for the internal-source field~$\Intsignal$; Earth, planetary or lunar radius&(\ref{scalarLaplaceequation})\strutt\\
$\satalt$&(average) radial position of the measurement satellite&(\ref{scalarLaplaceequation})\\
$\Outrad$&inner boundary radius for the external-source field~$\Extsignal$&(\ref{scalarLaplaceequation})\\
$\Omega$&unit sphere, a sphere of unit radius&(\ref{scalarLaplaceequation})\\
$r,\theta,\phi$&radial, colatitudinal, and longitudinal coordinates&(\ref{scalarLaplaceequation})\\
$\rvec,\thvec,\phvec$&radial, colatitudinal, and longitudinal unit vectors&(\ref{superposition})\\
$\Yfun_{lm}$&scalar surface spherical-harmonic; an orthonormal basis function for the potential fields $\Intsignal$ or $\Extsignal$&(\ref{Y_definition})\strutt\\
$l,m$ & spherical-harmonic degree, $l\ge 0$ and order $-l \leq m\leq l$&(\ref{Y_definition})\\
$\ssphcoefEarth_{lm}$&expansion coefficients of the internal-source potential $\Intsignal(\Earthrad\rvec)$ in the basis $\Yfun_{lm}(\rvec)$&(\ref{Vrr})\strutt\\
$\soutsphcoefEarth_{lm}$&expansion coefficients of the external-source potential $\Extsignal(\Outrad\rvec)$ in the basis $\Yfun_{lm}(\rvec)$&(\ref{Wrr})\\
$\Lin$&maximum spherical-harmonic degree (bandwidth) of the internal-source field~$\Intsignal$ being modeled&(\ref{crud})\strutt\\
$\Lout$&maximum spherical-harmonic degree (bandwidth) of the external field~$\Extsignal$ being modeled&(\ref{crud})\\
$\sphcoefEarth$&$\dimin$-dimensional vector of expansion coefficients $\ssphcoefEarth_{lm}$, for $0\leq l\leq \Lin$&(\ref{crud})\\
$\outsphcoefEarth$&$[\dimout]$-dimensional vector of expansion coefficients $\soutsphcoefEarth_{lm}$, for $1\leq l\leq \Lout$&(\ref{crud})\\
$\Yfunvec$&infinite-dimensional vector of all scalar spherical harmonics $\Yfun_{lm}$, for $0\leq l\le\infty$&(\ref{ass})\strutt\\
$\Yfunvec_\Lin$&$\dimin$-dimensional vector of the scalar spherical harmonics $\Yfun_{lm}$ for $0\leq l \leq \Lin$&(\ref{yfunvec})\\
$\Yfunvec_{>\Lin}$&infinite-dimensional vector of the scalar spherical harmonics $\Yfun_{lm}$ for $l\ge \Lin+1$&(\ref{yfunvec})\\
$\Yfunvec_\Lout$&$[\dimout]$-dimensional vector of the scalar spherical harmonics $\Yfun_{lm}$ for $1\leq l \leq \Lout$&(\ref{yfunvec})\\
$\Yfunvec_{>\Lout}$&infinite-dimensional vector of all scalar spherical harmonics $\Yfun_{lm}$ for $l\ge \Lout+1$&(\ref{yfunvec})\\\hline
$\Bfun$&vector-field superposition of the gradients of the internal- and external-source potentials, $\vecIntsignal+\vecExtsignal$&(\ref{superposition})\strutt\\
$\Efun_{lm}$&vector surface spherical-harmonic; orthonormal internal-source basis function for $\vecIntsignal$, for $0\leq l$, $-l\leq m\leq l$&(\ref{gradientvectorsphericalharmonics})\\
$\Ffun_{lm}$&vector spherical-harmonic; orthonormal external-source basis function for $\vecExtsignal$, for $1\leq l$ and $-l\leq m\leq l$&(\ref{bobbly})\\
$\Efunvec,\Ffunvec$&infinite-dimensional vector of all vector spherical-harmonics $\Efun_{lm}$ for $0\leq l\leq\infty$, or $\Ffun_{lm}$  for $1\leq l\leq\infty$ &(\ref{joob})\strutt\\
$\Efunvec_{\Lin}$&$\dimin$-dimensional vector of the internal-source vector spherical-harmonics $\Efun_{lm}$ for $0\leq l\leq \Lin$&(\ref{dotdefine})\\
$\Efunvec_{>\Lin}$&infinite-dimensional vector of the internal-source vector spherical-harmonics $\Efun_{lm}$ for $l\ge \Lin+1$&(\ref{joob})\\
$\Ffunvec_{\Lout}$&$[\dimout]$-dimensional vector of the external-source vector spherical-harmonics $\Ffun_{lm}$ for $1\leq l\leq \Lout$&(\ref{dotdefine})\\
$\Ffunvec_{>\Lout}$&infinite-dimensional vector of the external-source vector spherical-harmonics $\Ffun_{lm}$ for $l\ge \Lout+1$&(\ref{joob})\\
$\InUpelm_l(r)$&upward-continuation factor for internal-source fields, from planetary surface $\Earthrad$ up to $r$, for $l\ge 0$ &(\ref{gradv})\strutt\\
$\OutUpelm_l(r)$&downward-continuation factor for external-source fields, from inner boundary  $\Outrad$ down to $r$, for  $l\ge 1$ &(\ref{gradw})\\
$\InUpmat(r)$&diagonal matrix containing the $\InUpelm_l(r)$, for a context-dependent range of $l\ge 0$ &(\ref{InUpelements})\\
$\OutUpmat(r)$&diagonal matrix containing the  $\OutUpelm_l(r)$, for a context-dependent range of $l\ge 1$ &(\ref{OutUpelements})\\
$\InUpmat$&silent notation, whereby is meant  $\InUpmat(\satalt)$, evaluated at the (radial average of the) satellite position &(\ref{InUpels})\\
$\OutUpmat$&silent notation, whereby is meant $\OutUpmat(\satalt)$, evaluated at the (radial average of the) satellite position &(\ref{OutUpels})\\\hline
$\datavec(r\rvec)$&vector-valued function with the data collected by the satellite&(\ref{dataonlyinregionout})\strutt\\
$\vecnoise(r\rvec)$&vector-valued  function with observational noise contaminating the data&(\ref{dataonlyinregionout})\\
$\datavec,\vecnoise$&shorthand for the data and noise functions at average satellite altitude, $\datavec=\datavec(\satalt\rvec)$ and $\vecnoise=\vecnoise(\satalt\rvec)$&(\ref{dataonlyinregion})\\
$\region$&region of data availability on the unit sphere~$\Omega$; target region for Slepian-function concentration; model domain&(\ref{dataonlyinregionout})\\
$\estsphcoef^\Earthrad_\Lin$&estimated internal-source coefficient vector; minimizer of internal-source least-squares data functional on $\region$  &(\ref{inner-sourceoptimizationproblem})\strutt\\
$\estoutsphcoefEarth$&estimated external-source coefficient vector; minimizer of internal/external data misfit functional over $\region$ &(\ref{inner-outer-sourceoptimizationproblem})\\\hline
$\Kmat$&inner-product (over $\region$, hence ``localization'') matrix of upward-continued inner-source vector harmonics $\InUpmat\Efunvec_{\Lin}$&(\ref{defKin})\strutt\\
$\OutKmat$&inner-product matrix of up/downward-continued internal/external-source vector harmonics $\InUpmat\Efunvec_{\Lin}$ and $\OutUpmat\Ffunvec_{\Lout}$&(\ref{kernel_inout})\\
$\Lamat$&diagonal matrix containing the $\dimin$ eigenvalues of $\Kmat$: the ``optimization'' or ``concentration'' factors $\lambda_\alpha$ &(\ref{eigenprobleminner})\\
$\OutLamat$& diagonal matrix containing the $[\dimin + \dimout]$ eigenvalues of $\OutKmat$: the optimization factors  $\Outlambda_\alpha$ &(\ref{eigenprobleminnerouter})\\
$\Lamat_\nSlepfun$&diagonal matrix containing the $\nSlepfun$ largest eigenvalues $\lambda_\alpha$ of $\Kmat$;  $\nSlepfun\times\nSlepfun$-dimensional truncation of $\Lamat$ &(\ref{ugh})\\
$\OutLamat_\nSlepfun$&diagonal matrix containing the $\nSlepfun$ largest eigenvalues $\Outlambda_\alpha$ of $\OutKmat$;  $\nSlepfun\times\nSlepfun$-dimensional truncation of $\OutLamat$ &(\ref{ugh2})\\
$\Gmat$&eigenvector matrix of $\Kmat$; columns $\slepfuncoef_\alpha$ contain the $\Yfun_{lm}$-expansion coefficients $\sslepfuncoef_{lm,\alpha}$ of scalar functions $\Gfun_\alpha$&(\ref{eigenprobleminner})\\
$\OutGmat$&eigenvector matrix of $\OutKmat$; columns $\Outslepfuncoef_\alpha$ with $\Yfun_{lm}$-coefficients $\sOutslepfuncoef_{i\,lm,\alpha}$ and $\sslepfuncoef_{o\,lm,\alpha}$ of scalar $\OutGfun_{i\alpha}$ and $\OutGfun_{o\alpha}$&(\ref{eigenprobleminnerouter})\\ 
$\Gmat_\nSlepfun$&$\dimin\times\nSlepfun$ matrix with th $\nSlepfun$ best eigenvectors of $\Kmat$; restriction of $\Gmat$ to its first $\nSlepfun$ columns &(\ref{ugh})\\
$\OutGmat_\nSlepfun$&$[\dimin+\dimout]\times\nSlepfun$ matrix with the $\nSlepfun$ best eigenvectors of $\OutKmat$; column-restriction of $\OutGmat$ &(\ref{bo})\\
$\OutGmat_i$&$\dimin\times [\dimin+\dimout]$-dimensional matrix of columns $\Outslepfuncoef_{i\alpha}$ with coefficients  $\sOutslepfuncoef_{i\,lm,\alpha}$ of  $\OutGfun_{i\alpha}$  &(\ref{OutGmatstructure})\\
$\OutGmat_o$&$[\dimout]\times [\dimin+\dimout]$-dimensional matrix of columns  $\Outslepfuncoef_{o\alpha}$ with the  $\sslepfuncoef_{o\,lm,\alpha}$ of  $\OutGfun_{o\alpha}$ &(\ref{OutGmatstructure})\\
$\OutGmat_{i\nSlepfun},\OutGmat_{o\nSlepfun}$&restrictions of $\OutGmat_i$ and $\OutGmat_o$ to their first $\nSlepfun$ columns &(\ref{OutGmatstructure})\\
$\lambda_\alpha$&eigenvalues of $\Kmat$, in descending order, for $1\leq\alpha\leq\dimin$; the diagonal elements of $\Lamat$&(\ref{eigenprobleminner})\strutt\\
$\Outlambda_\alpha$&eigenvalues of $\OutKmat$, in descending order, for $1\leq\alpha\leq[\dimin+\dimout]$; the diagonal elements of $\OutLamat$&(\ref{eigenprobleminnerouter})\\
$\slepfuncoef_\alpha$&$\dimin$-dimensional column vector, the $\alpha$-th column of $\Gmat$, contains the $\sslepfuncoef_{lm,\alpha}$ &(\ref{intslepcoef})\strutt\\
$\Outslepfuncoef_\alpha$&$[\dimin+\dimout]$-dimensional column vector, the $\alpha$-th column of $\OutGmat$; also $\Outslepfuncoef_\alpha = \begin{pmatrix} \Outslepfuncoef_{i\alpha}^\matT &   \Outslepfuncoef_{o\alpha}^\matT \end{pmatrix}^\matT$&(\ref{bo})\\
$\Outslepfuncoef_{i\alpha}$&$\dimin$-dimensional column vector, the $\dimin$ first elements of $\Outslepfuncoef_\alpha$, contains the $\sOutslepfuncoef_{i\,lm,\alpha}$&(\ref{OutGmatstructure})\\
$\Outslepfuncoef_{o\alpha}$&$[\dimout]$-dimensional column vector, the $(\Lout+1)^2-1$ last elements of of $\Outslepfuncoef_\alpha$, contains the $\sOutslepfuncoef_{o\,lm,\alpha}$&(\ref{OutGmatstructure})\\
$\sslepfuncoef_{lm,\alpha}$& spherical-harmonic expansion coefficients of the $\Gfun_\alpha$: the elements of $\slepfuncoef_\alpha$ &(\ref{intonlypowerspec})\strutt\\
$\sOutslepfuncoef_{i\,lm,\alpha}$& spherical-harmonic expansion coefficients of the $\OutGfun_{i\alpha}$: first $\dimin$ scalar elements of $\Outslepfuncoef_\alpha$; the elements of $\Outslepfuncoef_{i\alpha}$&(\ref{OutGmatstructure})\\
$\sOutslepfuncoef_{o\,lm,\alpha}$& spherical-harmonic expansion coefficients of the $\OutGfun_{o\alpha}$: last $\dimout$ elements of $\Outslepfuncoef_\alpha$; the elements of $\Outslepfuncoef_{o\alpha}$&(\ref{OutGmatstructure})\\
AC-\aGVSF{} &  abbreviation of `altitude-cognizant \GVSFs{}' in their various forms& \textit{passim} \strutt\\
$\gpower_\alpha(l)$&spherical-harmonic power spectrum of $\Gfun_\alpha(\rvec)$ at the degree $l$; mean-squared value of the coefficients $\sslepfuncoef_{lm,\alpha}$ &(\ref{intonlypowerspec})\strutt\\
$\Gfun_\alpha$& scalar AC-\aGVSF{}; spherical-harmonic expansions of the eigenvectors of the internal-field matrix $\Kmat$&(\ref{intonlypotslepfun})\strutt\\
$\OutGfun_{i\alpha}$& scalar internal-source AC-\aGVSF{}; expansions of internal-source eigenvectors $\OutGmat_i$ of the full-field matrix $\OutKmat$ &(\ref{jo})\\
$\OutGfun_{o\alpha}$&scalar external-source AC-\aGVSF{}; expansions of external-source eigenfunctions $\OutGmat_o$ of the full-field matrix $\OutKmat$ &(\ref{ju})\\
$\Gfunup_\alpha$&vector internal-source AC-\aGVSF{} for the internal-source matrix $\Kmat$, evaluated on the planetary surface $\Earthrad$&(\ref{intonlyvecslepfun})\strutt\\
$\OutGfunup_{i\alpha}$&vector internal-source AC-\aGVSF{} for the full-field matrix $\OutKmat$, evaluated on the planetary surface $\Earthrad$&(\ref{blablablabla})\\
$\OutGfunup_{o\alpha}$&vector external-source AC-\aGVSF{} for the full-field matrix $\OutKmat$, at inner boundary of the external-source field $\Outrad$&(\ref{blablablabla2})\\
$\Gfunup_{\uparrow\alpha}$&vector internal-source AC-\aGVSF{}: the $\Gfunup_\alpha$ after upward-continuation from $\Earthrad$ to average satellite altitude $\satalt$&(\ref{maria})\\
$\OutGfunup_{i\uparrow\alpha}$&vector internal-source AC-\aGVSF{}: the $\OutGfunup_{i\alpha}$ after upward-continuation from $\Earthrad$ to average satellite altitude $\satalt$&(\ref{chit})\\
$\OutGfunup_{o\uparrow\alpha}$&vector external-source AC-\aGVSF{}: the $\OutGfunup_{o\alpha}$ after downward-continuation from $\Outrad$ to average satellite altitude $\satalt$&(\ref{chat})\\
$\OutGfunup_{\uparrow\alpha}$&vector full-field AC-\aGVSF{}, the summed $\OutGfunup_{i\alpha}$ and $\OutGfunup_{o\alpha}$, evaluated at average satellite altitude $\satalt$&(\ref{jbo})\\
$\Gfunvec,\Gfunvec_\nSlepfun$& vectors with the $\Gfun_\alpha$ for $1\leq\alpha \leq \dimin$ or $1\leq\alpha\leq \nSlepfun$; their complement is $\Gfunvec_{>\nSlepfun}$ &(\ref{cox})\strutt\\
$\OutGfunvec_i,\OutGfunvec_{i\nSlepfun}$ & vectors with the $\OutGfun_{i\alpha}$ for $1\leq \alpha\leq \dimin+\dimout$ or $1\leq\alpha\leq \nSlepfun$; their complement is $\OutGfunvec_{i>\nSlepfun}$&(\ref{blo})\\
$\OutGfunvec_o,\OutGfunvec_{o\nSlepfun}$ & vectors with the $\OutGfun_{o\alpha}$ for $1\leq \alpha\leq \dimin+\dimout$ or $1\leq\alpha\leq \nSlepfun$; their complement is $\OutGfunvec_{o>\nSlepfun}$ &(\ref{bli})\\
$\Gfunvecup,\Gfunvecup_\nSlepfun$& vectors with the $\Gfunup_\alpha$ for $1\leq \alpha\leq \dimin$ or $1\leq \alpha\leq \nSlepfun$, complemented by $\Gfunvecup_{>\nSlepfun}$&(\ref{cix})\strutt\\
$\OutGfunvecup_{i},\OutGfunvecup_{i\nSlepfun}$& vectors with the  $\OutGfunup_{i\alpha}$ for $1\leq \alpha\leq \dimin+\dimout$ or $1\leq \alpha\leq \nSlepfun$, complemented by $\OutGfunvecup_{i>\nSlepfun}$&(\ref{cixout})\\
$\OutGfunvecup_{o},\OutGfunvecup_{o\nSlepfun}$&vectors with the $\OutGfunup_{o\alpha}$ for $1\leq \alpha\leq \dimin+\dimout$ or $1\leq \alpha\leq \nSlepfun$, complemented by $\OutGfunvecup_{o>\nSlepfun}$ &(\ref{cixout2})\\
$\Gfunvecup_\uparrow,\Gfunvecup_{\uparrow\nSlepfun}$&vectors with the $\Gfunup_{\uparrow\alpha}$ for $1\leq \alpha\leq \dimin$ or $1\leq \alpha\leq \nSlepfun$, complemented by $\Gfunvecup_{\uparrow>\nSlepfun}$ &(\ref{maria})\\
$\OutGfunvecup_{i\uparrow},\OutGfunvecup_{i\uparrow\nSlepfun}$& vectors with the $\OutGfunup_{i\uparrow\alpha}$ for $1\leq \alpha\leq \dimin+\dimout$ or $1\leq \alpha\leq \nSlepfun$, complemented by $\OutGfunvecup_{i\uparrow>\nSlepfun}$&(\ref{chit})\\
$\OutGfunvecup_{o\uparrow},\OutGfunvecup_{o\uparrow\nSlepfun}$&  vectors with the $\OutGfunup_{o\uparrow\alpha}$ for $1\leq \alpha\leq \dimin+\dimout$ or $1\leq \alpha\leq \nSlepfun$, complemented by $\OutGfunvecup_{o\uparrow>\nSlepfun}$&(\ref{chat})\\
$\Gfunvecup_{\downarrow},\Gfunvecup_{\downarrow\nSlepfun}$& vectors containing downward-continued vector internal-source AC-\aGVSF{}&(\ref{pola})\strutt\\
$\OutGfunvecup_{i\downarrow},\OutGfunvecup_{i\downarrow\nSlepfun}$&vectors containing downward-continued internal-source functions from the set of full-field AC-\aGVSF{}&(\ref{dobi})\\
$\OutGfunvecup_{o\downarrow},\OutGfunvecup_{o\downarrow\nSlepfun}$& vectors containing downward-continued external-source functions from the set of full-field AC-\aGVSF{}&(\ref{doba})\\\hline
$s_{\alpha}$&expansion coefficients of internal-source potential $\Intsignal(\Earthrad\rvec)$ in the scalar AC-\aGVSF{} basis $\Gfun_{\alpha}(\rvec)$&(\ref{otherneedit})\strutt\\
$\outs_{i\alpha}$&expansion coefficients of internal-source potential $\Intsignal(\Earthrad\rvec)$ in the scalar AC-\aGVSF{} basis $\OutGfun_{i\alpha}(\rvec)$ &(\ref{alltheJoutinstuff0})\\
$\outs_{o\alpha}$&expansion coefficients of external-source potential $\Extsignal(\Outrad\rvec)$ in the scalar  AC-\aGVSF{} basis $\OutGfun_{o\alpha}(\rvec)$ &(\ref{alltheJoutinstuff2})\\
$t_\alpha$& sum of the coefficients $\outs_{i\alpha}$ and $\outs_{o\alpha}$ &(\ref{needitalso})\\
$\Slepcoef$&$\dimin$-dimensional vector of expansion coefficients $s_\alpha$, for $1\leq \alpha\leq \dimin$&(\ref{transfo})\strutt\\ 
$\outSlepcoef_i$&$[\dimin+\dimout]$-dimensional vector of coefficients $\outs_{i\alpha}$, for $1\leq \alpha\leq \dimin+\dimout$&(\ref{transfo2})\\
$\outSlepcoef_o$&$[\dimin+\dimout]$-dimensional vector of coefficients $\outs_{o\alpha}$, for $1\leq \alpha\leq \dimin+\dimout$&(\ref{transfo2p})\\
$\OutSlepcoefEarth$&$[\dimin+\dimout]$-dimensional vector of coefficients $\outs_{i\alpha}+\outs_{o\alpha}$&(\ref{needitalso})\\
$\Slepcoef_\nSlepfun$&vector containing the coefficients $s_\alpha$ for $1\leq \alpha\leq \nSlepfun$, approximating  $\Intsignal(\Earthrad\rvec)$ over $\region\subset\Omega$&(\ref{needit})\strutt\\
$\outSlepcoef_{i\nSlepfun}$&vector containing the coefficients $\outs_{i\alpha}$ for $1\leq \alpha\leq \nSlepfun$, approximating $\Intsignal(\Earthrad\rvec)$ over $\region\subset\Omega$&(\ref{alltheJoutinstuff1})\\
$\outSlepcoef_{o\nSlepfun}$&vector containing the coefficients $\outs_{o\alpha}$ for $1\leq \alpha\leq \nSlepfun$, approximating  $\Extsignal(\Outrad\rvec)$ over $\region\subset\Omega$&(\ref{needit2})\\
$\OutSlepcoefEarth_\nSlepfun$&vector containing the coefficients $t_\alpha$ for $1\leq \alpha\leq \nSlepfun$&(\ref{needitalso})\\
$\estSlepcoef_{\nSlepfun}$&truncated estimator of the $\nSlepfun$ first terms in the coefficient vector $\Slepcoef$, using $\nSlepfun$ terms of the inner-source AC-\aGVSF{} &(\ref{linearsysteminnerregularized})\strutt\\
$\estOutSlepcoef_\nSlepfun$&truncated estimator of the $\nSlepfun$ first terms in the coefficient vector $\OutSlepcoefEarth$, using $\nSlepfun$ terms of the full-field AC-\aGVSF{}   &(\ref{linearsysteminoutregularized})\\
$\sestSlepcoef_\alpha$& an element of $\estSlepcoef_{\nSlepfun}$; expansion coefficient of the truncated estimate  $\estIntsignal_\nSlepfun(\Earthrad\rvec)$ in the AC-\aGVSF{} basis $\Gfun_{\alpha}(\rvec)$&(\ref{slepianfunctionexpansioninner})\\
$\sestOutSlepcoef_{\alpha}$&an element of $\estOutSlepcoef_\nSlepfun$; expansion coefficient of the estimate $\estIntsignal_\nSlepfun(\Earthrad\rvec)$ in the internal-source AC-\aGVSF{} basis $\OutGfun_{i\alpha}(\rvec)$;&(\ref{slepianfunctionexpansionin})\\& 
also: expansion coefficient of the estimate $\estExtsignal_\nSlepfun(\Outrad\rvec)$ in the external-source AC-\aGVSF{} basis $\OutGfun_{o\alpha}(\rvec)$&(\ref{slepianfunctionexpansionout})\\
$\estIntsignal_\nSlepfun(\Earthrad\rvec)$&estimate of the internal-source field $\Intsignal$, in the truncated AC-\aGVSF{} basis, considering internal-source fields; &(\ref{slepianfunctionexpansioninner})\strutt\\
& estimate of the internal-source field $\Intsignal$, in the truncated AC-\aGVSF{} basis, considering internal/external fields &(\ref{slepianfunctionexpansionin})\\
$\estExtsignal_\nSlepfun(\Outrad\rvec)$& estimate of the external field $\Extsignal$, in the truncated AC-\aGVSF{} basis, considering internal and external fields&(\ref{slepianfunctionexpansionout})\\\hline
$\npoints$&number of vector-valued observations available: a discrete data set on $\region\subset\Omega$; for a total of $3\npoints$ components&(\ref{elmic})\strutt\\
$\dpoints$& vector with the three spherical vector components of the field observed at each of the positions $r_i\rvec_i$,  $1\leq i\leq\npoints$&(\ref{inlinsyspoints})\\
$\Epoints_\Uparrow$&$[\dimin\times 3\npoints]$-dimensional matrix with the  upward-continued $\Efun_{lm}$ harmonics evaluated at the data &(\ref{elmic})\\
$\Fpoints_\Uparrow$&$[\dimin\times 3\npoints]$-dimensional matrix with the downward-continued $\Ffun_{lm}$ harmonics at the data locations &(\ref{flmic})\\
$\GpointsUPJ$&$[\nSlepfun \times 3\npoints]$-dimensional matrix containing the internal-source AC-\aGVSF{} $\Gfunvecup_{\uparrow\nSlepfun}$ evaluated at the data locations&(\ref{yup})\\
$\OutGpointsUPJ$&$[\nSlepfun \times 3\npoints]$-dimensional matrix containing the full-field AC-\aGVSF{}  $\OutGfunvecup_{\uparrow\nSlepfun}$ evaluated at the data locations&(\ref{biz})\\
$\estSlepcoefi_\nSlepfun$&discrete-data truncated estimator of the $\nSlepfun$ first terms in the coefficient vector $\Slepcoef$; the $\sestSlepcoefi_{\alpha}$ for $1\leq\alpha\leq\nSlepfun$&(\ref{inlinsyspoints})\strutt\\
$\estOutSlepcoefi_\nSlepfun$&discrete-data truncated estimator of the $\nSlepfun$ first terms in the coefficient vector $\OutSlepcoefEarth$; the $\sestOutSlepcoefi_{\alpha}$ for $1\leq\alpha\leq\nSlepfun$&(\ref{bizbiz})\\
$\sestSlepcoefi_\alpha$&an element of $\estSlepcoefi_\nSlepfun$; expansion coefficient of $\estIntsignali_\nSlepfun(\Earthrad\rvec)$ in the AC-\aGVSF{} basis $\Gfun_{\alpha}(\rvec)$&(\ref{inestpotfield})\\
$\sestOutSlepcoefi_\alpha$&an element of $\estOutSlepcoefi_\nSlepfun$; expansion coefficient of $\estIntsignali_\nSlepfun(\Earthrad\rvec)$ and $\estExtsignali_\nSlepfun(\Outrad\rvec)$ in the AC-\aGVSF{} basis $\OutGfun_{i\alpha}(\rvec)$ and $\OutGfun_{o\alpha}(\rvec)$&(\ref{extfieldexpansion})\\
$\respoints$&vector with  data residuals obtained after comparison with predictions derived from the truncated estimator $\estSlepcoefi_\nSlepfun$  &(\ref{respoints1})\\
$\mathring{\respoints}$& vector with data residuals obtained after comparison with predictions derived from the truncated estimator $\estOutSlepcoefi_\nSlepfun$&(\ref{theres})\\
$\resmat$&$[3\npoints\times 3\npoints]$-dimensional diagonal matrix containing (thresholded) residuals $\respoints$ &(\ref{initerativeReweight})\\
$\mathring{\resmat}$&$[3\npoints\times 3\npoints]$-dimensional diagonal matrix containing (thresholded) residuals $\mathring{\respoints}$&(\ref{reweightedresidual2})\\
$\estIntsignali_\nSlepfun(\Earthrad\rvec)$& truncated-AC-\aGVSF{} estimate of the internal-source field $\Intsignal$, variable-altitude discrete data, inner-source fields; &(\ref{inestpotfield})\strutt\\
&truncated-AC-\aGVSF{}  estimate of internal-source field $\Intsignal$, variable-altitude discrete data, internal/external fields &(\ref{intfieldexpansion})\\
$\estExtsignali_\nSlepfun(\Outrad\rvec)$& truncated-AC-\aGVSF{} estimate of external field $\Extsignal$, from variable-altitude discrete data, internal/external fields&(\ref{extfieldexpansion})\\\hline
$\Hfuno_{\uparrow \alpha, >\Lin}$&spacelimited, infinitely-wideband AC-\aGVSF{} made from the degrees $l>\Lin$ after truncation of the $\Gfunup_{\uparrow \alpha}$ to $\region$&(\ref{InUpTruncRelation})\strutt\\
$\Hfunvec_{\uparrow J,>\Lin}$&vector collecting the $\Hfuno_{\uparrow \alpha, >\Lin}$ for $1\leq \alpha \leq \nSlepfun$&(\ref{InUpTruncRelation})\\
$\vslepfuncoef_{\uparrow \alpha,>\Lin}$&vector containing the $\Efun_{lm}$ expansion coefficients of the spacelimited $\Hfuno_{\uparrow \alpha J,>\Lin}$ &(\ref{inuptrunc})\\
$\Hfunor^{\Efunvec}_{\uparrow\alpha,>\Lin}$&spacelimited, wideband AC-\aGVSF{} made from the $\Efun_{lm}$ components, for $l>\Lin$, of the $\OutGfunup_{\uparrow \alpha}$ truncated to $\region$&(\ref{Hrelationsint})\strutt\\
$\OutHfunvec^{\Efunvec}_{\uparrow\nSlepfun,>\Lin}$&vector collecting the $\Hfunor^{\Efunvec}_{\uparrow\alpha,>\Lin}$ for $1\leq\alpha\leq \nSlepfun$&(\ref{Hrelationsint})\\
$\Outvslepfuncoef^{\Efunvec}_{\uparrow\alpha,>\Lin}$&vector containing the $\Efun_{lm}$ expansion coefficients of the spacelimited $\Hfunor^{\Efunvec}_{\uparrow \alpha J,>\Lin}$ &(\ref{blipE})\strutt\\
$\Hfunor^{\Ffunvec}_{\uparrow\alpha,>\Lout}$&spacelimited, wideband AC-\aGVSF{} made from the $\Ffun_{lm}$ components, for  $l>\Lin$, of the $\OutGfunup_{\uparrow \alpha}$ truncated to $\region$&(\ref{Hrelationsout})\strutt\\
$\OutHfunvec^{\Ffunvec}_{\uparrow\nSlepfun,>\Lout}$&vector collecting the $\Hfunor^{\Ffunvec}_{\uparrow\alpha,>\Lout}$ for $1\leq\alpha\leq \nSlepfun$&(\ref{Hrelationsout})\strutt\\
$\Outvslepfuncoef^{\Ffunvec}_{\uparrow\alpha,>\Lout}$&vector containing the $\Ffun_{lm}$ expansion coefficients of the spacelimited $\Hfunor^{\Ffunvec}_{\uparrow \alpha J,>\Lin}$&(\ref{blipF})\\\hline
$b$&spatial bias of the truncated AC-\aGVSF{} estimator $\estIntsignal_\nSlepfun(\Earthrad\rvec)$ considering internal-source fields only&(\ref{knot})\strutt\\
$b_\Intsignal$&first portion of the spatial bias of the truncated AC-\aGVSF{} estimator $\estIntsignal_\nSlepfun(\Earthrad\rvec)$ solving the full-field problem&(\ref{bibiint1})\\
$b_{\Intsignal\Extsignal}$&second portion of the spatial bias of the truncated AC-\aGVSF{} estimator $\estIntsignal_\nSlepfun(\Earthrad\rvec)$ solving the full-field problem&(\ref{bibiint2})\\
$b_\Extsignal$&first portion of the spatial bias of the truncated AC-\aGVSF{} estimator $\estExtsignal_\nSlepfun(\Outrad\rvec)$ solving the full-field problem&(\ref{bibiext1})\\
$b_{\Extsignal\Intsignal}$&second portion of the spatial bias of the truncated AC-\aGVSF{} estimator $\estExtsignal_\nSlepfun(\Outrad\rvec)$ solving the full-field problem&(\ref{bibiext2})\\
$\sigpower$&spatial two-point covariance of the scalar internal-source signal $\Intsignal(\Earthrad\rvec)$&(\ref{pipi})\strutt\\
$\sigpowerw$&spatial two-point covariance of the scalar external-source signal $\Extsignal(\Outrad\rvec)$ &(\ref{papa2})\\
$\noisepower$&spatial two-point covariance of the vector-valued data noise $\vecnoise(\satalt\rvec)$ &(\ref{popo})\\
$\mathrm{mse}$&mean-squared estimation error of the truncated AC-\aGVSF{} estimate $\estIntsignal_\nSlepfun(\Earthrad\rvec)$ for the internal-source field &(\ref{mse})\strutt\\
$\mathrm{mse}_\Intsignal$&mean-squared error of the truncated AC-\aGVSF{} internal-source estimate $\estIntsignal_\nSlepfun(\Earthrad\rvec)$  solving the full-field problem&(\ref{mseV})\\
$\mathrm{mse}_\Extsignal$&mean-squared error of the truncated AC-\aGVSF{} external-source estimate $\estExtsignal_\nSlepfun(\Outrad\rvec)$  solving full-field problem&(\ref{mseW})\\
\end{longtable}
\end{center}

\end{document}